\documentclass{aa}  

\usepackage[switch]{lineno}
\usepackage{graphicx}
\usepackage{natbib}
\usepackage{txfonts}
\usepackage[draft]{hyperref}

\usepackage[usenames]{color}

\newcommand{\bcep}{$\beta$~Cep }
\newcommand{\dsct}{$\delta$~Sct }
\newcommand{\gdor}{$\gamma$~Dor }
\newcommand{\Kepler}{\textit{Kepler} }

\begin{document} 


   \title{K2 space photometry reveals rotational modulation and stellar pulsations in chemically peculiar A and B stars}
   \titlerunning{K2 photometry of chemically peculiar A and B stars}

   \author{D. M. Bowman \inst{1} 
          \and
          B. Buysschaert \inst{1,2}
          \and
          C. Neiner \inst{2}
          \and
          P. I. P{\'a}pics \inst{1}
          \and
          M. E. Oksala\inst{3,2}
          \and
          C. Aerts \inst{1,4}
          }

    \institute{Instituut voor Sterrenkunde, KU Leuven, Celestijnenlaan 200D, 3001 Leuven, Belgium \\
              \email{dominic.bowman@kuleuven.be} 
         \and
                LESIA, Observatoire de Paris, PSL Research University, CNRS, Sorbonne Universit{\'e}s, UPMC Univ. Paris 06, Univ. Paris Diderot, Sorbonne Paris Cit{\'e}, 5 place Jules Janssen, F-92195 Meudon, France
        \and
                Department of Physics, California Lutheran University, 60 West Olsen Road 3700, Thousand Oaks, CA, 91360, USA
        \and
                Department of Astrophysics, IMAPP, Radboud University Nijmegen, NL-6500 GL Nijmegen, The Netherlands
         }

   \date{Received March, 19 2018; accepted May 4, 2018}

 
  \abstract
   {The physics of magnetic hot stars and how a large-scale magnetic field affects their interior properties is largely unknown. Few studies have combined high-quality observations and modelling of magnetic pulsating stars, known as magneto-asteroseismology, primarily because of the dearth of detected pulsations in stars with a confirmed and well-characterised large-scale magnetic field.}
   {We aim to characterise observational signatures of rotation and pulsation in chemically peculiar candidate magnetic stars using photometry from the K2 space mission. Thus, we identify the best candidate targets for ground-based, optical spectropolarimetric follow-up observations to confirm the presence of a large-scale magnetic field.}
   {We employed customised reduction and detrending tools to process the K2 photometry into optimised light curves for a variability analysis. We searched for the periodic photometric signatures of rotational modulation caused by surface abundance inhomogeneities in 56 chemically peculiar A and B stars. Furthermore, we searched for intrinsic variability caused by pulsations (coherent or otherwise) in the amplitude spectra of these stars.}
   {The rotation periods of 38 chemically peculiar stars are determined, 16 of which are the first determination of the rotation period in the literature. We confirm the discovery of high-overtone roAp pulsation modes in HD~177765 and find an additional 3 Ap and Bp stars that show evidence of high-overtone pressure modes found in roAp stars in the form of possible Nyquist alias frequencies in their amplitude spectra. Furthermore, we find 6 chemically peculiar stars that show evidence of intrinsic variability caused by gravity or pressure pulsation modes.}
   {The discovery of pulsations in a non-negligible fraction of chemically peculiar stars make these stars high-priority targets for spectropolarimetric campaigns to confirm the presence of their expected large-scale magnetic field. The ultimate goal is to perform magneto-asteroseismology and probe the interior physics of magnetic pulsating stars.}


   \maketitle


\section{Introduction}
\label{section: intro}

Main-sequence stars of spectral type A and B are possibly the most diverse group of stars in the Hertzsprung--Russell (HR) diagram, as they exhibit many different aspects of physics including rotation, binarity, pulsation, and the possible presence of a large-scale magnetic field. The synergy of these properties have been studied to varying levels of success, yet the interaction of physical processes in the deep radiative interiors of magnetic hot stars remain largely unknown even though they are important for stellar structure and evolution \citep{Maeder_rotation_BOOK, Meynet_rotation_BOOK}. The shortcomings in our knowledge of how a large-scale magnetic field affects the internal properties of a star, such as rotation and convective core overshooting, are largest for stars on the upper main sequence, which consequently produce large uncertainties in theoretical models and need to be mitigated. 

The study of multi-periodic stellar pulsations is known as asteroseismology and represents a unique methodology for studying the interior physics of a star using its pulsations. The two main types of pulsation modes in stars of spectral type A and B are gravity (g) modes and pressure (p) modes, which are most sensitive to the near-core and near-surface regions in a star, respectively \citep{ASTERO_BOOK}. Among the B stars are the beta Cephei ($\beta$~Cep) and slowly pulsating B (SPB) stars, which predominantly pulsate in p- and g-modes, respectively \citep{ASTERO_BOOK}. Similarly, further down the main sequence amongst the A and F stars are the delta~Scuti ($\delta$~Sct) and gamma~Doradus ($\gamma$~Dor) stars, which also predominately pulsate in p- and g-modes, respectively. \citet{Baglin1973a}, \citet{Waelkens1991c}, \citet{Breger1996b}, \citet{Handler1999b}, \citet{Breger2000b}, \citet{Guzik2000a}, \citet{Rod2001}, \citet{ASTERO_BOOK}, and \citet{Bowman_BOOK} provide reviews of pulsations in B, A, and F stars. From high-quality continuous photometric data sets, such as those provided by the CoRoT \citep{Baglin2006, Auvergne2009} and \Kepler \citep{Borucki2010, Koch2010} space telescopes, individual pulsation modes can be identified and used to probe the largely unknown interior physics of pulsating stars on the upper main sequence. 

The coherent pulsations in early-type stars are typically driven by a heat-engine (opacity) operating in partial ionisation zones and/or a flux blocking driving mechanism, although recent work has shown that other types of mode excitation are also possible \citep{Neiner2012d, Antoci2014b, Houdek2015, Xiong2016a}. For example, stochastically excited gravito-inertial modes have been observed in the early-B star HD~51452 \citep{Neiner2012d}, and this type of mode excitation is likely important for rapidly rotating early-type stars \citep{Mathis2014}. Also, evidence for stochastically excited gravity waves has been detected in a handful of O stars \citep{Aerts2015c, Aerts2017a, Aerts2018a, Simon-Diaz2018a}, which are predicted to effectively distribute angular momentum and provide a constraint of interior rotation within a star \citep{Rogers2013b, Rogers2015}. 

Approximately 10$\%$ of intermediate- and high-mass stars host a detectable large-scale magnetic field at their surface \citep{Power_MASTER, Power2007, Grunhut2015b, Neiner2015d, Wade2016a, Villebrun2016, Grunhut2017, Sikora2018a}. The magnetic field in these stars typically resembles a dipole inclined to the rotation axis, which is known as the oblique rotator model \citep{Stibbs1950}. The total number of magnetic field detections continues to increase thanks to dedicated observing campaigns using high-resolution and high signal-to-noise ($S/N$) optical spectropolarimetry --- for example, MiMeS \citep{Wade2016a}, the BOB campaign \citep{Morel2015b}, and the BRITE spectropolarimetric survey \citep{Neiner2017b}.

Currently, how a large-scale magnetic field affects the physical processes deep within a star's interior, such as convective core overshooting and the radial chemical mixing and rotation profiles, is poorly constrained using observations. To date, only a dozen pulsating magnetic upper main-sequence stars are known \citep{Buysschaert2017e}, and only three of these have been studied using forward seismic modelling -- i.e. magneto-asteroseismology -- specifically, \bcep \citep{Shibahashi2000b, Henrichs2013}, V2052~Oph \citep{Neiner2012a, Handler2012a, Briquet2012}, and HD~43317 \citep{Papics2012a, Buysschaert2017b, Buysschaert2018b*}. This is primarily because of a lack of high-precision and long-term photometric observations of confirmed magnetic stars, which are essential to resolve and extract individual pulsation mode frequencies in pulsating stars and perform forward seismic modelling. 

Theory and numerical simulations predict that the internal component of a sufficiently strong large-scale magnetic field should instigate uniform rotation within the radiative layers of early-type stars \citep{Moss1992b, Browning2004d, Browning2004a, Mathis2005a, Zahn2011a}. On the other hand, uniform to weak differential rotation was determined using asteroseismology for 67 intermediate-mass stars \citep{Aerts2017b}; the majority of these stars are not known to host a (detectable) large-scale magnetic field, which suggests that other physical processes also lead to (quasi-)rigid rotation within early-type stars. Additionally, theoretical models predict that a large-scale magnetic field affects and can even suppress the excitation of waves (e.g. \citealt{Saio2005, Saio2014a, Lecoanet2017a}). The exceptions to this are the high-overtone p-mode pulsations observed in the rapidly oscillating Ap stars. The uniform rotation and a large-scale magnetic field are also predicted by theory to lead to a smaller convective core overshooting region (e.g. \citealt{Press1981a, Browning2004d, Browning2004a}). This has been inferred observationally for the \bcep star V2052~Oph \citep{Briquet2012} by comparing the overshooting value with that of the non-magnetic \bcep star $\theta$~Oph \citep{Briquet2007e, Briquet2012}. 

Similarly, from the forward seismic modelling of the magnetic SPB star HD~43317, \citet{Buysschaert2018b*} were able to constrain convective core overshooting to relatively small values, yet the lack of g-mode period spacing pattern spanning a large range of consecutive radial orders could not exclude moderate values of convective core overshooting that have been found for non-magnetic B stars \citep{Moravveji2015b, Moravveji2016b}. Clearly, additional observational studies of pulsating magnetic stars are needed to constrain the interior physics of these stars.


Chemically peculiar (CP) stars on the upper main sequence have been historically grouped into different categories (e.g. \citealt{Sargent1964, Jaschek1974a, Preston1974, Wolff1983a, Smith_K_1996}), in which the presence of the observed chemical peculiarities is often, but not always, related to the presence of a large-scale magnetic field \citep{Michaud1970, Vauclair1979}. The various flavours of CP stars on the upper main sequence were condensed into four main groups, CP\,1 to CP\,4, by \citet{Preston1974}.

The CP\,1 stars (also known as Am stars) were first characterised by \citet{Titus1940} and later by \citet{Smith_M_1971a, Smith_M_1971b} as slowly rotating non-magnetic A stars with a difference of at least five spectral subclasses between their Ca~K and metallic line strengths. The Am stars are common with approximately 50$\%$ of stars of spectral type A8 classified as Am \citep{Smith_M_1973}. Stars that show less than five subclasses between their Ca~K and metallic line strengths are classified as marginal Am (Am:) stars \citep{Cowley1969, Kurtz1978b}, since they are significantly different from chemically normal A stars. The majority of Am stars are typically found in short-period ($1 \leq P_{\rm orb} \leq 10$~d) binary systems \citep{Roman1948, Abt1961, Abt1967, Abt1985c, Smalley2014}, which is significantly higher than the binary fraction of chemically normal A stars \citep{Abt2009, Duchene2013b, Moe2017a}. The short orbital periods are sufficient to tidally brake Am stars into being slow rotators and allow gravitational settling and radiative diffusion, and different metallic species thereby rise from radiative levitation or sink from gravity \citep{Breger1970, Baglin1973a, Vauclair1979, Turcotte2000c}. This leads to transition metals forming clouds near to the stellar surface, which are observed as overabundance anomalies \citep{Breger1970, Kurtz2000b}.

The CP\,2 stars, also known as Ap and Bp (or collectively known as ApBp) stars, constitute approximately 10~per~cent of all A and B stars \citep{Wolff1968a}. The ApBp stars have spectroscopic overabundances of elements such as Cr, Eu, Si, and Sr, which can be as large as 10$^{6}$ times solar values. These stars are also known to host large-scale magnetic fields, which range from a few hundred Gauss (e.g. \citealt{Auriere2004}) to as large as 34~kG (e.g. HD~215441; \citealt{Babcock1960}), and are responsible for the observed chemical peculiarities. The magnetic field in an ApBp star is typically inclined to the rotation axis, which is known as the oblique rotator model; the angle between the magnetic and rotation axes is known as the angle of obliquity \citep{Stibbs1950}.

The first studies of binarity amongst ApBp stars were carried out by \citet{Abt1973a}, who found that approximately $20\%$ of ApBp stars were in binary systems. They also found a dearth of ApBp stars in close binary systems, yet a binary fraction for wide binary systems that is typical of normal A and B stars. Later works by \citet{Gerbaldi1985}, \citet{North1998a}, \citet{Carrier2002f}, \citet{Mathys2017a}, and \citet{Landstreet2017a} have confirmed the dearth of ApBp stars in close-binary systems, which are typically defined as orbital periods less than 3~d. The only known exception being HD~200405, which has a binary orbital period of 1.635~d \citep{Carrier2002f}.

The CP\,3 stars, also known as HgMn stars, are non-magnetic stars with enhanced Hg and Mn absorption lines in their spectra. However CP\,4 stars, also known as He-weak stars, have weaker {\sc He~I} lines than expected for their photometric {\it UBV} colours and are thought to be an extension of the CP\,2 and CP\,3 stars to higher temperatures in the HR~diagram, since they also host a large-scale magnetic field and show strong Si, Ti, and Sr absorption lines in their spectra \citep{Sargent1964, Preston1974}. Another subgroup within CP stars are He-strong stars, which also host a large-scale magnetic field \citep{Borra1979}.

Magnetic CP stars are typically slow rotators with rotation periods of order a day, but can be as long as a few hundred years \citep{Mathys2015}; the fastest rotators have rotation periods of about 0.5~d (e.g. \citealt{Adelman2002d, Mathys2004a}). This distinctly different distribution in the rotation periods of magnetic stars compared to non-magnetic A and B stars is caused by magnetic braking during the pre-main-sequence contraction phase \citep{Abt1995, Stepien2000}. Although a large range spanning several orders of magnitude exists in the rotation periods of magnetic stars, the change in the rotation period of a star is relatively small during the main-sequence stage of evolution, so the magnetic field must be generated prior to the main sequence \citep{Kochukhov2006b, Neiner2015d, Villebrun2016, Alecian2017a*}. The slow rotation and a strong large-scale magnetic field cause atomic diffusion and stratification in the atmosphere, which lead to surface abundance inhomogeneities. For such rotationally variable CP stars (known as $\alpha^{2}$~CVn stars) and/or stars with a detectable magnetosphere (e.g. \citealt{Shultz2018b}), the rotation period can be determined from the rotational modulation observed in the light curve \citep{Stibbs1950}. These surface abundance inhomogeneities and the topology of the large-scale magnetic field can be studied in great detail using tomographic imaging techniques (see e.g. \citealt{Kochukhov2014a, Kochukhov2015a, Kochukhov2016a, Kochukhov2017a, Oksala2017a}).


Rapidly oscillating Ap (roAp) stars are a rare subgroup of Ap stars that were first discovered by \citet{Kurtz1978a, Kurtz1982}. Currently, only 61 confirmed roAp stars are known, which represents a few per cent of all the known Ap stars \citep{Smalley2015, Joshi2016}. Similarly to Ap stars, roAp stars are typically not found in close-binary or multiple systems (e.g. \citealt{Scholler2012}). These CP magnetic stars pulsate in high-overtone p~modes with periods between 5 and 24 min \citep{Kurtz1982, Kurtz1990b, Martinez1994d, Cunha2002d, Kurtz2007d, Alentiev2012, Smalley2015, Joshi2016} and have photometric peak-to-peak amplitudes as large as 34~mmag in the Johnson $B$ filter \citep{Kurtz1990b, Holdsworth2018a}. Most of the known roAp stars were discovered using high-cadence ground-based photometry, although some were detected using spectroscopic radial velocity studies (e.g. \citealt{Kochukhov2002e, Elkin2005b}).

Unlike chemically normal pulsating A stars, i.e. \dsct stars, the strong magnetic field in an Ap star is believed to suppress the excitation of low-overtone p~modes \citep{Saio2005}. The exact nature of the excitation of the high-overtone p~modes in roAp stars is not known, but both the $\kappa$-mechanism operating in the hydrogen ionisation zone \citep{Cunha2002a} and turbulent pressure are thought to be responsible \citep{Cunha2013}. The pulsation symmetry axis of a roAp star is nearly aligned with the magnetic symmetry axis, both of which are inclined to the rotation axis. This is known as the oblique pulsator model and allows the pulsations to be viewed from various orientations as the star rotates \citep{Kurtz1982, Dziembowski1985b, Shibahashi1993, Takata1995b, Dziembowski1996b, Bigot2000, Bigot2002c, Bigot2011b}. 

With so few pulsating ApBp stars known and fewer still with high-quality continuous photometry, more observational studies searching for intrinsic variability in these stars are needed. In this paper, we perform a detailed study of {56} CP stars using high-precision photometry obtained by the K2 space mission \citep{Howell2014}. This statistically significant number of stars allows us to determine previously unknown rotation periods and carry out a systematic search for variability caused by stellar pulsations (coherent modes and/or travelling waves). Our goal is to identify pulsating stars for follow-up with spectropolarimetry to confirm the presence of a large-scale magnetic field and to perform magneto-asteroseismology to probe the physics of their interiors.


\section{Observations of CP stars by K2 }
\label{section: observations}

Typically, a star is classified to host a large-scale magnetic field if the Zeeman signature is detected in spectropolarimetric observations using a false alarm probability (FAP) criterion \citep{Donati1992b, Donati1997c}. Also, strong large-scale magnetic fields can induce detectable Zeeman splitting of various spectral absorption lines (e.g. \citealt{Mathys2017a}), whereas stellar pulsations are detected using high-precision (and ideally high-cadence) photometric or spectroscopic time series (see chapter~4 from \citealt{ASTERO_BOOK}). The CP\,2 and 4 stars are expected to host a large-scale magnetic field, but before the advent of space photometry, few CP stars had high-quality uninterrupted photometric observations available to investigate their intrinsic variability. 

Recently, \citet{Buysschaert2018a*} analysed a subset of 16 CP stars using high-resolution and high-$S/N$ ground-based optical spectropolarimetry combined with photometry from the K2 space mission. This analysis led to the confirmation of a large-scale magnetic field in 12 of these 16 stars,  10 stars of which are the first such detection of a large-scale magnetic field. In this work, we study a larger sample of CP stars observed by the K2 space mission and identify the best candidate pulsating stars for ground-based spectropolarimetric follow-up campaigns to confirm the presence (or absence) of a large-scale magnetic field.


        \subsection{Target selection}
        \label{subsection: target}
        
        We compiled a sample of {50} ApBp (CP\,2), 1 HgMn (CP\,3) star, {2} He weak (CP\,4) stars using the catalogue of \citet{Renson2009} and an additional 3 CP stars identified by \citet{Crawford_D_1955} and \citet{McCuskey1967}, for which space photometry from Campaigns 00--12 of the K2 mission was requested. The total of {56} stars within our sample are listed in Table~\ref{table: star names}, which provides the K2 EPIC identification and HD numbers, the $B$ and $V$ band magnitudes from SIMBAD\footnote{SIMBAD website: \url{http://simbad.u-strasbg.fr/simbad/}}, and the spectral type listed in \citet{Renson2009}. The star EPIC~202061263 is not resolved using SIMBAD, so its $B$- and $V$-mag values have been extracted using its coordinates ($\alpha = 06\,03\,08.82$, $\delta = +23\,47\,23.71$; J2000.0).

\begin{table}
\caption{Fifty-six CP stars in our sample with $B$- and $V$-mag values from SIMBAD and spectral type from \citet{Renson2009}.} 
\begin{center}
\begin{tabular}{l r r r l}
\hline \hline
\multicolumn{1}{c}{K2 EPIC ID} & \multicolumn{1}{c}{HD} & \multicolumn{1}{c}{$B$~mag} & \multicolumn{1}{c}{$V$~mag} & \multicolumn{1}{c}{Sp.\,Type} \\
\hline
201667495       &       107000  &       8.21            &       8.02            &       A2 Sr              \\
201777342       &       97859   &       9.24            &       9.35            &       B9 Si              \\
{202060145}\tablefootmark{a}    &       $-$             &       12.00   &       11.30   &       B9 {\sc IV}\,p     \\
{202061199}\tablefootmark{a}    &       255134  &       9.46            &       9.18            &       B1 {\sc IV}\,p     \\ 
{202061263}\tablefootmark{b}    &       $-$             &       13.4            &       11.8            &       F2 p                       \\
203092367       &       150035  &       8.92            &       8.71            &       A3 CrEuSr  \\
203749199       &       152366  &       8.11            &       8.08            &       B8 Si              \\
203814494       &       142990  &       5.34            &       5.43            &       B6 He weak \\
203917770       &       145792  &       6.45            &       6.41            &       B6 He weak \\
204964091       &       147010  &       7.56            &       7.40            &       B9 SiCrSr          \\
206026652       &       215766  &       5.65            &       5.68            &       B9 Si              \\
206120416       &       210424  &       5.30            &       5.42            &       B6 Si              \\
206326769       &       211838  &       5.29            &       5.34            &       B8 MgMn            \\
210964459       &       26571   &       6.31            &       6.12            &       B8 Si              \\
213786701       &       173657  &       7.48            &       7.41            &       B9 SiCr            \\
214133870       &       177562  &       7.39            &       7.38            &       B8 Si              \\
214197027       &       173857  &       10.44   &       10.40   &       A0 SiCr            \\
214503319       &       177765  &       9.60            &       9.15            &       A5 SrEuCr  \\
214775703       &       178786  &       10.20   &       10.24   &       A0 Si              \\
215357858       &       174356  &       9.37            &       9.18            &       B9 Si              \\
215431167       &       174146  &       10.56   &       10.03   &       F Si                      \\
215584931       &       172480  &       10.22   &       10.06   &       A0 Si              \\
215876375       &       184343  &       10.10   &       9.74            &       A5 SrCrEu  \\
216005035       &       182459  &       9.52            &       9.59            &       A0 Si              \\
216956748       &       181810  &       10.65   &       10.66   &       A0 EuCrSr  \\
217323447       &       180303  &       9.50            &       9.48            &       A0 SrCrEu  \\
217437213       &       177016  &       9.67            &       9.34            &       F0 EuSrCr  \\
218676652       &       173406  &       7.51            &       7.43            &       B9 Si              \\
218818457       &       176330  &       10.54   &       10.39   &       A0 Si              \\
219198038       &       177013  &       9.34            &       9.04            &       A2 EuCrSr  \\
219353144       &       181004  &       9.54            &       9.64            &       B9 Si              \\
223573464       &       161851  &       8.72            &       8.61            &       A0 Si              \\
224206658       &       165972  &       9.02            &       8.96            &       B9 Si              \\
224332430       &       162759  &       10.41   &       10.10   &       A0 Si              \\
224351176       &       167700  &       9.87            &       9.54            &       A0 SrEu            \\
224487047       &       165321  &       9.13            &       8.97            &       A0 Si              \\
224947037       &       162814  &       10.86   &       10.24   &       A2 Si              \\
225191577       &       164068  &       10.04   &       9.90            &       A0 Si              \\
225382260       &       161342  &       8.95            &       8.66            &       B8 Si              \\
225990054       &       158596  &       9.17            &       8.94            &       B9 Si              \\
226097699       &       166190  &       10.01   &       9.81            &       B9 Si              \\
226241087       &       164224  &       8.67            &       8.49            &       B9 CrEu            \\
227108971       &       164190  &       9.37            &       9.13            &       B9 Si              \\
227231984       &       158336  &       9.49            &       9.36            &       B9 Si              \\
227305488       &       166542  &       9.92            &       9.94            &       A0 Si              \\
227365417       &       168187  &       10.06   &       10.02   &       B9 Si              \\
227373493       &       166804  &       8.84            &       8.88            &       B9 Si              \\
227825246       &       164085  &       10.42   &       10.19   &       A0 Si              \\
228293755       &       165945  &       9.67            &       9.41            &       A5 SrEuCr  \\
230753303       &       153997  &       9.56            &       9.50            &       A0 Si              \\
232147357       &       153192  &       10.12   &       10.08   &       A2 EuCr            \\
232176043       &       152834  &       8.85            &       8.83            &       A0 Si              \\
232284277       &       155127  &       8.59            &       8.38            &       B9 EuCrSr  \\
246016562       &       219831  &       10.20   &       9.99            &       A2 Sr              \\
246152326       &       220556  &       10.20   &       9.85            &       A2 SrEuCr  \\
247729177       &       284639  &       10.13   &       9.68            &       A0 SiCr            \\
\hline \hline
\end{tabular}
\end{center}
\tablefoot{Alternate literature spectral types.}
\tablefoottext{a}{\citet{Crawford_D_1955},}
\tablefoottext{b}{\citet{McCuskey1967}}
\label{table: star names}
\end{table}


        \subsection{K2 photometry}
        \label{subsection: K2}
        
        The nominal \Kepler mission was originally designed to achieve a primary goal of observing transits of Earth-like planets orbiting Sun-like stars \citep{Borucki2010}. Observations are available in two modes: long cadence (LC; 29.5~min) and short cadence (SC; 58.5~s). Only about 500 stars can  be observed in SC at any given epoch \citep{Gilliland2010}. The primary 4 yr \Kepler mission came to an end in May 2013 after the spacecraft lost the ability to use three reaction wheels, which were necessary to maintain the pointing of the telescope. An ingenious solution was to redefine the mission parameters and point the telescope in the direction of the ecliptic where torque forces caused by solar radiation pressure are minimised \citep{Howell2014}. The new mission, K2, has provided high-quality photometry of many different aspects of astronomy, and the data from this mission are grouped into various campaigns that are each approximately 80~d in length \citep{Howell2014}. The K2 mission has proven extremely valuable for investigating pulsating stars, including some of the most massive stars (e.g. \citealt{Buysschaert2015, Johnston2017a, White2017b, Aerts2018a}), stars at different stages of evolution (e.g. \citealt{Kurtz2016a, Hermes2017e}), and stars in clusters (e.g. \citealt{Lund2016a, Stello2016c}).

        \subsubsection{K2 data reduction}
        \label{subsubsection: data reduction}

        We downloaded the K2 target pixel files (TPF), which are available from Mikulski Archive for Space Telescopes (MAST\footnote{MAST website: \url{http://archive.stsci.edu/kepler/}}), to investigate the photometric variability of {56} CP stars. For each available K2 (sub-)campaign for each star, we determined an optimum (non-circular) aperture by stacking all TPFs and including the pixels that capture the flux of the target star during a campaign. This is a necessary step since a target star moves on the telescope CCD because of the (quasi-)periodic K2 thruster firing events that occur approximately every 6~hr \citep{Howell2014}.
        
        The resultant light curve for each (sub-)campaign was corrected for the mean background flux and detrended using the \texttt{k2sc} software package, which uses Gaussian processes to preserve dominant periodicity in a light curve and remove instrumental systematics \citep{Aigrain2015a, Aigrain2016a, Aigrain2017a}. Finally, we excluded data points with a bad quality flag output from \texttt{k2sc}, combined all available data for each star into a single light curve, converted the light curve into magnitudes, and subtracted the mean to produce a light curve with a mean magnitude of zero.


\section{Extracting rotational modulation}      
\label{section: rot mod}

For a CP star with surface abundance inhomogeneities, rotational modulation can be extracted from the Fourier transform of its light curve, i.e. an amplitude spectrum. The (often non-sinusoidal) rotational modulation in the light curve of a CP star is observed as a series of integer harmonics ($n\,\nu_{\rm rot}$) of the surface rotation frequency ($\nu_{\rm rot}$) of the star in its amplitude spectrum. The number of harmonics range from unity (purely sinusoidal) to dozens dependent on the number, size, and location of the abundance inhomogeneities on  the surface of a star and the viewing angle of the observer \citep{Stibbs1950}. By extension, the number of detectable harmonics depend on the data quality, specifically the noise level in the amplitude spectrum.

From the analysis of the Ap star KIC~2569073, \citet{Drury2017a} found a significant change in the amplitude of the rotational modulation signal when comparing new and literature ground-based photometric observations separated by two decades, although no measurable change in the rotation period was found. Physically, this can be interpreted as a change in the size of the surface abundance inhomogeneities on this star, but not their location since the viewing angle had not changed. We expect the rotation periods of CP stars to be constant for time spans much longer than the length of our K2 observations (see e.g. \citealt{Mathys2015}), although a recent study by \citet{Kriticka2017a} investigated how torsional oscillations within a star can explain the observed periodic rotation period variations in the CP stars CU~Vir and HD~37776.
        
Thus, it is reasonable to extract a series of harmonics from within the typical frequency range of the rotation periods of CP stars, i.e. $P_{\rm rot} \gtrsim 0.5$~d ($0 < \nu_{\rm rot} \lesssim 2$~d$^{-1}$), and interpret this as rotational modulation signal. However, due care and attention is necessary to ensure that the correct peak is selected as the rotation frequency, where the highest peak in the amplitude spectra of some CP stars does not represent the true rotation frequency (see Fig.~\ref{figure: EPIC216956748} for an example). Although we used the \texttt{k2sc} software package to remove instrumental systematics when creating our optimised K2 light curves, variance inevitably remains in the output light curves. To determine the highest accuracy in the rotation periods for our sample of CP stars and an improved amplitude spectrum noise level, we used a two-step approach. An amplitude spectrum was calculated via a discrete Fourier transform (DFT; \citealt{Deeming1975, Kurtz1985b}) with an oversampling of (at least) ten for a star. We extracted the low-frequency peak in the amplitude spectrum that represents the rotation frequency for each star and optimised its frequency, amplitude, and phase with a linear (i.e. at fixed frequency) and subsequently a non-linear least-squares fit to the light curve using the equation
\begin{equation}
\Delta m = A \cos(2\pi\nu(t - t_0)+\phi) ~ ,
\label{equation: cosine}
\end{equation}

\noindent where $A$ is the amplitude (mmag), $\nu$ is the frequency (d$^{-1}$), $t$ is the time (d), and $\phi$ is the phase (rad). For each star, the midpoint of the K2 light curve was selected as the zero-point of the timescale, $t_0$, in Eq.~(\ref{equation: cosine}) to reduce frequency uncertainties (see e.g. \citealt{Montgomery1999, Kurtz2015b, Bowman_BOOK}) and determine the rotational modulation model. 

Uncertainties for frequency, amplitude, and phase were derived using the formulae provided by \citet{Montgomery1999}, which are consistent with the 1$\sigma$ uncertainties output from a least-squares fit. It should be noted that these uncertainties are underestimates of the true uncertainty because the data in photometric observations can be correlated producing non-white noise in an amplitude spectrum (see e.g. \citealt{Bowman2015a, Holdsworth2018b}). The ratio of the noise levels at low (where the noise is not white) and high frequencies (where the noise is white) in the residual amplitude spectra of our CP stars is typically between a value of two and three, so to be conservative we multiplied the frequency uncertainty obtained using the formula from \citet{Montgomery1999} by a factor of three before propagating it into a rotation period uncertainty.

Next, we determined the number of frequency harmonics to include in the rotational modulation model by using a least-squares fit and including a series of consecutive integer harmonics of the rotation frequency that have an amplitude larger than $3\sigma$ of the amplitude error obtained using the formula from \citet{Montgomery1999}. For the stars in our sample where rotational modulation was detected, the above method produced a preliminary rotational modulation model, which we subtracted from the light curve of a star and we used a locally weighted scatterplot smoothing (LoWeSS; \citealt{Cleveland1979, Seabold2010}) filter to determine any remaining systematics in the residual light curve. The smoothing of the employed LoWeSS filter is similar to the low-order polynomial instrumental artefacts often seen in \Kepler mission data (e.g. \citealt{Papics2017a}) and likely represents small changes in the temperature of the CCD or background flux during a K2 campaign. We chose to perform this further detrending since we are interested in searching for pulsation modes in these stars after extracting the rotational modulation. The rotation frequency extraction, optimisation, and multi-frequency model was repeated a second time with the newly improved detrended light curve to gain the most accurate rotational modulation models. 

In all stars for which rotational modulation was detected, this two-step approach did not alter the extracted rotation frequency (or period) by more than its uncertainty, but did improve the quality of the light curves in the majority of cases. For example, we calculated the noise level in frequency windows at low ($1 \leq \nu \leq 5~$d$^{-1}$) and high ($21 \leq \nu \leq 24~$d$^{-1}$) frequency in the resultant amplitude spectra and found that in the majority of all cases that the noise level was slightly lower after applying the LoWeSS filter. Furthermore, we calculated the amplitude $S/N$ ratio of the extracted rotation frequency peak and found that it was slightly higher in practically all cases. Thus, this two-step approach is justified for producing higher quality light curves and improved amplitude spectra in order for us to reach the goal of detecting pulsations in stars with rotational modulation.

It is important to note that any variance in an amplitude spectrum that remains within the Rayleigh frequency resolution criterion of an extracted peak after pre-whitening the rotational modulation model is likely an artefact of the data reduction. The remaining variance that is within the Rayleigh frequency resolution is caused by changes in the peak-to-peak amplitudes in the light curve, which is consequently observed as small amplitude and/or frequency modulation in the rotation signal and its harmonics in an amplitude spectrum. In the standard procedure of pre-whitening, frequencies are extracted as purely periodic (co)sinusoids and leave remaining variance in the residual amplitude spectrum if the signal is non-periodic (see e.g. \citealt{Degroote2009a, Papics2017a, Bowman_BOOK}). Thus, any signal within the Rayleigh resolution of an extracted frequency in a residual amplitude spectrum should not be claimed as astrophysical since it could be instrumental. Similarly, due to the inherent quasi-periodic nature of the thruster firing, which occurs approximately every $5.9$~hr \citep{Howell2014}, and the quasi-periodic missing data points that occur approximately every 2~d in K2 data, a complex aliasing structure at and around $\nu_{\rm thrus} = 4.08$~d$^{-1}$ can be seen in an amplitude spectrum. The combined effect can produce side lobes at $\nu_{\rm thrust} \pm 0.5$~d$^{-1}$, or aliases caused by high-amplitude peaks around the thruster frequency, which also should not be considered as astrophysical.

In the subsequent sections, we discuss CP stars in our sample for which we detect no rotational modulation in section~\ref{section: no results}, CP stars with rotational modulation in section~\ref{section: rotation results}, CP stars with rotational modulation and additional variability indicative of stellar pulsations in section~\ref{section: rotation and pulsation results}, and CP stars with no rotational modulation yet variability caused by pulsations in section~\ref{section: pulsation results}.


\section{Results: CP stars that lack rotational modulation}
\label{section: no results}
        
The method for extracting rotation periods clearly does not work for stars that (i) are not rotationally variable and have no rotational modulation signal in their light curve or amplitude spectra; or (ii) have a rotation period that is similar or longer than the length of the K2 observations; and/or (iii) the available K2 photometry is of poor quality. For CP stars in our sample that show no evidence of rotational modulation we were unable to perform our two-step methodology. Since no rotation periods were derived for these stars, the step of smoothing the residual light curve using a LoWeSS filter was not applicable. This is the case for {12} stars in our sample, whose EPIC and HD numbers and details of the K2 observations are given in Table~\ref{table: non-rotation results}. Amongst these stars are two interesting cases: the SB2 binary system EPIC~206026652 (HD~215766) and the known roAp star EPIC~214503319 (HD~177765), which are discussed in more detail below.

\begin{table}
\caption{Properties of the {12} CP stars for which rotation periods could not be extracted using K2 observations, including the EPIC and HD numbers, K2 campaign number, length of useable K2 data $\Delta T$, and number of data points $N$.} 
\begin{center}
\begin{tabular}{l r r r r}
\hline \hline
\multicolumn{1}{c}{EPIC ID} & \multicolumn{1}{c}{HD} & \multicolumn{1}{c}{Camp.} & \multicolumn{1}{c}{$\Delta T$} & \multicolumn{1}{c}{$N$} \\
\multicolumn{1}{c}{} & \multicolumn{1}{c}{} & \multicolumn{1}{c}{} & \multicolumn{1}{c}{(d)} & \multicolumn{1}{c}{} \\
\hline
202061199       &       255134  &       00              &       36.18   &       1594            \\
202061263       &       $-$             &       00              &       36.18   &       1602            \\
206026652\tablefootmark{a}      &       215766  &       03              &       69.16   &       2785            \\
214133870\tablefootmark{b}      &       177562  &       07              &       81.30   &       3494            \\
214503319\tablefootmark{c}      &       177765  &       07              &       81.24   &       3432            \\
219353144       &       181004  &       07              &       81.32   &       3196            \\
224332430       &       162759  &       09              &       71.34   &       3002            \\
224351176       &       167700  &       09              &       71.34   &       2924            \\
224487047       &       165321  &       09              &       71.34   &       3034            \\
225382260       &       161342  &       09              &       71.34   &       2927            \\
227365417       &       168187  &       09              &       71.34   &       2919            \\
246016562\tablefootmark{d}      &       219831  &       12              &       78.83   &       3089            \\
\hline \hline
\end{tabular}
\end{center}
\tablefoot{Literature values for rotation and magnetic field strength.}
\small
\tablefoottext{a}{A possible rotation period of $P_{\rm rot} = 5.2310 \pm 0.0087$~d was measured by \citet{Wraight2012a}, but we find no evidence of this in the K2 photometry.} \\
\tablefoottext{b}{Non-detection of a magnetic field by \citet{Buysschaert2018a*}, and poor quality K2 photometry caused by significant loss of flux in the sub-raster of the TPF.} \\
\tablefoottext{c}{A known magnetic star with a literature rotation period of $P_{\rm rot} \gg 5$~yr \citep{Mathys1997a, Mathys2017a}, and roAp pulsations at $\nu = 61.02$~d$^{-1}$ \citep{Alentiev2012}.} \\
\tablefoottext{d}{No obvious rotation period found by \citet{Wraight2012a}.}
\label{table: non-rotation results}
\end{table}             

                        
        \subsection{EPIC~206026652 -- HD~215766}
        \label{subsection: EPIC206026652}
\begin{figure*}
        \centering
        \includegraphics[width=0.95\textwidth]{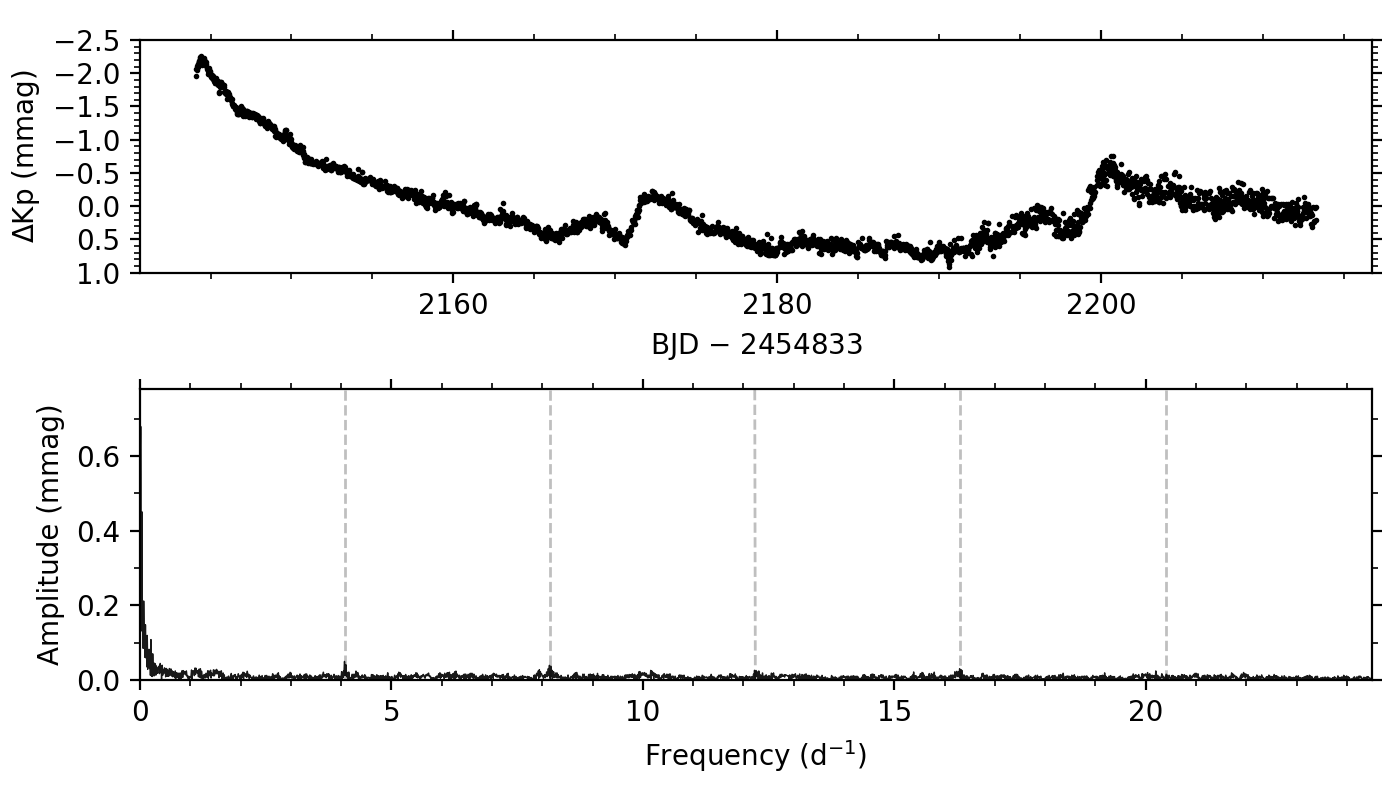}
        \caption{K2 light curve and amplitude spectrum for EPIC~206026652 (HD~215766) are given in the top and bottom panels, respectively. The vertical dashed grey lines indicate the location of integer multiples of the K2 thruster firing frequency of $\nu_{\rm thrus} = 4.08$~d$^{-1}$.}
        \label{figure: EPIC206026652}
        \end{figure*}

        EPIC~206026652 (HD~215766) has been identified as a spectroscopic (SB2) binary system by \citet{Renson2009}, a projected surface rotational velocity of $v\,\sin\,i = 80 \pm 12$~km\,s$^{-1}$ \citep{Chini2012}, and a possible rotation period of $P_{\rm rot} = 5.2310 \pm 0.0087$~d \citep{Wraight2012a}. However, \citet{Wraight2012a} commented that the detection of this rotation period is uncertain since their observations suffer from significant blending and systematics that severely affect the detection of a reliable period. The K2 light curve and amplitude spectrum of HD~215766 are shown in the top and bottom panels of Fig.~\ref{figure: EPIC206026652}, respectively. We do not find a significant rotation period of HD~215766 using the available {$\sim$}70~d of K2 photometry. The measured period of $5.2310 \pm 0.0087$~d by \citet{Wraight2012a} has an amplitude of less than 60~$\mu$mag in our K2 observations and is not significant given the comparable noise level for this star.
        
        \subsection{Known roAp star: EPIC~214503319 (HD~177765)}
        \label{subsection: EPIC214503319}

\begin{figure*}
        \centering
        \includegraphics[width=0.95\textwidth]{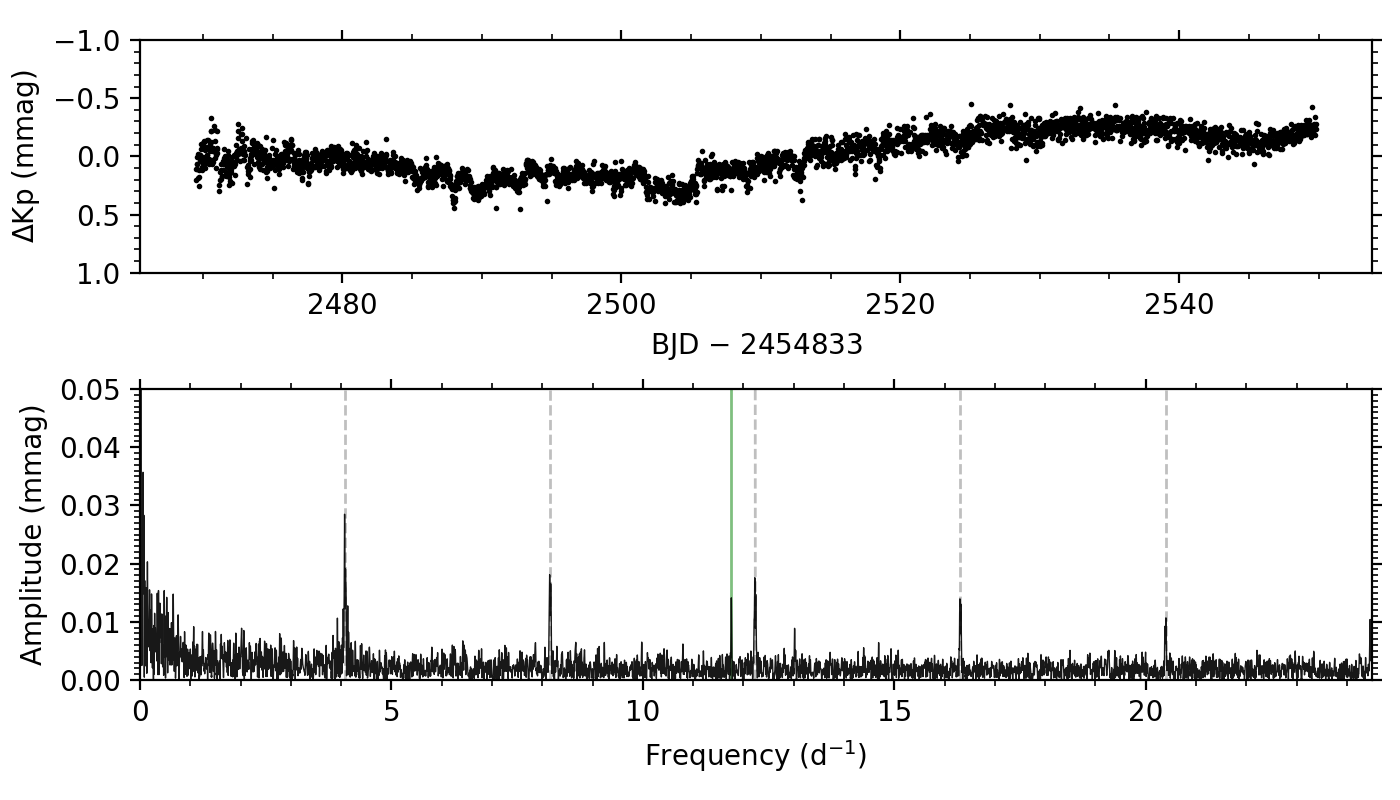}
        \caption{K2 light curve and amplitude spectrum for EPIC~214503319 (HD~177765) are given in the top and bottom panels, respectively. The vertical dashed grey lines indicate the location of integer multiples of the K2 thruster firing frequency of $\nu_{\rm thrus} = 4.08$~d$^{-1}$. HD~177765 is a confirmed roAp star, and a Nyquist alias of its high-frequency pulsation modes can be seen at $11.760 \pm 0.002$~d$^{-1}$, indicated by the solid green line.}
        \label{figure: EPIC214503319}
        \end{figure*}

        EPIC~214503319 (HD~177765) is the only previously confirmed roAp star in our sample. Although null detections (NDs) of high-overtone pulsations in this star appear in the literature (e.g. \citealt{Martinez1994e}), the discovery of its roAp pulsations was first made by \citet{Alentiev2012}. The exact rotation period of HD~177765 is not known but has been constrained to $P_{\rm rot} \gg 5$~yr by \citet{Mathys2017a}, which explains our non-detection using {$\sim$}80~d of K2 space photometry. The K2 light curve and amplitude spectrum of HD~177765 are shown in the top and bottom panels of Fig.~\ref{figure: EPIC214503319}, respectively. HD~177765 has also been confirmed as a magnetic CP star with a mean field modulus of $<B> \simeq 3.2$~kG \citep{Mathys1997a, Renson2009, Mathys2017a},  which is compatible with the minimum polar strength of $B_p \simeq 3.5$~kG from \citet{Buysschaert2018a*}, and a projected surface rotational velocity of $v\,\sin\,i = 5 \pm 2$~km\,s$^{-1}$ \citep{Buysschaert2018a*}. We note that the LC \Kepler sampling frequency of $\nu_{\rm samp} = 48.94$~d$^{-1}$ can be used to demonstrate that the peak at $11.760 \pm 0.002$~d$^{-1}$ with an amplitude of $14 \pm 4$~$\mu$mag in our K2 observations is the Nyquist alias of its high-frequency pulsation with a period of 23.6~min ($\nu = 61.02$~d$^{-1}$; \citealt{Alentiev2012}), as shown in the summary figure in Fig.~\ref{figure: EPIC214503319}. Further analysis of this star has also been performed in which multiple roAp pulsation modes have been detected \citep{Holdsworth2016b}.
        
\section{Results: CP stars with measured rotational modulation but no other variability}
\label{section: rotation results}
        
For the {38} stars in which rotational modulation was detected, the measured and literature (if available) rotation periods are given in Table~\ref{table: rotation results}, and rotational modulation figures are given in Appendix~\ref{section: appendix: rotation}. In our sample, {10} stars are almost entirely absent from the scientific literature with the exception of a spectral type from the \citet{Renson2009} catalogue. For {16} stars in our sample, we report the first measurement of rotation periods.

\begin{figure*}
\centering
\includegraphics[width=0.95\textwidth]{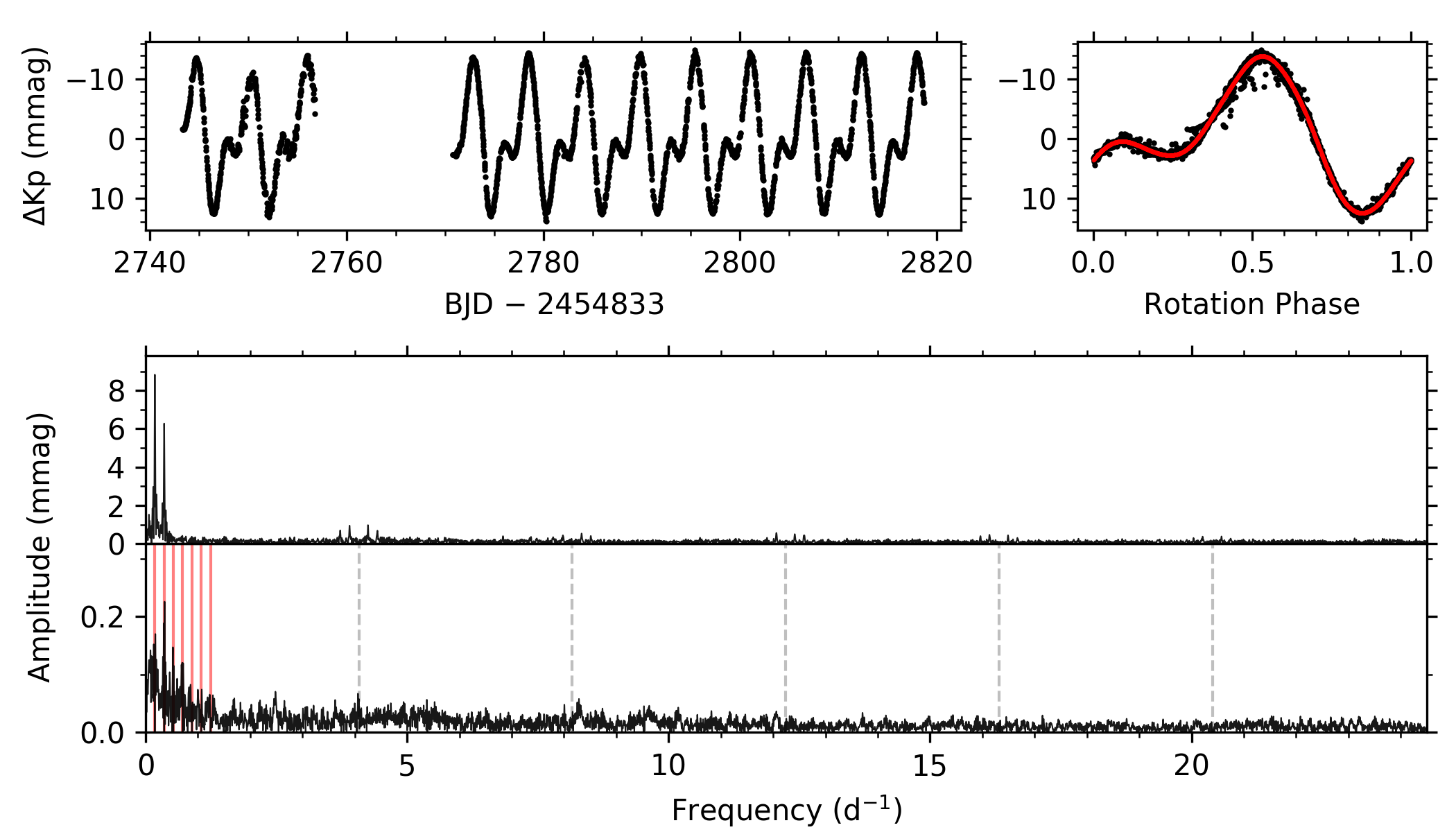}
\caption{Rotational modulation in EPIC~201667495 (HD~107000). The detrended K2 light curve is shown in the top left panel. The bottom panel shows the amplitude spectrum calculated up to the K2 LC Nyquist frequency of 24.47~d$^{-1}$; the lower part shows the residual amplitude spectrum calculated after the rotational modulation signal was removed for comparison. We note the change in ordinate scale. The vertical red lines indicate the location of the extracted rotation frequency and of its significant harmonics, which were used to construct the rotational modulation frequency model. The vertical dashed grey lines indicate the location of integer multiples of the K2 thruster firing frequency of $\nu_{\rm thrus} = 4.08$~d$^{-1}$. The phase-folded light curve using the derived rotation period is shown in the top right panel with black circles indicating the K2 observations and the red line indicating the rotational modulation frequency model.}
\label{figure: example}
\end{figure*}

For each star for which we measure a rotation period, the results of our analysis of the available K2 photometry are summarised in a multi-panel figure; the example for EPIC~201667495 (HD~107000) is shown in Fig.~\ref{figure: example}. The multi-panel summary figures for the detected rotational modulation in the other CP stars discussed in this paper are given in Appendices~\ref{section: appendix: rotation} and \ref{section: appendix: rotation and pulsation}. Each summary figure contains the detrended K2 light curve, the phase-folded light curve, and the amplitude spectrum before and after pre-whitening the rotational modulation model.
        
\begin{table*}
\caption{Properties of the CP stars for which variability was detected using K2 observations, including the EPIC and HD numbers; K2 campaign number; length of available K2 data, $\Delta T$; number of data points, $N$;  measured rotation period, $P_{\rm rot}$; and its 3$\sigma$ uncertainty using the formulae from \citet{Montgomery1999}; the literature rotation period (if available); and the first reference of the magnetic field detection (if available), where DD indicates a definite detection, MD indicates a marginal detection, and ND indicates a non-detection. A full discussion of each star and discrepancies in literature values for rotation periods are given in the text.} 
\begin{center}

\resizebox{0.99\textwidth}{!}{

\begin{tabular}{r r r r r l l l c}
\hline \hline
\multicolumn{1}{c}{EPIC ID} & \multicolumn{1}{c}{HD} & \multicolumn{1}{c}{Camp.} & \multicolumn{1}{c}{$\Delta T$} & \multicolumn{1}{c}{$N$} & \multicolumn{1}{c}{$P_{\rm rot}$ (d)} &\multicolumn{1}{c}{$P_{\rm rot}$ (d)} & \multicolumn{1}{c}{$B$-field} & \multicolumn{1}{c}{Fig.~No.} \\
\multicolumn{1}{c}{} & \multicolumn{1}{c}{} & \multicolumn{1}{c}{} & \multicolumn{1}{c}{(d)} & \multicolumn{1}{c}{} & \multicolumn{1}{c}{This work} & \multicolumn{1}{c}{Literature} & \multicolumn{1}{c}{} & \multicolumn{1}{c}{} \\
\hline
\multicolumn{8}{l}{Stars with rotational modulation (figures in Appendix~\ref{section: appendix: rotation}):} \vspace{0.1cm} \\

201667495       &       107000  &       10              &       75.35   &       2392            &       $5.641 \pm 0.001$              &       $5.638$\tablefootmark{l}        &       DD\tablefootmark{k}     &       \ref{figure: example}        \\

201777342       &       97859   &       01              &       78.74   &       3155            &       $0.792122 \pm 0.000004$   &       $0.7921 \pm 0.0002$\tablefootmark{p}    &       DD\tablefootmark{f}     &       \ref{figure: EPIC201777342}  \\

203092367       &       150035  &       02              &       62.13   &       2145            &       $2.403 \pm 0.002$              &       $2.3389 \pm 0.0013$\tablefootmark{p}    &       ND\tablefootmark{n}             &       \ref{figure: EPIC203092367}  \\

203814494       &       142990  &       02              &       67.96   &       2244            &       $0.97892 \pm 0.00002$    &       $0.978832 \pm 0.000002$\tablefootmark{m}        &       DD\tablefootmark{e}             &       \ref{figure: EPIC203814494}  \\

204964091       &       147010  &       02              &       77.25   &       2302            &       $3.9216 \pm 0.0003$     &       $3.9207 \pm 0.0003$\tablefootmark{c}    &       DD\tablefootmark{i}     &       \ref{figure: EPIC204964091}  \\

210964459       &       26571   &       04              &       70.82   &       2826            &       $15.733 \pm 0.002$              &       $15.7505 \pm 0.0004$\tablefootmark{p}   &       $-$     &       \ref{figure: EPIC210964459}  \\

213786701       &       173657  &       07              &       81.28   &       3361            &       $1.93810 \pm 0.00006$    &       $1.93789 \pm 0.00005$\tablefootmark{g}, {$1.94 \pm 0.05$}\tablefootmark{f} &       ND\tablefootmark{f}             &       \ref{figure: EPIC213786701}  \\

214197027       &       173857  &       07              &       81.32   &       3389            &       $1.7348 \pm 0.0003$     &       $-$     &       $-$     &       \ref{figure: EPIC214197027}     \\

214775703       &       178786  &       07              &       80.87   &       3366            &       $1.76413 \pm 0.00004$    &       $-$     &       $-$     &       \ref{figure: EPIC214775703}     \\

215357858       &       174356  &       07              &       81.32   &       3406            &       $4.045 \pm 0.003$              &       $4.0431 \pm 0.0003$\tablefootmark{g}    &       $-$     &       \ref{figure: EPIC215357858}  \\

215431167       &       174146  &       07              &       81.32   &       3416            &       $11.1820 \pm 0.0005$     &       $11.185 \pm 0.001$\tablefootmark{g}     &       $-$     &       \ref{figure: EPIC215431167}  \\

215584931       &       172480  &       07              &       81.30   &       3422            &       $0.87133 \pm 0.00001$    &       $0.87134 \pm 0.00001$\tablefootmark{g}  &       $-$     &       \ref{figure: EPIC215584931}  \\

215876375       &       184343  &       07              &       81.07   &       3364            &       $7.1987 \pm 0.0006$     &       $-$     &       $-$     &       \ref{figure: EPIC215876375}     \\

216005035       &       182459  &       07              &       81.32   &       3426            &       $1.26055 \pm 0.00002$    &       $-$     &       $-$     &       \ref{figure: EPIC216005035}     \\

217323447       &       180303  &       07              &       81.32   &       3372            &       $2.7927 \pm 0.0002$     &       $-$     &       $-$     &       \ref{figure: EPIC217323447}     \\

218676652       &       173406  &       07              &       81.32   &       3230            &       $4.562 \pm 0.002$              &       $5.095$\tablefootmark{o}, {$4.6 \pm 0.3$}\tablefootmark{f}         &       ND\tablefootmark{f}     &       \ref{figure: EPIC218676652}     \\

218818457       &       176330  &       07              &       81.26   &       3193            &       $13.26 \pm 0.01$               &       $-$     &       $-$     &       \ref{figure: EPIC218818457}  \\

219198038       &       177013  &       07              &       81.28   &       3293            &       $4.8730 \pm 0.0001$     &       $4.873$\tablefootmark{o}, {$4.9 \pm 0.3$}\tablefootmark{f}      &       MD\tablefootmark{f}     &       \ref{figure: EPIC219198038}  \\

224206658       &       165972  &       09              &       71.34   &       2938            &       $2.7596 \pm 0.0003$     &       {$2.8 \pm 0.1$}\tablefootmark{f}        &       DD\tablefootmark{f}     &       \ref{figure: EPIC224206658}  \\

226097699       &       166190  &       09              &       70.40   &       2939            &       $1.1696 \pm 0.0001$     &       $-$     &       $-$     &       \ref{figure: EPIC226097699}     \\

226241087       &       164224  &       09              &       71.12   &       2947            &       $0.73156 \pm 0.00001$    &       ${0.73 \pm 0.01}$\tablefootmark{f}      &       DD\tablefootmark{f}     &       \ref{figure: EPIC226241087}  \\

227108971       &       164190  &       09              &       68.64   &       2811            &       $4.657 \pm 0.002$              &       $-$     &       $-$     &       \ref{figure: EPIC227108971}  \\

227373493       &       166804  &       09              &       71.14   &       2919            &       $3.7024 \pm 0.0005$     &       $3.7035 \pm 0.0002$\tablefootmark{g}, {$3.7 \pm 0.2$}\tablefootmark{f}         &       DD\tablefootmark{f}     &       \ref{figure: EPIC227373493}     \\

227825246       &       164085  &       09              &       71.34   &       3052            &       $1.22079 \pm 0.00002$    &       $1.22072 \pm 0.00002$\tablefootmark{g}  &       $-$     &       \ref{figure: EPIC227825246}  \\

232147357       &       153192  &       11              &       68.55   &       2928            &       $11.393 \pm 0.003$              &       $11.357 \pm 0.002$\tablefootmark{d}     &       $-$     &       \ref{figure: EPIC232147357}  \\

232176043       &       152834  &       11              &       67.83   &       2791            &       $4.2919 \pm 0.0007$     &       $4.2928 \pm 0.0003$\tablefootmark{g}, {$4.4 \pm 0.3$}\tablefootmark{f}  &       DD\tablefootmark{f}     &       \ref{figure: EPIC232176043}  \\

232284277       &       155127  &       11              &       74.13   &       2949            &       $5.5171 \pm 0.0007$     &       $5.5243 \pm 0.0004$\tablefootmark{g}, {$5.5 \pm 0.4$}\tablefootmark{f}  &       DD\tablefootmark{f}     &       \ref{figure: EPIC232284277}  \\

247729177       &       284639  &       13              &       80.52   &       3618            &       $9.155 \pm 0.001$              &       $-$     &       $-$     &       \ref{figure: EPIC247729177}  \\

\hline
\multicolumn{8}{l}{Stars with rotational modulation and additional variability indicative of stellar pulsations (figures in Appendix~\ref{section: appendix: rotation and pulsation}):} \vspace{0.1cm} \\

203749199       &       152366  &       02, 11  &       835.99  &       5209            &       $3.2308 \pm 0.0001$     &       $3.23 \pm 0.01$\tablefootmark{f}        &       DD\tablefootmark{f}     &       \ref{figure: EPIC203749199}  \\

216956748       &       181810  &       07              &       80.83   &       3415            &       $13.509 \pm 0.009$              &       $-$     &       $-$     &       \ref{figure: EPIC216956748}  \\

217437213       &       177016  &       07              &       81.32   &       3397            &       $3.276 \pm 0.004$              &       $-$     &       $-$     &       \ref{figure: EPIC217437213}  \\

223573464       &       161851  &       09              &       71.34   &       2901            &       $1.6209 \pm 0.0007$     &       $-$     &       $-$     &       \ref{figure: EPIC223573464}     \\

225191577       &       164068  &       09              &       71.34   &       3031            &       $2.4130 \pm 0.0003$     &       $-$     &       $-$     &       \ref{figure: EPIC225191577}     \\

225990054       &       158596  &       09, 11  &       222.77  &       5796            &       $2.02208 \pm 0.00003$    &       $2.02206 \pm 0.00005$\tablefootmark{d}, {$2.02 \pm 0.02$}\tablefootmark{f} &       DD\tablefootmark{f}     &       \ref{figure: EPIC225990054}  \\

227231984       &       158336  &       11              &       73.21   &       3015            &       $1.7609 \pm 0.0008$     &       $1.76009 \pm 0.00004$\tablefootmark{d}  &       $-$     &       \ref{figure: EPIC227231984}  \\

227305488       &       166542  &       09              &       68.64   &       2869            &       $3.6331 \pm 0.0003$     &       $-$     &       $-$     &       \ref{figure: EPIC227305488}     \\

228293755       &       165945  &       09              &       70.40   &       2858            &       $2.1352 \pm 0.0002$     &       $-$     &       $-$     &       \ref{figure: EPIC228293755}     \\

230753303       &       153997  &       11              &       74.13   &       3031            &       $5.4966 \pm 0.0006$     &       $-$     &       $-$     &       \ref{figure: EPIC230753303}     \\

\hline
\multicolumn{8}{l}{Stars that show variability indicative of stellar pulsations yet lack clear rotational modulation (figures in Appendix~\ref{section: appendix: pulsation}):} \vspace{0.1cm} \\

202060145       &       $-$             &       00              &       36.18   &       1685            &       $-$     &       1.1472\tablefootmark{a} &       $-$     &       \ref{figure: EPIC202060145}  \\

203917770       &       145792  &       02              &       65.73   &       2258            &       $-$     &       0.8478\tablefootmark{t}, $1.6956 \pm 0.0005$\tablefootmark{p}    &       DD\tablefootmark{n}     &       \ref{figure: EPIC203917770}  \\

206120416       &       210424  &       03              &       67.73   &       2702            &       $2.111 \pm 0.003$\tablefootmark{$\dagger$}     &       $3.9613 \pm 0.0033$\tablefootmark{p}, 1.4465\tablefootmark{b}         &       $-$     &       \ref{figure: EPIC206120416}     \\

206326769       &       211838  &       03              &       69.16   &       2771            &       $-$     &       $6.5633 \pm 0.0063$\tablefootmark{j}    &       ND\tablefootmark{h}     &       \ref{figure: EPIC206326769}  \\

224947037       &       162814  &       09              &       71.34   &       3067            &       $8.60 \pm 0.05$\tablefootmark{$\dagger$}      &       $-$     &       $-$     &       \ref{figure: EPIC224947037}  \\

246152326       &       220556  &       12              &       78.42   &       3258            &       $-$     &       $-$     &       $-$     &       \ref{figure: EPIC246152326}  \\

\hline \hline
\end{tabular} }
\end{center}
\tablefoot{References for literature rotation periods and magnetic field detections.}
\tablefoottext{$\dagger$}{We identify this as a possible rotation period for this star, but it does not represent a unique solution.} \\
\tablefoottext{a}{\citet{Armstrong2015a},}
\tablefoottext{b}{\citet{Armstrong2016b},}
\tablefoottext{c}{\citet{Bailey2013b},}
\tablefoottext{d}{\citet{Bernhard2015a},}
\tablefoottext{e}{\citet{Borra1983},}
\tablefoottext{f}{\citet{Buysschaert2018a*},}
\tablefoottext{g}{\citet{Hummerich2016a},}
\tablefoottext{h}{\citet{Makaganiuk2011a},}
\tablefoottext{i}{\citet{Mathys1995b},}
\tablefoottext{j}{\citet{Paunzen2013a},}
\tablefoottext{k}{\citet{Romanyuk2008},}
\tablefoottext{l}{\citet{Romanyuk2015d},}
\tablefoottext{m}{\citet{Shultz2018b},}
\tablefoottext{n}{\citet{Thompson1987},}
\tablefoottext{o}{\citet{Watson2006},}
\tablefoottext{p}{\citet{Wraight2012a}.}
\label{table: rotation results}
\end{table*}

        
        \subsection{EPIC~201667495 -- HD~107000}
        \label{subsection: EPIC201667495}
        The rotation period of EPIC~201667495 (HD~107000) is somewhat disputed in the literature; that is $P_{\rm rot} = 2.8187 \pm 0.0024$~d was measured by \citet{Wraight2012a}, yet twice this value, i.e. $5.638$~d, was measured by \citet{Romanyuk2015d}. Therefore, it is likely that \citet{Wraight2012a} measured twice the rotation frequency, i.e. half the rotation period. The longer rotation has since been confirmed and was determined to be {$P_{\rm rot} = 5.6 \pm 0.7$~d} and a projected surface rotational velocity of $v\,\sin\,i = 20 \pm 5$~km\,s$^{-1}$ by \citet{Buysschaert2018a*}, and this star is known to host a large-scale magnetic field with a polar strength of order $B_p \simeq 750$~G \citep{Romanyuk2008, Romanyuk2014b, Romanyuk2015d, Buysschaert2018a*}. In this work, we measured a rotation period of $5.641 \pm 0.001$~d with K2 photometry, which is more precise than previous literature values. The summary figure of EPIC~201667495 is shown in Fig.~\ref{figure: example}.

                
        \subsection{EPIC~201777342 -- HD~97859}
        \label{subsection: EPIC201777342}
        The rotation period of this star is known with values of $P_{\rm rot} = 0.7921 \pm 0.0002$ from \citet{Wraight2012a} and {$P_{\rm rot} = 0.792 \pm 0.008$~d} from \citet{Buysschaert2018a*}. HD~97859 is also known to host a large-scale magnetic field with a polar strength of order $B_p \simeq 1.7$~kG and has a projected surface rotational velocity of $v\,\sin\,i = 83 \pm 1$~km\,s$^{-1}$ \citep{Buysschaert2018a*}. In this work, we measure a rotation period of $P_{\rm rot} = 0.792122 \pm 0.000004$~d, which is consistent yet more precise than previous literature values. A significant period of 1.5848~d (i.e. twice the rotation period) for this star was found by \citet{Armstrong2016b} using their automated classification method, but this frequency is absent from our K2 photometry after removing the rotational modulation signal including all its harmonics. The summary figure of EPIC~201777342 is shown in Fig.~\ref{figure: EPIC201777342}.

        
        \subsection{EPIC~203092367 -- HD~150035}
        \label{subsection: EPIC203092367}
        This star was investigated by \citet{Thompson1987} for possibly having a magnetic field, although the detection of a few hundred Gauss field was deemed not to be statistically significant. Modern observations using spectropolarimetry will more conclusively determine the strength and significance of the possible magnetic field in this star. HD~150035 has a literature rotation period of $P_{\rm rot} = 2.3389 \pm 0.0013$~d from \citet{Wraight2012a} using STEREO photometry. In this work, we measure a rotation period of $P_{\rm rot} = 2.403 \pm 0.002$~d, which is significantly different from the literature value given by \citet{Wraight2012a}. The summary figure of EPIC~203092367 is shown in Fig.~\ref{figure: EPIC203092367}, in which a peak at $6.108 \pm 0.002$~d$^{-1}$ lies close to $1.5\,\nu_{\rm thrus}$ and so we exclude this frequency since it is likely instrumental. HD~150035 is a bright star and its necessarily large pixel mask may include neighbouring stars. Short cadence data are also available for HD~150035, which provided the ability to exclude the possibility that this star is a roAp star since we found no variability indicative of stellar pulsations up to the SC Nyquist frequency of 714~d$^{-1}$.

                                                                                                        
        \subsection{EPIC~203814494 -- HD~142990}
        \label{subsection: EPIC203814494}
        This star is a known magnetic star with a polar strength of $B_p \simeq 1$~kG \citep{Borra1983, Bohlender1993, Wade2016a, Shultz2018b}. Recently, \citet{Shultz2018b} identified that this star has a significant non-dipolar magnetic field structure. Measurements of the projected surface rotational velocity of HD~142990 include $v\,\sin\,i = 125 \pm 64$~km\,s$^{-1}$ \citep{Aerts2014a} and $v\,\sin\,i = 122 \pm 2$~km\,s \citep{Shultz2018b}. Previous measurements of the rotation period of this star are $P_{\rm rot} = 0.9789 \pm 0.0003$~d \citep{Wraight2012a} and $P_{\rm rot} = 0.978832 \pm 0.000002$~d \citep{Shultz2018b}. In this work, we measure a rotation period of $P_{\rm rot} = 0.97892 \pm 0.00002$~d, which is consistent with the value from \citet{Wraight2012a} yet significantly different from the value from \citet{Shultz2018b}. The summary figure of EPIC~203814494 is shown in Fig.~\ref{figure: EPIC203814494}.

                        
        \subsection{EPIC~204964091 -- HD~147010}
        \label{subsection: EPIC204964091}
        This star is part of the Scorpio-Centaurus association and has a strong magnetic field with a mean magnetic field modulus of $<B> \simeq 16$~kG \citep{Thompson1987, Mathys1995b, Mathys1997c, Bailey2013b, Mathys2017a}. HD~147010 has a rotation period of $P_{\rm rot} = 3.9209 \pm 0.0048$~d \citep{Wraight2012a}. Similarly, \citet{Bailey2013b} measured a projected surface rotational velocity of $v\,\sin\,i = 15 \pm 2$~km\,s$^{-1}$ and a rotation period of $P_{\rm rot} = 3.9207 \pm 0.0003$~d. In this work, we measure a rotation period of $P_{\rm rot} = 3.9216 \pm 0.0003$~d, which is consistent with the value from \citet{Wraight2012a} yet significantly different from the value given by \citet{Bailey2013b}, who determined the rotation period from the periodic changes in the measured longitudinal field strength. The summary figure of EPIC~204964091 is shown in Fig.~\ref{figure: EPIC204964091}.

                                                        
        \subsection{EPIC~210964459 -- HD~26571}
        \label{subsection: EPIC210964459}
        This star is confirmed as a slow rotator with $P_{\rm rot} = 15.7505 \pm 0.0004$~d via ground-based photometry \citep{Adelman2008c} and has a projected surface rotational velocity of $v\,\sin\,i = 20.0 \pm 3.0$~km\,s$^{-1}$ \citep{Wraight2012a}. However, the rotation period of HD~26571 is somewhat disputed in the literature, where a value of $1.06$~d was given by \citet{Catalano1998a}, although these authors commented that this value is dubious and could arise from a 1-day alias, and \citet{Wraight2012a} classifying the star as non-variable. In this work, we measure a rotation period of $P_{\rm rot} = 15.733 \pm 0.002$~d, which is significantly different from the literature value from \citet{Adelman2008c}. The summary figure of EPIC~210964459 is shown in Fig.~\ref{figure: EPIC210964459}.

        
        \subsection{EPIC~213786701 -- HD~173657}
        \label{subsection: EPIC213786701}
        This star has a literature rotation period of $P_{\rm rot} = 1.93789 \pm 0.00005$~d derived from ASAS-3 photometry \citep{Hummerich2016a}. Recently, a non-detection of a large-scale magnetic field with an upper limit of $\sim$80~G, a projected surface rotational velocity of $v\,\sin\,i = 91 \pm 5$~km\,s$^{-1}$ and a rotation period of {$P_{\rm rot} = 1.94 \pm 0.05$~d} for HD~173657 was reported by \citet{Buysschaert2018a*}. In this work, we measure a rotation period of $P_{\rm rot} = 1.93810 \pm 0.00006$~d, which is significantly different from the value from \citet{Hummerich2016a}. The summary figure of EPIC~213786701 is shown in Fig.~\ref{figure: EPIC213786701}.

        
        \subsection{EPIC~214197027 -- HD~173857}
        \label{subsection: EPIC214197027}
        We report the first measurement of the rotation period of $P_{\rm rot} = 1.7349 \pm 0.0003$~d for HD~173857. The summary figure of EPIC~214197027 is shown in Fig.~\ref{figure: EPIC214197027}.

        
        \subsection{EPIC~214775703 -- HD~178786}
        \label{subsection: EPIC214775703}
        We report the first measurement of the rotation period of $P_{\rm rot} = 1.76413 \pm 0.00004$~d for HD~178786. The summary figure of EPIC~214775703 is shown in Fig.~\ref{figure: EPIC214775703}.

                
        \subsection{EPIC~215357858 -- HD~174356}
        \label{subsection: EPIC215357858}
        This star has a literature rotation period of $P_{\rm rot} = 4.0431 \pm 0.0003$~d derived from ASAS-3 photometry \citep{Hummerich2016a}. In this work, we measure a rotation period of $P_{\rm rot} = 4.045 \pm 0.003$~d, which is consistent with the literature value from \citet{Hummerich2016a}. The summary figure of EPIC~215357858 is shown in Fig.~\ref{figure: EPIC215357858}. However, a second period is detected in the amplitude spectrum of HD~174356, which is not well-resolved from the harmonic of the measured rotation period. This suggests that HD~174356 may be a binary or multiple system similar to the magnetic Bp star Atlas \citep{White2017b}; this has not been noted in previous studies. The light curve of HD~174356 has the characteristic signature of migrating spots on the surface of a cool star with a large convective envelope like the Sun, for which differential latitudinal surface rotation is common. However, HD~174356 has a spectral type of B9~Si \citep{Renson2009} and migrating spots on the surface of a hot and candidate magnetic star is not expected and would imply a large amount of differential latitudinal surface rotation. Clearly, further study of this system is needed to resolve the intrinsic variability of this star.

        
        \subsection{EPIC~215431167 -- HD~174146}
        \label{subsection: EPIC215431167}
        This star has a literature rotation period of $P_{\rm rot} = 11.185 \pm 0.001$~d derived from ASAS-3 photometry \citep{Hummerich2016a}. In this work, we measure a rotation period of $P_{\rm rot} = 11.1820 \pm 0.0005$~d, which is significantly different yet more precise than the value from \citet{Hummerich2016a}. The summary figure of EPIC~215431167 is shown in Fig.~\ref{figure: EPIC215431167}.

        
        \subsection{EPIC~215584931 -- HD~172480}
        \label{subsection: EPIC215584931}
        This star has a literature rotation period of $P_{\rm rot} = 0.87134 \pm 0.00001$~d derived from ASAS-3 photometry \citep{Hummerich2016a}. In this work, we measure a rotation period of $P_{\rm rot} = 0.87133 \pm 0.00001$~d, which is consistent with the literature value from \citet{Hummerich2016a}. The summary figure of EPIC~215584931 is shown in Fig.~\ref{figure: EPIC215584931}.

        
        \subsection{EPIC~215876375 -- HD~184343}
        \label{subsection: EPIC215876375}
        We report the first measurement of the rotation period of $P_{\rm rot} = 7.1987 \pm 0.0006$~d for HD~184343. The summary figure of EPIC~215876375 is shown in Fig.~\ref{figure: EPIC215876375}.

        
        \subsection{EPIC~216005035 -- HD~182459}
        \label{subsection: EPIC216005035}
        We report the first measurement of the rotation period of $P_{\rm rot} = 1.26055 \pm 0.00002$~d for HD~182459. The summary figure of EPIC~216005035 is shown in Fig.~\ref{figure: EPIC216005035}.

                                        
        \subsection{EPIC~217323447 -- HD~180303}
        \label{subsection: EPIC217323447}       
        We report the first measurement of the rotation period of $P_{\rm rot} = 2.7927 \pm 0.0002$~d for HD~180303. The summary figure of EPIC~217323447 is shown in Fig.~\ref{figure: EPIC217323447}.

        
        \subsection{EPIC~218676652 -- HD~173406}
        \label{subsection: EPIC218676652}
        This star is reported to have unresolved periodic variability by \citet{Koen2002c}, although a rotation period of $P_{\rm rot} = 5.095$~d is measured by \citet{Watson2006} using photometry from the American Association of Variable Star Observers (AAVSO). A non-detection of a large-scale magnetic field with an upper limit of $\sim$50~G, a rotation period of $P_{\rm rot} = 4.6 \pm 0.3$~d and a projected surface rotational velocity of $v\,\sin\,i = 38 \pm 2$~km\,s$^{-1}$ were measured by \citet{Buysschaert2018a*} for HD~173406. In this work, we measure a rotation period of $P_{\rm rot} = 4.562 \pm 0.002$~d, which is significantly different from the literature value of 5.095~d from \citet{Watson2006}. From inspection of the light curve and amplitude spectrum, it is clear that HD~173406 exhibits periodic behaviour over the K2 observations, but the cause of the 0.5~d difference in rotation period compared to the analysis by \citet{Watson2006} is unclear. The summary figure of EPIC~218676652 is shown in Fig.~\ref{figure: EPIC218676652}.

                                        
        \subsection{EPIC~218818457 -- HD~176330}
        \label{subsection: EPIC218818457}       
        We report the first measurement of the rotation period of $P_{\rm rot} = 13.26 \pm 0.01$~d for HD~176330. The summary figure of EPIC~218818457 is shown in Fig.~\ref{figure: EPIC218818457}.

                
        \subsection{EPIC~219198038 -- HD~177013}
        \label{subsection: EPIC219198038}
        This star has a literature rotation period of $P_{\rm rot} = 4.873$~d via AAVSO photometry \citep{Watson2006}, and a marginal detection of a magnetic field with a polar strength of $B_p \simeq 600$~G, a projected surface rotational velocity of $v\,\sin\,i = 24 \pm 6$~km\,s$^{-1}$, and rotation period of $P_{\rm rot} = 4.9 \pm 0.3$~d from \citet{Buysschaert2018a*}. In this work, we measure a rotation period of $P_{\rm rot} = 4.8730 \pm 0.0001$~d, which is consistent with the literature values. The summary figure of EPIC~219198038 is shown in Fig.~\ref{figure: EPIC219198038}.

        
        \subsection{EPIC~224206658 -- HD~165972}
        \label{subsection: EPIC224206658}
        This star has a literature rotation period of $P_{\rm rot} = 2.7596 \pm 0.0001$ derived from ASAS-3 photometry \citep{Hummerich2016a}, and a magnetic polar field strength of $B_p \simeq 1$~kG, a projected surface rotational velocity of $v\,\sin\,i = 29 \pm 4$~km\,s$^{-1}$ and a rotation period of $P_{\rm rot} = 2.8 \pm 0.1$~d from \citet{Buysschaert2018a*}. In this work, we measure a rotation period of $P_{\rm rot} = 2.7596 \pm 0.0003$~d, which is consistent with the literature value from \citet{Hummerich2016a}. The summary figure of EPIC~224206658 is shown in Fig.~\ref{figure: EPIC224206658}.

                
        \subsection{EPIC~226097699 -- HD~166190}
        \label{subsection: EPIC226097699}
        We report the first measurement of the rotation period of $P_{\rm rot} = 1.1696 \pm 0.0001$~d for HD~166190. The summary figure of EPIC~226097699 is shown in Fig.~\ref{figure: EPIC226097699}.


        \subsection{EPIC~226241087 -- HD~164224}
        \label{subsection: EPIC226241087}
        This star has a confirmed large-scale magnetic field with a polar strength of $B_p \simeq 1.7$~kG, a rotation period of $P_{\rm rot} = 0.73 \pm 0.01$~d and projected surface rotational velocity of $v\,\sin\,i = 22 \pm 4$~km\,s$^{-1}$ from \citet{Buysschaert2018a*}. In this work, we measure a rotation period of $P_{\rm rot} = 0.73156 \pm 0.00001$~d, which is consistent with the literature value. The summary figure of EPIC~226241087 is shown in Fig.~\ref{figure: EPIC226241087}.


        \subsection{EPIC~227108971 -- HD~164190}
        \label{subsection: EPIC227108971}       
        We report the first measurement of the rotation period of $P_{\rm rot} = 4.657 \pm 0.002$~d for HD~164190. The summary figure of EPIC~227108971 shown in Fig.~\ref{figure: EPIC227108971}.


        \subsection{EPIC~227373493 -- HD~166804}
        \label{subsection: EPIC227373493}
        This star has literature rotation periods of $P_{\rm rot} = 1.76009 \pm 0.00004$~d \citep{Bernhard2015a} and $P_{\rm rot} = 3.7035 \pm 0.0002$~d \citep{Hummerich2016a}, both derived using ASAS-3 photometry. More recently, \citet{Buysschaert2018a*} confirmed the presence of a large-scale magnetic field with a polar strength of $B_p \simeq 1.4$~kG, a projected surface rotational velocity of $v\,\sin\,i = 45 \pm 3$~km\,s$^{-1}$, and a rotation period of $P_{\rm rot} = 3.7 \pm 0.2$~d for HD~166804. We measure a rotation period of $P_{\rm rot} = 3.7024 \pm 0.0005$~d, which is significantly different from the literature value from \citet{Hummerich2016a}. The summary figure of EPIC~227373493 is shown in Fig.~\ref{figure: EPIC227373493}.

        
        \subsection{EPIC~227825246 -- HD~164085}
        \label{subsection: EPIC227825246}
        This star has a literature rotation period of $P_{\rm rot} = 1.22072 \pm 0.00002$~d derived from ASAS-3 photometry \citep{Hummerich2016a}. We measure a rotation period of $P_{\rm rot} = 1.22079 \pm 0.00002$~d, which is significantly different from the literature value from \citet{Hummerich2016a}. The summary figure of EPIC~227825246 is shown in Fig.~\ref{figure: EPIC227825246}.

        
        \subsection{EPIC~232147357 -- HD~153192}
        \label{subsection: EPIC232147357}
        This star has a literature rotation period of $P_{\rm rot} = 11.357 \pm 0.002$~d derived using ASAS-3 photometry \citep{Bernhard2015a}. We measure a rotation period of $P_{\rm rot} = 11.393 \pm 0.003$~d, which is significantly different from the literature value from \citet{Bernhard2015a}. The summary figure of EPIC~232147357 is shown in Fig.~\ref{figure: EPIC232147357}.

        
        \subsection{EPIC~232176043 -- HD~152834}
        \label{subsection: EPIC232176043}
        This star has a literature rotation period of $P_{\rm rot} = 4.2928 \pm 0.0003$~d derived from ASAS-3 photometry \citep{Hummerich2016a}. Recently, \citet{Buysschaert2018a*} confirmed the presence of a large-scale magnetic field with a polar strength of $B_p \simeq 700$~G, a projected surface rotational velocity of $v\,\sin\,i = 13 \pm 1$~km\,s$^{-1}$, and a rotation period of $P_{\rm rot} = 4.4 \pm 0.3$~d. We measure a rotation period of $P_{\rm rot} = 4.2919 \pm 0.0007$~d, which is significantly different from the literature value from \citet{Hummerich2016a}. The summary figure of EPIC~232176043 is shown in Fig.~\ref{figure: EPIC232176043}.

        
        \subsection{EPIC~232284277 -- HD~155127}
        \label{subsection: EPIC232284277}
        This star has a literature rotation period of $P_{\rm rot} = 5.5243 \pm 0.0004$~d derived from ASAS-3 photometry \citep{Hummerich2016a}. Recently, \citet{Buysschaert2018a*} confirmed the presence of a large-scale magnetic field with a polar strength of $B_p \simeq 1.3$~kG, and measured a projected surface rotational velocity of $v\,\sin\,i = 37 \pm 3$~km\,s$^{-1}$ and a rotation period of $P_{\rm rot} = 5.5 \pm 0.4$~d. We measure a rotation period of $P_{\rm rot} = 5.5171 \pm 0.0007$~d, which is significantly different from the literature value from \citet{Hummerich2016a}. The summary figure of EPIC~232284277 is shown in Fig.~\ref{figure: EPIC232284277}.

        
        \subsection{EPIC~247729177 -- HD~284639}
        \label{subsection: EPIC247729177}
        We report the first measurement of the rotation period of $P_{\rm rot} = 9.155 \pm 0.001$~d for HD~284639. Although \citet{Wraight2012a} commented HD~284639 has no obvious $P_{\rm rot}$ using STEREO photometry, we clearly detect rotational modulation in the K2 light curve and amplitude spectrum for HD~284639. The summary figure of EPIC~247729177 is shown in Fig.~\ref{figure: EPIC247729177}.


\section{Results: CP stars with rotational modulation and additional variability indicative of pulsations}
\label{section: rotation and pulsation results}
        
In this section, we discuss the stars that have measured rotation periods determined using K2 photometry and that also show additional variability indicative of stellar pulsations. The rotational modulation figures are given in Appendix~\ref{section: appendix: rotation and pulsation}.

        
        \subsection{EPIC~203749199 -- HD~152366}
        \label{subsection: EPIC203749199}
        This star has recently been confirmed to have a large-scale magnetic field with a polar strength of $B_p \simeq 250$~G, a measured rotation period of $P_{\rm rot} = 3.23 \pm 0.01$~d, and a projected surface rotational velocity of $v\,\sin\,i = 23 \pm 2$~km\,s$^{-1}$ by \citet{Buysschaert2018a*}. In this work, we measure a rotation period of $P_{\rm rot} = 3.2308 \pm 0.0001$~d. The summary figure of EPIC~203749199 is shown in Fig.~\ref{figure: EPIC203749199}. We also classify this star as a possible pulsating star because a significant and resolved peak at $\nu = 0.5079 \pm 0.0002$~d$^{-1}$ can be seen in the residual amplitude spectrum that is not associated with harmonics of the rotation frequency. However, if separate amplitude spectra for each of the two subsections of the light curve are examined, we find that the amplitude of this peak varies during the two subsections, which span a total of $\sim$\,960~d. Therefore, we cannot exclude the possibility that this frequency is instrumental. We find no evidence of contamination from a nearby source for this star.

        
        \subsection{EPIC~216956748 -- HD~181810}
        \label{subsection: EPIC216956748}       
        We report the first measurement of the rotation period of $P_{\rm rot} = 13.509 \pm 0.009$~d for HD~181810. The summary figure of EPIC~216956748 is shown in Fig.~\ref{figure: EPIC216956748}. We also classify this star as a candidate pulsating star because a significant and isolated peak can be seen at $5.9331 \pm 0.0005$~d$^{-1}$ in the residual amplitude spectrum, which cannot be associated with harmonics of the rotation frequency or multiples of the K2 thruster firing frequency. The peak at $5.9331 \pm 0.0005$~d$^{-1}$ could be a Nyquist alias frequency of a high-frequency roAp pulsation mode frequency or a \dsct pulsation mode frequency, therefore further observations at a high cadence are needed to confirm this star as a roAp star. A subharmonic of this peak found at $2.967 \pm 0.001$~d$^{-1}$ is also visible in the residual amplitude spectrum of HD~181810, but it has a $S/N = 1.4$ and so we do not consider it significant. Contamination from a nearby source is unlikely for this star, as the nearest possible source of contamination was not included in our pixel mask.

        
        \subsection{EPIC~217437213 -- HD~177016}
        \label{subsection: EPIC217437213}
        We report the first measurement of the rotation period of $P_{\rm rot} = 3.276 \pm 0.004$~d for HD~177016. The summary figure of EPIC~217437213 is shown in Fig.~\ref{figure: EPIC217437213}. We also classify this star as a candidate pulsating star because multiple resolved frequencies can be seen in the g-mode frequency regime in the residual amplitude spectrum, which are not associated with harmonics of the rotation frequency or multiples of the K2 thruster firing frequency. It is possible that the measured rotation period of HD~177016 is a g-mode pulsation frequency, but we conclude this to be unlikely because of the six significant harmonics. We find no evidence of contamination from a nearby source for this star. Regardless, this CP star clearly has variability indicative of pulsations and warrants further study.

        
        \subsection{EPIC~223573464 -- HD~161851}
        \label{subsection: EPIC223573464}
        We report the first measurement of the rotation period of $P_{\rm rot} = 1.6209 \pm 0.0007$~d for HD~161851. The summary figure of EPIC~223573464 is shown in Fig.~\ref{figure: EPIC223573464}. We also classify this star as a candidate pulsating star because a significant low-frequency power excess can be seen between $0.1 \leq \nu \leq 3.0$~d$^{-1}$ in its original and residual amplitude spectra, and its light curve exhibits multi-periodic variability. The low-frequency power excess, shown in Fig.~\ref{figure: EPIC223573464}, could be caused by unresolved g-mode pulsation frequencies making HD~161851 a candidate SPB star since it has a spectral type of A0~Si \citep{Renson2009}. The observed power-law nature is similar to that seen in numerical simulations of stochastically excited gravity waves \citep{Rogers2013b, Rogers2015, Aerts2015c, Aerts2017a, Aerts2018a, Simon-Diaz2018a}. However, contamination from a neighbouring cool star, for example a \gdor star, is a likely explanation since HD~161851 lies within a crowded region. On the other hand, the low-frequency power excess could be caused by the quality of the K2 photometry and reduction, but we find this unlikely because of the broad range and high amplitude of the observed power excess.

                        
        \subsection{EPIC~225191577 -- HD~164068}
        \label{subsection: EPIC225191577}               
        We report the first measurement of the rotation period of $P_{\rm rot} = 2.4130 \pm 0.0003$~d for HD~164068. The summary figure of EPIC~225191577 is shown in Fig.~\ref{figure: EPIC225191577}. We also classify this star as a candidate pulsating star because multiple significant peaks can be seen in the residual amplitude spectrum that are not associated with harmonics of the rotation frequency or multiples of the K2 thruster firing frequency. We interpret these peaks as possible g- and p-mode pulsation frequencies that are within the typical frequency range for \dsct stars \citep{Bowman_BOOK, Bowman2018a}. On the other hand, contamination from a background or nearby and possible companion star seems a likely explanation since HD~164068 lies within a crowded region. The nearest significant source of contamination to HD~164068 is J18002338-2258032, which is 15~arcsec away and 2.5~mag fainter in $V$ so flux from this star may be bleeding into the aperture mask of the target star. If the pulsation modes originate in this neighbouring star, their amplitudes are being diluted by a factor of 10 in flux, making them of order 5~mmag, which is within the typical p-mode amplitude range observed in \dsct stars \citep{Bowman_BOOK, Bowman2018a}. Therefore, it remains unclear if the observed pulsation modes originate in HD~164068, a companion, or contaminating star.

                                                
        \subsection{EPIC~225990054 -- HD~158596} 
        \label{subsection: EPIC225990054}
        This star has a literature rotation period of $P_{\rm rot} = 2.02206 \pm 0.00005$~d derived using ASAS-3 photometry \citep{Bernhard2015a}. Furthermore, this star was recently confirmed to have a large-scale magnetic field with a polar strength of $B_p \simeq 1.8$~kG, a rotation period of $P_{\rm rot} = 2.02 \pm 0.02$~d, and a projected surface rotational velocity of $v\,\sin\,i = 60 \pm 3$~km\,s$^{-1}$ by \citet{Buysschaert2018a*}. In this work, we measure a rotation period of $P_{\rm rot} = 2.02208 \pm 0.00003$~d, which is consistent with the literature value from \citet{Bernhard2015a}. The summary figure of EPIC~225990054 is shown in Fig.~\ref{figure: EPIC225990054}.
        
        Similar to \citet{Buysschaert2018a*}, we detect a significant isolated peak in the residual amplitude spectrum of HD~158596 at $17.0074 \pm 0.0004$~d$^{-1}$, and interpret it as a Nyquist alias frequency of a high-frequency pulsation mode. This frequency is the only significant peak found in the residual amplitude spectrum of HD~158596 and is present in both subsections (i.e. C09 and C11) of the K2 light curve.  We postulate that this frequency is a Nyquist alias and not an intrinsic p-mode frequency since mono-periodic \dsct stars appear to be rare when using high-quality space photometry. For example, from a sample of 983 \dsct stars observed by the nominal \Kepler mission, \citet{Bowman2016a} find only four such stars, i.e. less than half a percentage as an occurrence rate. Furthermore, the spectral type of HD~158596 is B9\,Si, which places it outside of the \dsct instability strip. We note that HD~158596 was observed by the K2 mission in campaigns 09 and 11, such that a significant gap is present in the available observations. This results in a complex window pattern structure in its amplitude spectrum, especially around the the K2 thruster firing frequency. 
        
        Contamination from a neighbouring faint source is also a possible explanation. HD~158596 has a nearby star with an angular separation of 0.3~arcsec \citep{ESA1997, Fabricius2002a} making it a visual double star and possibly part of a close-binary system, although no evidence of the companion star was found in spectroscopy of HD~158596 by \citet{Buysschaert2018a*}. High-cadence photometry of HD~158596 is required to confirm that the peak at $17.0074 \pm 0.0004$~d$^{-1}$ is a Nyquist alias of a high-frequency roAp pulsation mode.

        
        \subsection{EPIC~227231984 -- HD~158336} 
        \label{subsection: EPIC227231984}
        This star has a literature rotation period of $P_{\rm rot} = 1.76009 \pm 0.00004$~d derived using ASAS-3 photometry \citep{Bernhard2015a}. We measure a rotation period of $P_{\rm rot} = 1.7609 \pm 0.0008$~d, which is consistent with the literature value from \citet{Bernhard2015a}. The summary figure of EPIC~227231984 is shown in Fig.~\ref{figure: EPIC227231984}. We also classify this star as a candidate pulsating star because a significant low-frequency power excess can be seen between $0.1 \leq \nu \leq 3.0$~d$^{-1}$ in its original and residual amplitude spectra. This power excess could be caused by unresolved g-mode pulsation frequencies, making HD~158336 a candidate SPB star since it has a spectral type of B9~Si \citep{Renson2009}. The observed power-law nature is similar to that seen in numerical simulations of stochastically excited gravity waves \citep{Rogers2013b, Rogers2015, Aerts2015c, Aerts2017a, Aerts2018a, Simon-Diaz2018a}. It is difficult to exclude an imperfect reduction as a possible cause of the low-frequency power excess, but we find this unlikely because of the broad range and high amplitude of the observed power excess.


        \subsection{EPIC~227305488 -- HD~166542}
        \label{subsection: EPIC227305488}
        We report the first measurement of the rotation period of $P_{\rm rot} = 3.6331 \pm 0.0003$~d for HD~166542. The summary figure of EPIC~227305488 is shown in Fig.~\ref{figure: EPIC227305488}. We also classify this star as a candidate pulsating star because an isolated peak at $16.429 \pm 0.004$~d$^{-1}$ can be seen in the residual amplitude spectrum, which is not associated with a harmonic of the rotation frequency or a multiple of the K2 thruster firing frequency. {Contamination from a neighbouring source is also possible since HD~166542 lies within in a crowded region.} This peak could be a Nyquist alias of a high-frequency pulsation mode, thus high-cadence photometry of HD~166542 is required to confirm this star as a roAp star and determine the frequency of its intrinsic pulsation modes.

                
        \subsection{EPIC~228293755 -- HD~165945}
        \label{subsection: EPIC228293755}
        We report the first measurement of the rotation period of $P_{\rm rot} = 2.13522 \pm 0.00008$~d for HD~165945. The summary figure of EPIC~228293755 is shown in Fig.~\ref{figure: EPIC228293755}. We also classify this star as a candidate pulsating star because two significant peaks can be seen in the residual amplitude spectrum that are not associated with harmonics of the rotation frequency or multiples of the K2 thruster firing frequency. Contamination is possible for HD~165945 since it lies within a region with multiple faint background sources. We interpret these peaks as g-mode pulsation frequencies, thus HD~165945 warrants further study to establish these pulsations originate in the target or a background source.


        \subsection{EPIC~230753303 -- HD~153997}
        \label{subsection: EPIC230753303}
        We report the first measurement of the rotation period of $P_{\rm rot} = 5.4966 \pm 0.0006$~d for HD~153997. The summary figure of EPIC~230753303 is shown in Fig.~\ref{figure: EPIC230753303}. We also classify this star as a candidate pulsating star because a significant peak can be seen in the residual amplitude spectrum that is not associated with a harmonic of the rotation frequency or a multiple of the K2 thruster firing frequency. {Contamination from a neighbouring source is also possible for HD~153997.} We interpret this peak as a g-mode pulsation frequency, thus HD~153997 warrants further study to establish if this variability originates in the target or a background source.


\section{Results: CP stars with intrinsic variability indicative of stellar pulsations yet lacking significant photometric rotational modulation}
\label{section: pulsation results}

In this section, CP stars in our sample that show no evidence of rotational modulation, but show variability indicative of pulsations are discussed. For pulsating stars, and especially for g modes, it is difficult to extract individual pulsation mode frequencies, because of the poor frequency resolution of a {$\sim$}80-d light curve being of order 0.01~d$^{-1}$. It is known that g and p modes can be spaced closer than 0.001~d$^{-1}$ in frequency \citep{Bowman2016a, Bowman_BOOK}, which produces complex beating patterns and an unresolved power excess in an amplitude spectrum instead of individual pulsation mode frequencies. 

In our sample of CP stars, we find {five} stars that show variability consistent with multiple g-mode pulsation frequencies and {one} star that shows variability consistent with multiple p-mode pulsation frequencies. In all cases, a rotation period was purposefully not determined for these stars because of the likelihood that peaks in the low frequency ($0 \leq \nu \leq 4$) are either (unresolved) g-mode pulsation frequencies or combination frequencies of p-mode pulsation frequencies (see e.g. \citealt{Degroote2009a, Degroote2012b, Thoul2013, Kurtz2015b, Bowman_BOOK}). The available K2 photometry and amplitude spectra are shown in Appendix~\ref{section: appendix: pulsation}; all the significant (pulsation mode) frequencies are identified by vertical blue lines in the residual amplitude spectrum and significant peaks are defined as those that have an amplitude $S/N \geq 4$ \citep{Breger1993b}.


        \subsection{EPIC~202060145}
        \label{subsection: EPIC202060145}
        This star is a member of the open cluster M67 and has been determined to be part of a double star system using speckle interferometry obtained at the US naval observatory, but a binary orbit solution is yet to be determined \citep{Mason2002, Mason2004, Hartkopf2011}. A dominant period of 1.1472~d was measured by \citet{Armstrong2015a} using an automated classification algorithm applied to the K2 photometry, but this period was labelled as quasi-periodic. In this work, we measure a dominant period of $1.1456 \pm 0.0008$~d using K2 photometry, but the amplitude spectrum of EPIC~202060145 clearly contains multiple low-frequency peaks that we interpret as g-mode pulsation frequencies. {Since EPIC~202060145 is part of close double-star system, it is possible that the observed pulsation mode frequencies originate from the unseen companion.} The summary figure of EPIC~202060145 is shown in Fig.~\ref{figure: EPIC202060145}.


        \subsection{EPIC~203917770 -- HD~145792}
        \label{subsection: EPIC203917770}
        This star has been confirmed to have a large-scale magnetic field \citep{Thompson1987} and part of a close-binary system with a projected surface rotational velocity of $v\,\sin\,i = 30 \pm 8$~km\,s$^{-1}$ by \citet{Zorec2012}. The spectral type of HD~145792 is B6~He~weak from the \citet{Renson2009} catalogue making this star a CP\,4 star. The rotation period of HD~145792 is somewhat disputed as \citet{Renson2009} measure a value of $P_{\rm rot}$ of 0.8478~d from spectroscopy and \citet{Wraight2012a} measure approximately twice this value, i.e. $P_{\rm rot} = 1.6956 \pm 0.0005$~d using STEREO photometry. In this work, we measure a dominant period of $0.84815 \pm 0.00006$~d, which is similar to the value found by \citet{Renson2009} and approximately half the rotation period from \citet{Wraight2012a}. The summary figure of EPIC~203917770 is shown in Fig.~\ref{figure: EPIC203917770}. A comparable dominant period of 0.847862~d was also measured by \citet{Armstrong2016b} using an automated classification algorithm applied to K2 photometry, although the physical cause of the variability was not discussed in detail. We interpret the variability in HD~145792, specifically the isolated peaks in its amplitude spectrum, as g-mode pulsation frequencies. None of the low-frequency peaks have a significant harmonic, thus are unlikely to represent rotational modulation caused by surface abundance inhomogeneities. Since this star is part of a close-binary system, we cannot exclude contamination as the source of the observed pulsational variability.

                
        \subsection{EPIC~206120416 -- HD~210424}
        \label{subsection: EPIC206120416}
        This star has a literature rotation period of $P_{\rm rot} = 3.9613 \pm 0.0033$~d derived from STEREO photometry \citep{Wraight2012a}. However, the corresponding frequency peak has an amplitude less than 80~$\mu$mag and $S/N < 1.0$ in our K2 photometry thus it is not significant. We find no evidence of contamination from a nearby source for this star. We measure a dominant period of $P = 1.4478 \pm 0.0005$~d ($\nu = 0.6907 \pm 0.0005$~d$^{-1}$); a comparable dominant period of 1.4465~d is also measured by \citet{Armstrong2016b} via an automated classification algorithm applied to K2 photometry, although the physical cause of this variability was not discussed by \citet{Armstrong2016b}. The summary figure of HD~210424 is shown in Fig.~\ref{figure: EPIC206120416}, in which three relatively high-amplitude peaks at $\nu_1= 0.4738 \pm 0.0006$~d$^{-1}$, $\nu_2 = 0.6907 \pm 0.0005$~d$^{-1}$ and $\nu_3 = 0.959 \pm 0.001$~d$^{-1}$ can be seen in the amplitude spectrum.
        
        Upon first inspection, it would appear that possible harmonics of a possible rotation frequency are within the amplitude spectrum (i.e. $2\nu_1 = \nu_3$) given the Rayleigh frequency resolution (0.0145~d$^{-1}$) of our data. However, given the large variance in the residual amplitude spectra of HD~210424 around these frequencies, it is clear that they are not well-resolved in the limited length of the 67.7-d K2 observations. This, coupled with the fact that none of these frequencies are compatible with the literature rotation period for this star from \citet{Wraight2012a}, led us to the conclusion that these low-frequency peaks and the low-frequency power excess is caused by the beating of multiple unresolved pulsation mode frequencies (see e.g. \citealt{Degroote2011, Bowman2016a, Bowman_BOOK}). 
        
        From inspection of the light curve and amplitude spectrum in Fig.~\ref{figure: EPIC206120416}, HD~210424 is clearly a multi-periodic pulsator, so caution is needed when identifying the correct rotation frequency caused by surface abundance inhomogeneities. To that end, we include $P_{\rm rot} = 2.111 \pm 0.003$~d using $\nu_1$ as a possible rotation period for HD~210424 in Table~\ref{table: rotation results}. If $\nu_1$ is indeed the rotation frequency and $\nu_3$ is its harmonic, this does not explain the dominant frequency $\nu_2$, nor does it explain the low-frequency power excess, which has amplitudes of order 200~$\mu$mag at frequencies between $0 < \nu < 1.5$~d$^{-1}$. We conclude that HD~210424 is a multi-periodic pulsator with an amplitude spectrum that can be explained by multiple unresolved g~modes with possible amplitude and/or frequency modulation (see e.g. \citealt{Degroote2011, Bowman2016a, Bowman_BOOK}), or travelling gravity waves \citep{Rogers2013b, Rogers2015, Aerts2015c, Aerts2017a, Aerts2018a, Simon-Diaz2018a}.

        
        \subsection{EPIC~206326769 -- HD~211838}
        \label{subsection: EPIC206326769}
        This star is reported as a spectroscopic binary (SB1) and has a  ND for a large-scale magnetic field \citep{Makaganiuk2011a}, as expected for a CP\,3 star with a spectral type of B8~MgMn \citep{Renson2009}. More recently, a study by \citet{Paunzen2013a} measured a rotation period of $P_{\rm rot} = 6.5633 \pm 0.0063$~d using STEREO photometry and a projected surface rotational velocity of $v\,\sin\,i = 65.0 \pm 6.9$~km\,s$^{-1}$ for HD~211838. The summary figure for our K2 observations of EPIC~206326769 is shown in Fig.~\ref{figure: EPIC206326769}. The rotation period measured by \citet{Paunzen2013a} is not statistically significant in our K2 photometry becasue it has an amplitude of less than 100~$\mu$Hz and $S/N < 4$. Most interestingly, \citet{Paunzen2013a} also report the presence of \gdor and \dsct pulsations in HD~211838. In our K2 photometry, we measure a dominant period of $P = 1.1203 \pm 0.0002$~d ($\nu = 0.892 \pm 0.0002$~d$^{-1}$). A comparable dominant period of 1.119692~d was also measured by \citet{Armstrong2016b} using an automated classification algorithm applied to K2 photometry, although the physical cause of this variability was not discussed in detail. We also detect three additional frequencies in the amplitude spectrum of HD~211838, which are $0.4746 \pm 0.0002$~d$^{-1}$, $0.3927 \pm 0.0005$~d$^{-1}$, and $0.3251 \pm 0.0005$~d$^{-1}$, but the latter two have $S/N = 3.6$ and so are not shown as vertical blue lines in Fig.~\ref{figure: EPIC206326769}. We interpret the variability in HD~211838, specifically the isolated peaks in its amplitude spectrum, as pulsation mode frequencies. The low-frequency peaks are likely g-mode pulsation frequencies and not rotational modulation caused by surface abundance inhomogeneities. Contamination from the companion binary star or background sources is also possible for HD~211838 since it is a bright star in a SB1 system and requires a large pixel mask that likely contains flux from other stars.

        
        \subsection{EPIC~224947037 -- HD~162814}
        \label{subsection: EPIC224947037}
        We measure a dominant period of $P = 0.66203 \pm 0.00008$~d ($\nu = 1.5105 \pm 0.0002$~d$^{-1}$) for HD~162814 and interpret its variability as multi-periodic pulsation mode frequencies. The summary figure of EPIC~224947037 is shown in Fig.~\ref{figure: EPIC224947037}. The low-frequency peaks are likely g-mode pulsation frequencies and not rotational modulation caused by surface abundance inhomogeneities because no significant harmonics of any frequencies were detected in the amplitude spectrum. We detect a possible rotation period of $P_{\rm rot} = 8.62 \pm 0.07$~d ($\nu = 0.116 \pm 0.001$~d$^{-1}$) for HD~162814, which has an amplitude of $166 \pm 9$~$\mu$mag and a $S/N = 5.7$ in the amplitude spectrum, but it has no significant harmonics. This frequency may be a (difference) combination frequency of the dominant frequency at $1.5105 \pm 0.0002$~d$^{-1}$ and a frequency peak at $1.39 \pm 0.01$~d$^{-1}$, the latter of which has $S/N \simeq 3.5$, so it is not significant. Contamination is also a possible explanation of the multiple pulsation mode frequencies in the light curve and amplitude spectrum of HD~162814, as it lies within a crowded field.

        
        \subsection{EPIC~246152326 -- HD~220556}
        \label{subsection: EPIC246152326}
        HD~220556 is one of the more interesting stars in our sample because of its rich pulsation frequency spectrum; its summary figure is shown in Fig.~\ref{figure: EPIC246152326}. We find no significant evidence of rotational modulation consistent with surface abundance inhomogeneities in the multi-periodic light curve of HD~220556. {From studying the pixel masks for HD~220556, we also find no evidence of contamination from a nearby or background source.} There are, however, multiple observed frequencies between $10 \leq \nu \leq 22$~d$^{-1}$ in the amplitude spectrum of HD~220556. 
        
        It should be noted that the spectral type of HD~220556 is given as A2~SrEuCr by \citet{Renson2009}, but this star is listed as Ap SrEuCr in the Michigan catalogue \citep{Houk1999}. Furthermore, HD~220556 is included in the AAVSO Photometric All-Sky Survey (APASS) of RAdial Velocity Experiment (RAVE) stars, and \citet{Munari2014f} derive an approximate effective temperature of $T_{\rm eff} \simeq 7500$~K. This places HD~220556 within the classical instability strip with theoretical pulsation models for main-sequence \dsct stars predicting p-mode pulsation frequencies between $5 \lesssim \nu \lesssim 70$~d$^{-1}$ \citep{Pamyat1999b, Pamyat2000a}. Therefore, we conclude that HD~220556 is a multi-periodic \dsct pulsator.
        
        With only LC K2 data, we are unable to probe frequencies above the LC Nyquist frequency of 24.47~d$^{-1}$. Furthermore, only {$\sim$}80~d of K2 data prevents us from applying the super-Nyquist technique developed by \citet{Murphy2013a}, which requires at least 1~yr of \Kepler observations to distinguish Nyquist alias and real frequencies in an amplitude spectrum. If the observed frequencies between $10 \lesssim \nu \lesssim 20$~d$^{-1}$ are in fact Nyquist alias frequencies of high-frequency p~modes, then the real pulsation modes would lie either between $27 \lesssim \nu \lesssim39$~d$^{-1}$ or between $52 \lesssim \nu \lesssim 64$~d$^{-1}$. The observed and two possible frequency regimes are plausible, yet high-frequency p~modes are expected and more commonly observed in hot \dsct stars \citep{Baglin1973a, Breger2000b, Bowman2018a}. The real pulsation modes would also have higher amplitudes than their Nyquist aliases \citep{Murphy2013a, Bowman2016a}. Therefore the detection of p~modes in a CP star, which is expected to have a large-scale magnetic field, is an interesting discovery and warrants further study since a strong large-scale magnetic field is predicted to damp low-overtone p modes \citep{Saio2005, Saio2014a}. If such a field exists in HD~220556, it is as yet to be determined.


\section{Discussion and conclusions}
\label{section: discussion}

We have searched for variability in the light curves of {56} CP\,2, 3 and 4 stars using high-quality optimised photometry from the K2 space mission. For {12} stars in our sample, we do not detect any significant photometric variability, so we conclude that if these stars are photometrically variable then their rotation periods must be comparable or longer than the available {$\sim$}80~d of K2 space photometry. This is certainly the case for EPIC~214503319 (HD~177765), which is a known magnetic star with a literature rotation period of $P_{\rm rot} \gg 5$~yr \citep{Mathys1997a, Mathys2017a}; our detection of the Nyquist alias of its high-frequency roAp pulsation modes is shown in Fig.~\ref{figure: EPIC214503319}.
        
\begin{figure}
\centering
\includegraphics[width=0.99\columnwidth]{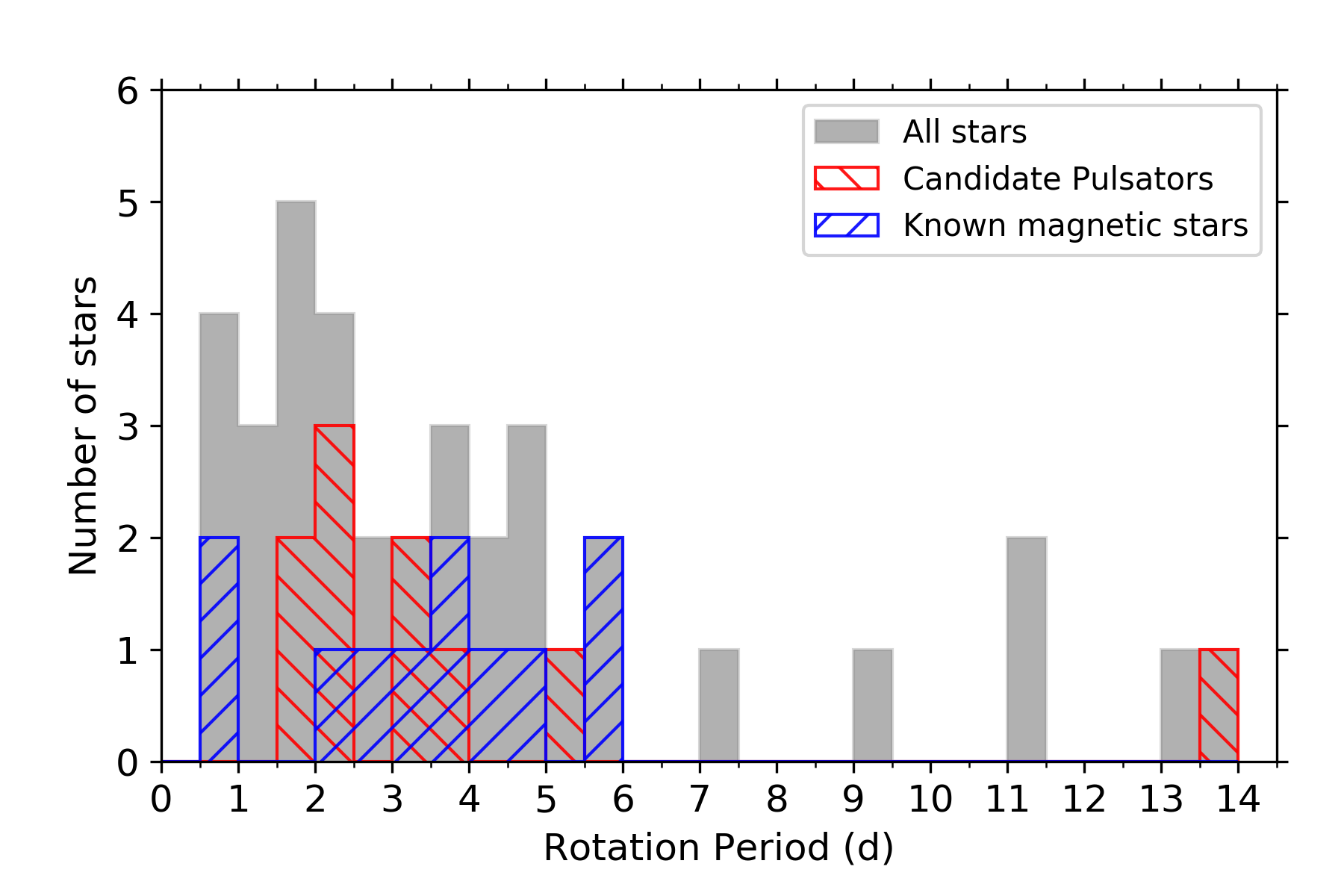}
\caption{Distribution of rotation periods for all {38} CP stars for which a rotation period is determined is shown in black. The distribution for the subset of candidate pulsating CP stars is shown in red, and those that have confirmed large-scale magnetic fields are shown in blue.}
\label{figure: rotation histogram}
\end{figure}

We detect rotational modulation caused by surface abundance inhomogeneities and measure the rotation period of {38} stars in our sample, of which 16 are new detections that were previously unknown in the literature. Our results for all CP stars for which a rotation period using K2 space photometry are summarised in Table~\ref{table: rotation results}, including a comparison to literature values if they are available. We also show the distribution of the rotation periods for the {38} CP stars in Fig.~\ref{figure: rotation histogram}, in which the grey region denotes all 38 stars in our sample with measured rotation periods, and the red and blue hatched regions denoting the candidate pulsators and known magnetic stars, respectively. The majority of our stars are moderate to slow rotators with rotation periods that range from $P_{\rm rot} = 0.73156 \pm 0.00001$~d in EPIC~226241087 (HD~164224) to $P_{\rm rot} = 15.733 \pm 0.002$~d in EPIC~210964459 (HD~26571), which are typical for ApBp stars (see e.g. \citealt{Adelman2002d, Mathys2004a}). However, since only {$\sim$}80~d of K2 photometry is available for a star in each K2 campaign, our results are limited to stars with rotation periods shorter than the length of the available K2 data.
        
The majority of the CP stars in our sample show clear rotational modulation in the K2 space photometry, and as theory predicts, these stars with surface abundance inhomogeneities should host a large-scale magnetic field \citep{Stibbs1950}. A precise determination of the rotation period of a magnetic CP star is important for determining the topology of the magnetic field in a star, including if the magnetic field is more complex than a simple dipolar field (see e.g. \citealt{Shultz2018b}). We have successfully identified pulsating CP stars that are preferred targets to follow up with high-resolution spectropolarimetry to confirm their spectral type, rotation period, and quantify the strength of the magnetic field at the stellar surface. 

However, there are notable examples of stars within our sample for which we determine a rotation period that is significantly different from other studies. For example, we calculate a rotation period of $2.403 \pm 0.002$~d using K2 photometry for EPIC~203092367 (HD~150035), whereas \citet{Wraight2012a} determined a rotation period of $2.3389 \pm 0.0013$~d using STEREO photometry. There are also stars in our sample for which we detect a clear rotation period yet other studies do not (e.g. EPIC~210964459; HD~26571), and vice versa. These differences only occur for a minority of stars in our sample, but highlight the need to investigate the stability of rotational modulation and its impact on determining stellar rotation periods in CP stars.
        
We identify {ten} CP stars with clear and periodic rotational modulation caused by stable surface abundance inhomogeneities, but also show variability caused by stellar pulsations. After extracting the rotational modulation in these stars, we find significant peaks in the residual amplitude spectra of these stars that are likely pulsation mode frequencies since they were not associated with harmonics of the rotation frequency or K2 thruster firing frequencies. It is true that contamination from a background star, binary companion, or a neighbouring star could be responsible for the significant frequency peaks in these stars, yet we made sure to minimise sources of contamination when creating our customised pixel masks for each star. There is also no obvious dependence of the detection of pulsations and the spectral types of these stars. Of course, our study is not the first detection of pulsations in intermediate-mass CP stars. For example, \dsct stars with chemical peculiarities include HD~41641 \citep{Escorza2016a} and HD~188774 \citep{Neiner2015f}, the latter of which has also been confirmed to host a magnetic field using spectropolarimetry. 
                
One star in our sample, EPIC~214503319 (HD~177765), is a known roAp pulsator \citep{Alentiev2012}, for which we detect the Nyquist alias frequency in our K2 observations as shown in Fig.~\ref{figure: EPIC214503319}. We also identify three stars as candidate roAp stars, EPIC~225990054 (HD~158596), EPIC~227305488 (HD~166542), and EPIC~216956748 (HD~181810), since only a single isolated peak is present in their residual amplitude spectra that could be a Nyquist alias of a high-frequency roAp pulsation mode. We note that EPIC~225990054 (HD~158596) was previously identified as a candidate roAp pulsator by \citet{Buysschaert2018a*}. In the future, we will obtain high-cadence ground-based photometry of these stars to confirm the presence of high-frequency roAp pulsation modes in these stars, which would represent a further increase in the members of this rare type of pulsating magnetic CP star. 
        
Furthermore, we find {six} stars that show multiple frequencies in their amplitude spectra that are indicative of stellar pulsations, yet lack rotational modulation: i.e. EPIC~202060145, EPIC~203917770 (HD~145792), EPIC~206120416 (HD~210424), EPIC~206326769 (HD~211838), EPIC~224947037 (HD~162814), and EPIC~246152326 (HD~220556). It should be noted that three of these stars have literature rotation periods, but we find no significant evidence that these literature rotation periods are correct according to our K2 photometry. This, coupled with the fact that all of these six stars are multi-periodic, lead us to the conclusion that these stars are intrinsically variable from pulsation modes. Unfortunately, a single {$\sim$}80-d campaign of K2 space photometry is insufficient to confidently extract pulsation mode frequencies, which can be spaced as close as 0.001~d$^{-1}$ \citep{Bowman2016a, Bowman_BOOK}, and our observations are not long enough to sufficiently resolve pulsation mode frequencies spaced closer than 0.013~d$^{-1}$ in the best of cases. In the near future, the {\it Transiting Exoplanet Survey Satellite} (TESS) will provide high-quality photometry across almost the entire sky with time spans up to 1~yr in its two continuous viewing zones \citep{Ricker2015}. The TESS mission will increase the available photometry of many previously studied stars, and allow us to vastly expand our search for pulsating magnetic stars.

The rapidly rotating B9 star HD~174648 was studied by \citet{Degroote2011} using {$\sim$}30~d of CoRoT photometry and these authors concluded that the significant frequencies detected in the amplitude spectrum were unlikely to be g~modes because the star has an effective temperature of $T_{\rm eff} = 11\,000 \pm 1000$~K and a surface gravity of $\log\,g = 4.2 \pm 0.3$~cm\,s$^{-2}$. Thus HD~174648 is in between the cool edge of the SPB g-mode instability region and the hot edge of the classical instability strip, which is similar to many of the CP stars in our study. \citet{Degroote2011} performed a detailed study to investigate whether closely spaced frequencies separated by 0.05~d$^{-1}$ could be explained by latitudinal surface differential rotation in HD~174648; the measured frequency splittings correspond to approximately 1~$\%$ difference in the equatorial and polar surface rotation periods. Clearly, distinguishing g-mode pulsation frequencies and differential latitudinal surface rotation is non-trivial, especially since pulsation mode frequencies can be spaced closer than 0.001~d$^{-1}$ \citep{Bowman2016a, Bowman_BOOK}. 

The CP stars in our sample, which unlike the fast-rotating B9 star HD~174648 \citep{Degroote2011}, have rotation periods of order days and are expected to have large-scale magnetic fields that instigate uniform rotation within their radiative envelopes \citep{Moss1992b, Browning2004d, Browning2004a, Mathis2005a, Zahn2011a}. Furthermore, the relatively short time spans of K2 mission data coupled with their slow rotation periods severely limits the ability to detect differential latitudinal surface rotation. In the best cases for stars with {$\sim$}80~d of K2 photometry is available; this corresponds to a Rayleigh frequency resolution of approximately 0.01~d$^{-1}$ meaning that rotationally split pulsation modes and/or differential latitudinal surface rotation is unlikely to be detectable for slowly rotating CP stars. As also stated by \citet{Degroote2011}, it is difficult to impossible to differentiate between unresolved pulsation mode frequencies and (latitudinal surface differential) rotational modulation with photometric data of limited length.

The discovery of pulsations in a non-negligible fraction of CP stars using K2 space photometry make these stars high-priority targets for spectropolarimetric follow-up campaigns to confirm the presence of a large-scale magnetic field. Magnetic pulsating stars are rare and provide the opportunity to test theoretical models of pulsation in the presence of a magnetic field (see e.g. \citealt{Shibahashi2000b, Neiner2012a, Handler2012a, Briquet2012, Henrichs2013, Buysschaert2017b, Buysschaert2018b*}). This is especially true for roAp stars with currently only 61 of these pulsators known; see \citet{Smalley2015} and \citet{Joshi2016} for recent catalogues. Almost all of the CP stars identified as pulsator candidates in our study are bright ($V \lesssim 10$ mag), making them preferred targets for ground-based follow-up spectropolarimetric campaigns to confirm their spectral type and rotation period, and the topology of their magnetic field, if present. This is especially true for the three stars identified in our study as candidate roAp stars, HD~158596, HD~166542, and HD~181810, and the candidate magnetic \dsct star HD~220556.

The relatively poor frequency resolution and quality of K2 photometry compared to the original 4-yr \Kepler mission, also has large implications for the successful application of asteroseismology in terms of forward seismic modelling (see \citealt{Buysschaert2018b*} and \citealt{Aerts2018b**}). Specifically, asteroseismic studies require multiple pulsation mode frequencies to be uniquely identified in terms of their geometry (i.e. radial order $n$, angular degree $\ell,$ and azimuthal order $m$). For g~modes in main-sequence B, A, and F stars, the commonly used method of mode identification is to use period spacing patterns (see e.g. \citealt{Degroote2010a, VanReeth2015b, Papics2015, Papics2017a}), which allow the physics of the near-core region to be probed. This method of mode identification is non-trivial even for stars with high-quality observations spanning years, such as the nominal \Kepler mission. The quantitative comparison of observed pulsation mode frequencies with theoretical predictions requires long-term data sets in order to achieve a good frequency resolution. Our study clearly demonstrates that a non-negligible fraction of CP stars are candidate pulsators. Light curves of sufficient frequency resolution to perform asteroseismic modelling for many CP stars are expected in the near future by the TESS mission \citep{Ricker2015} and later from the PLATO mission \citep{Rauer2014}.


\begin{acknowledgements}
We thank the referee for his or her comments that improved the manuscript, and the {\it Kepler}/K2 science teams for providing such excellent data. Funding for the {\it Kepler}/K2 mission is provided by NASA's Science Mission Directorate. The K2 data presented in this paper were obtained from the Mikulski Archive for Space Telescopes (MAST). Support for MAST for non-HST data is provided by the NASA Office of Space Science via grant NNX09AF08G and by other grants and contracts. The research leading to these results has received funding from the European Research Council (ERC) under the European Union's Horizon 2020 research and innovation programme (grant agreement N$^{\rm o}$670519: MAMSIE). This research has made use of the SIMBAD database, operated at CDS, Strasbourg, France; the SAO/NASA Astrophysics Data System; and the VizieR catalogue access tool, CDS, Strasbourg, France. 
\end{acknowledgements}


\bibliographystyle{aa}
\bibliography{/Users/Dom/Documents/RESEARCH/Bibliography/master_bib}



\begin{appendix}


\clearpage
\section{CP stars with rotational modulation}
\label{section: appendix: rotation}

In this section, the light curves and amplitude spectra of stars with rotational modulation caused by surface abundance inhomogeneities are provided.

\begin{figure*}
\centering
\includegraphics[width=0.95\textwidth]{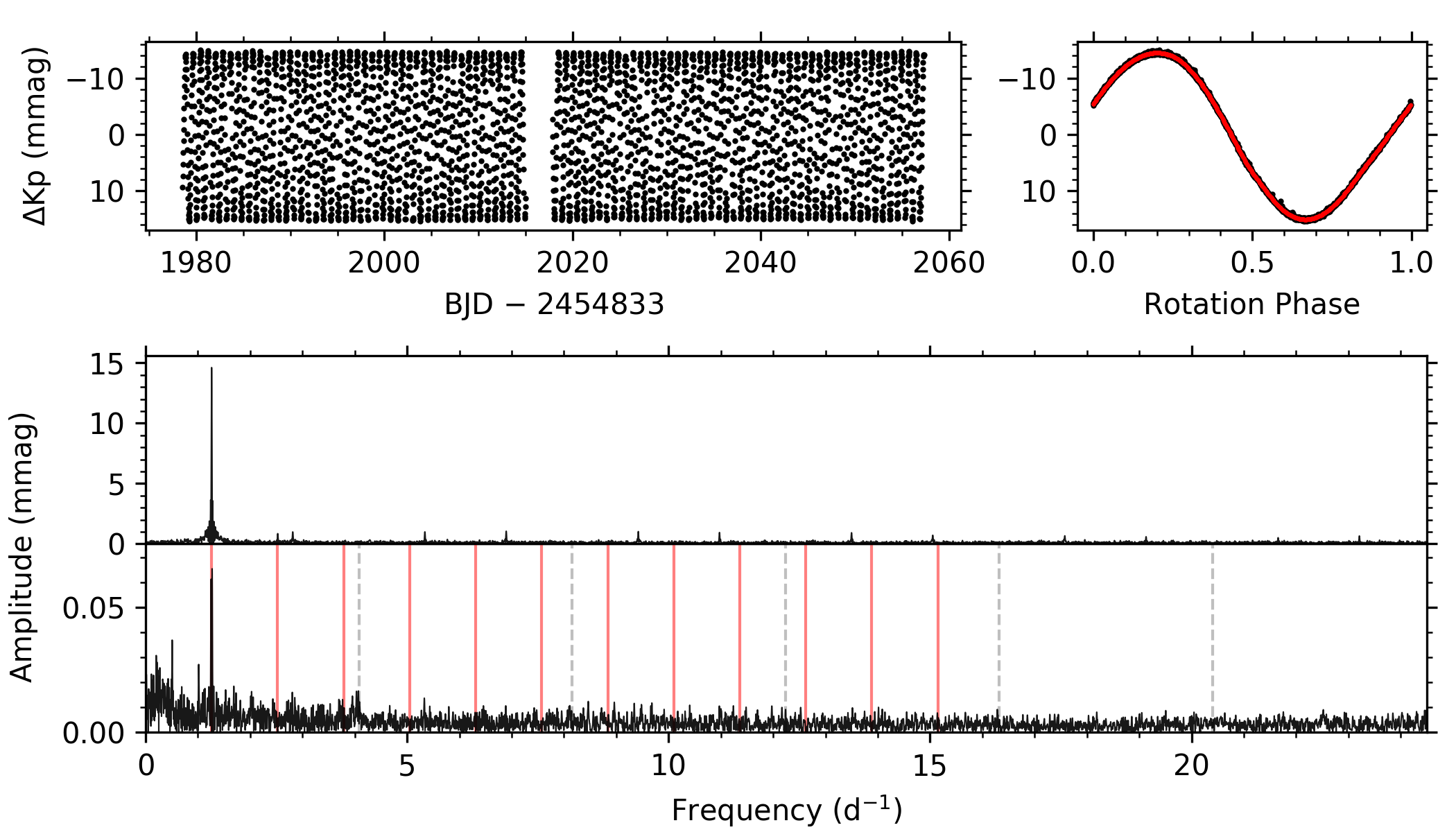}
\caption{Rotational modulation in EPIC~201777342 (HD~97859); same layout shown as in Fig.~\ref{figure: example}.}
\label{figure: EPIC201777342}
\end{figure*}

\begin{figure*}
\centering
\includegraphics[width=0.95\textwidth]{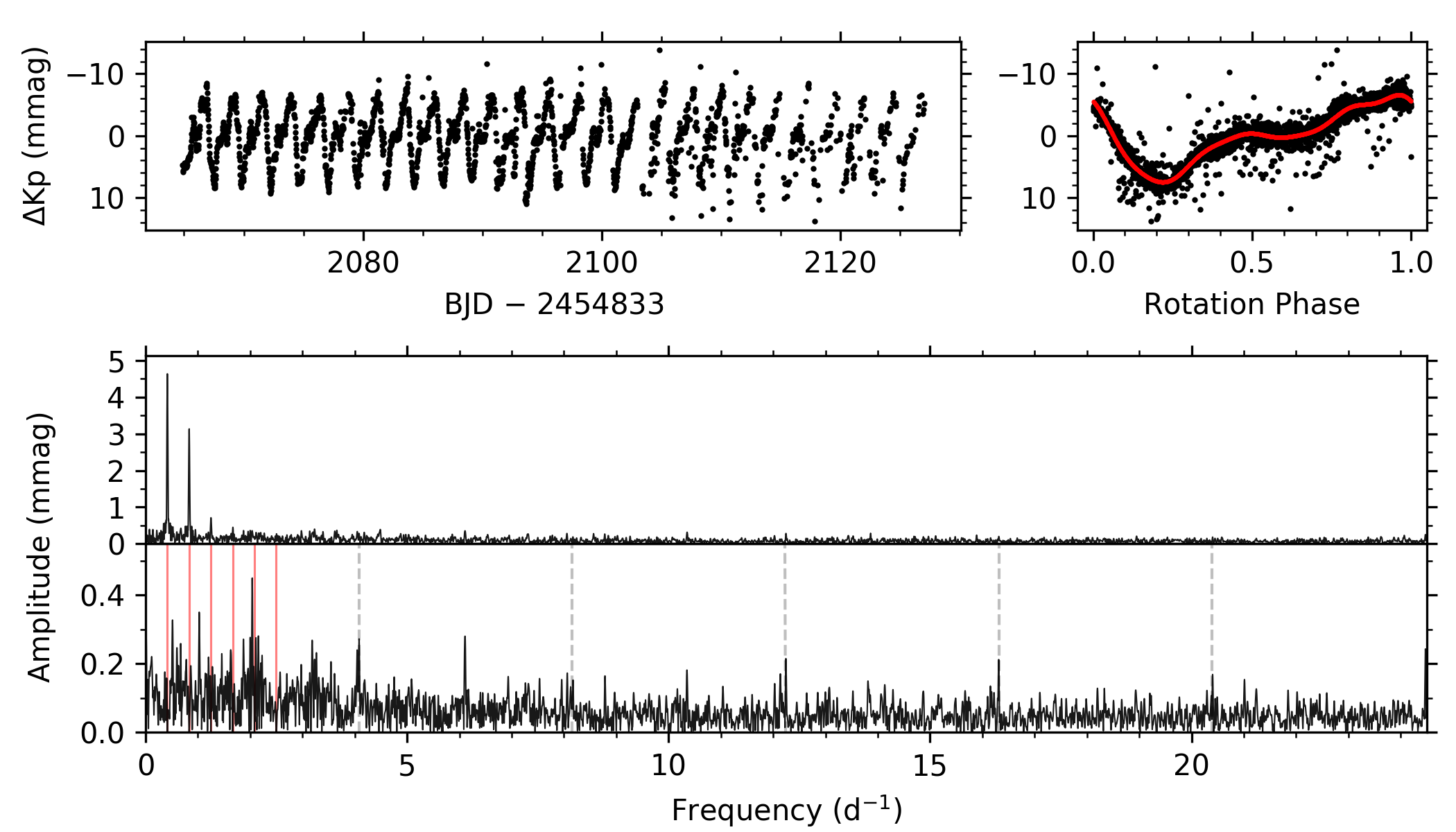}
\caption{Rotational modulation in EPIC~203092367 (HD~150035); same layout shown as in Fig.~\ref{figure: example}.}
\label{figure: EPIC203092367}
\end{figure*}

\clearpage 

\begin{figure*}
\centering
\includegraphics[width=0.95\textwidth]{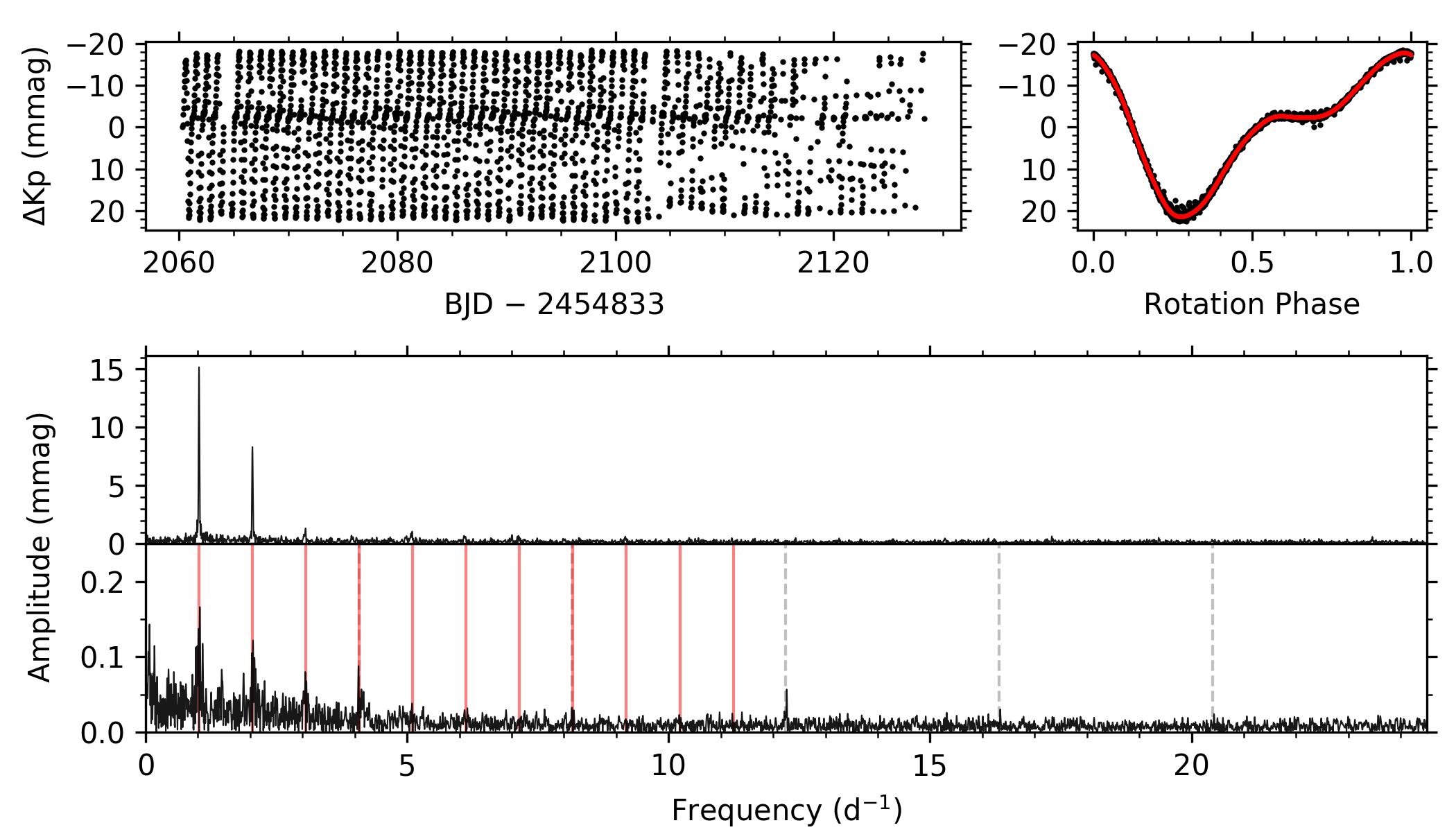}
\caption{Rotational modulation in EPIC~203814494 (HD~142990); same layout shown as in Fig.~\ref{figure: example}.}
\label{figure: EPIC203814494}
\end{figure*}

\begin{figure*}
\centering
\includegraphics[width=0.95\textwidth]{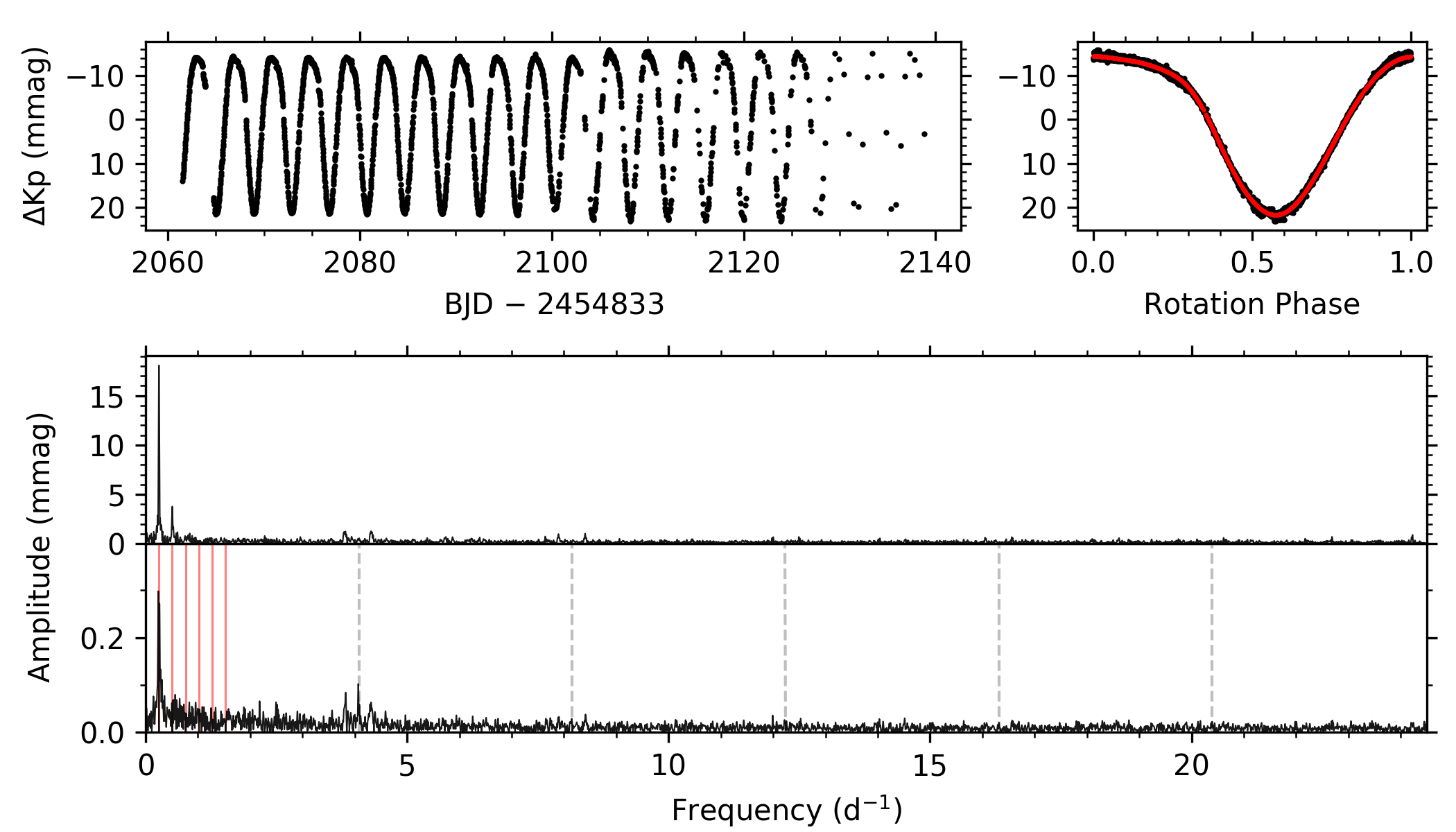}
\caption{Rotational modulation in EPIC~204964091 (HD~147010); same layout shown as in Fig.~\ref{figure: example}.}
\label{figure: EPIC204964091}
\end{figure*}

\clearpage 

\begin{figure*}
\centering
\includegraphics[width=0.95\textwidth]{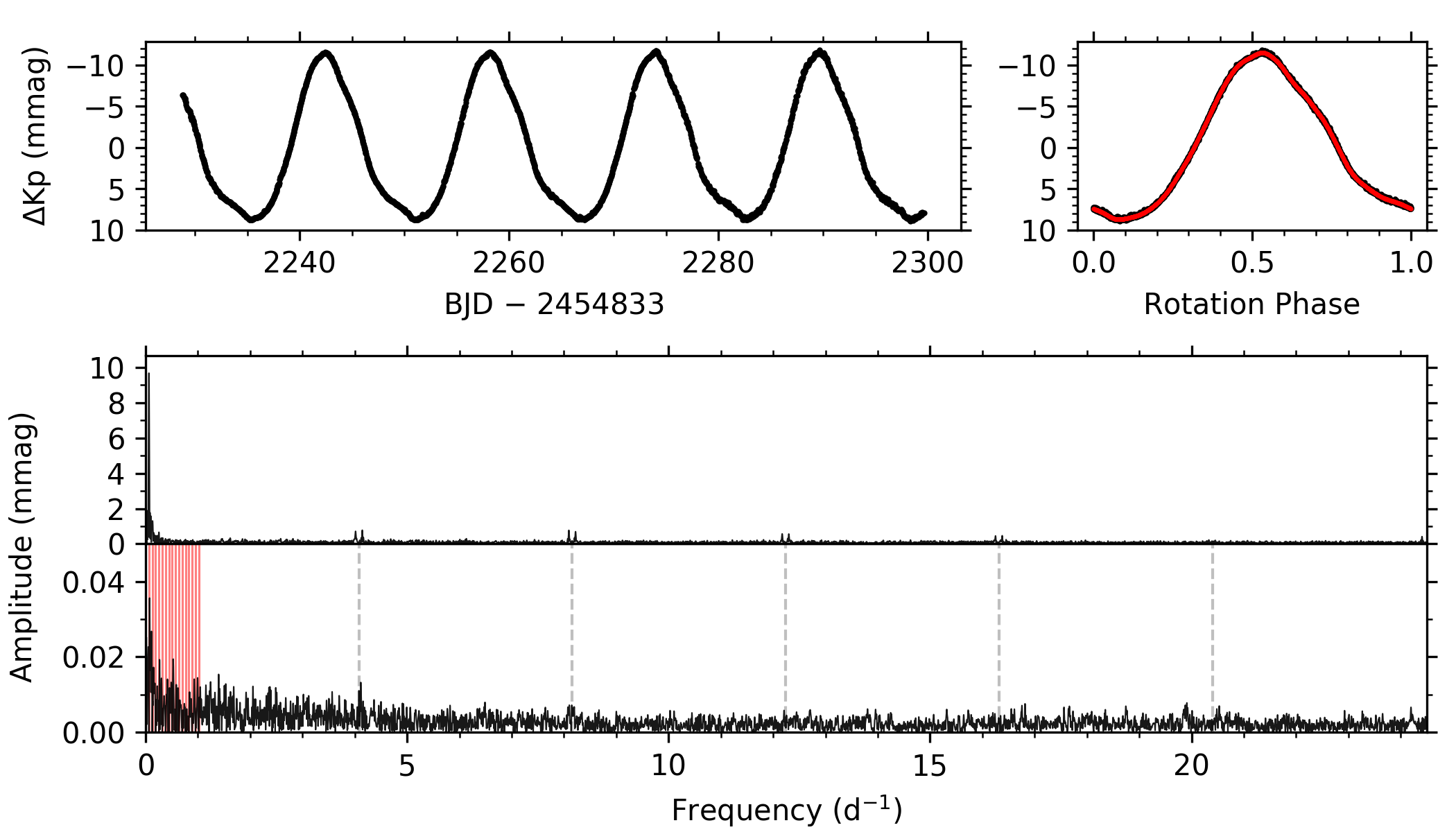}
\caption{Rotational modulation in EPIC~210964459 (HD~26571); same layout shown as in Fig.~\ref{figure: example}.}
\label{figure: EPIC210964459}
\end{figure*}

\begin{figure*}
\centering
\includegraphics[width=0.95\textwidth]{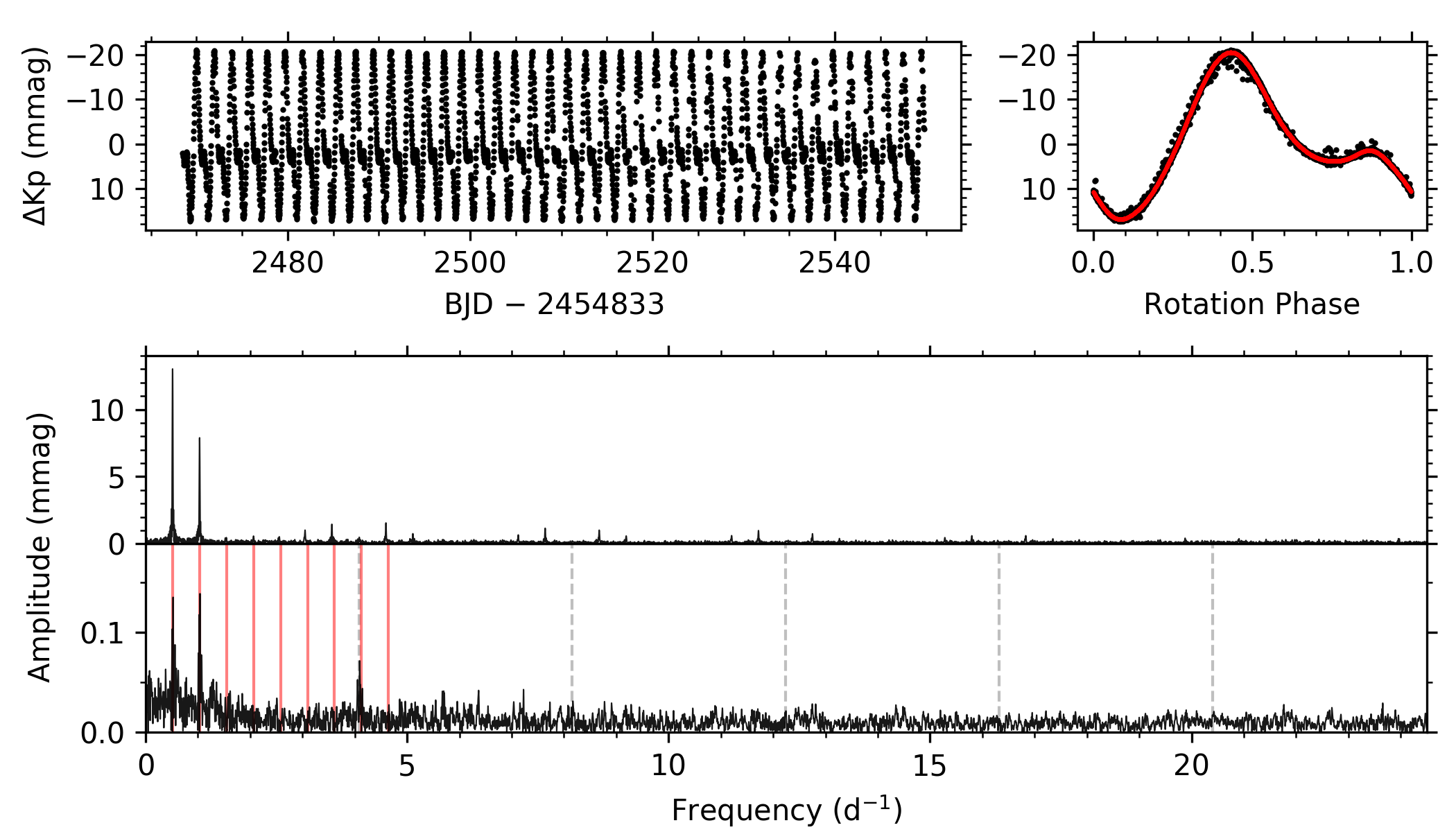}
\caption{Rotational modulation in EPIC~213786701 (HD~173657); same layout shown as in Fig.~\ref{figure: example}.}
\label{figure: EPIC213786701}
\end{figure*}

\clearpage 

\begin{figure*}
\centering
\includegraphics[width=0.95\textwidth]{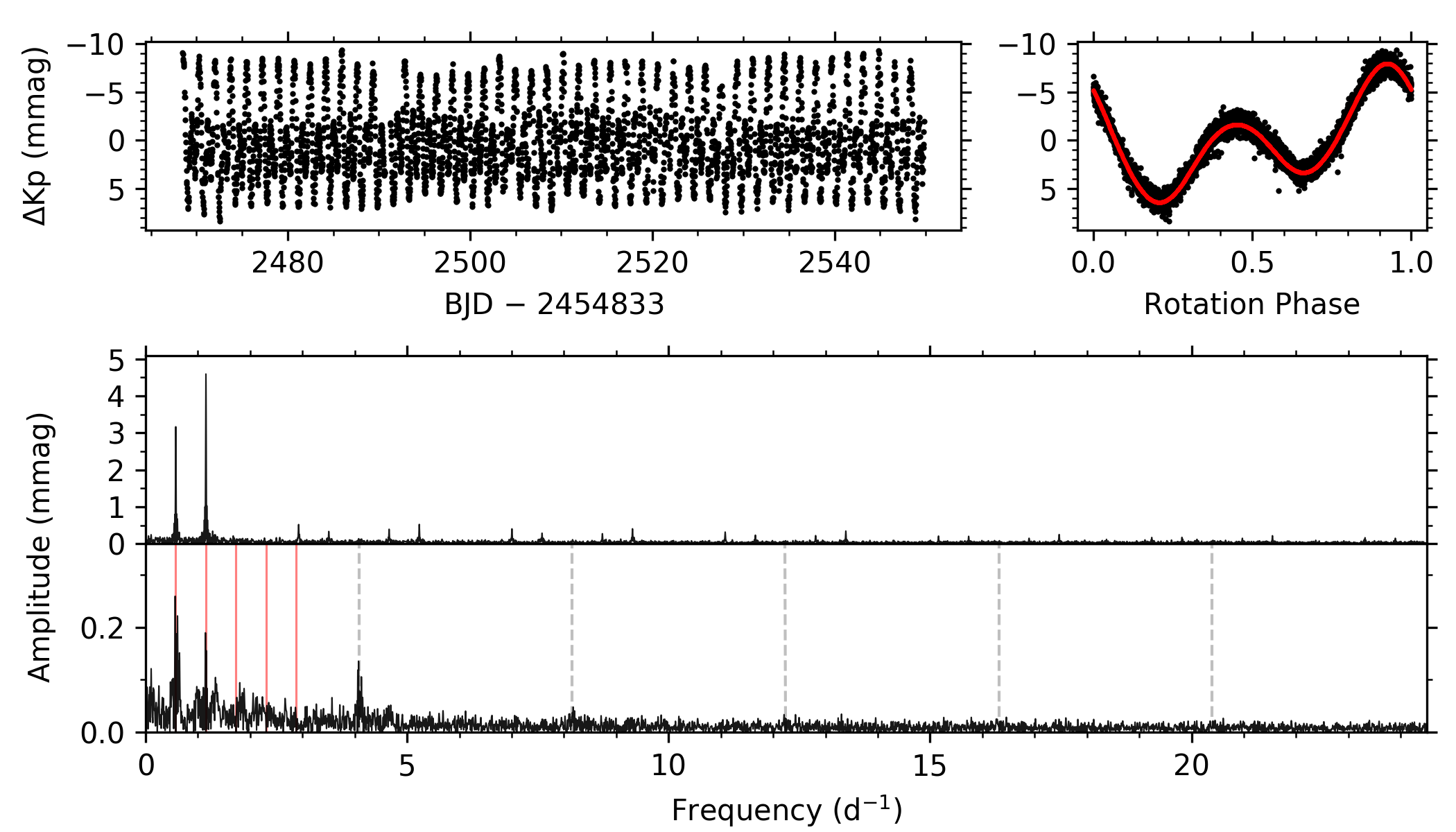}
\caption{Rotational modulation in EPIC~214197027 (HD~173857); same layout shown as in Fig.~\ref{figure: example}.}
\label{figure: EPIC214197027}
\end{figure*}

\begin{figure*}
\centering
\includegraphics[width=0.95\textwidth]{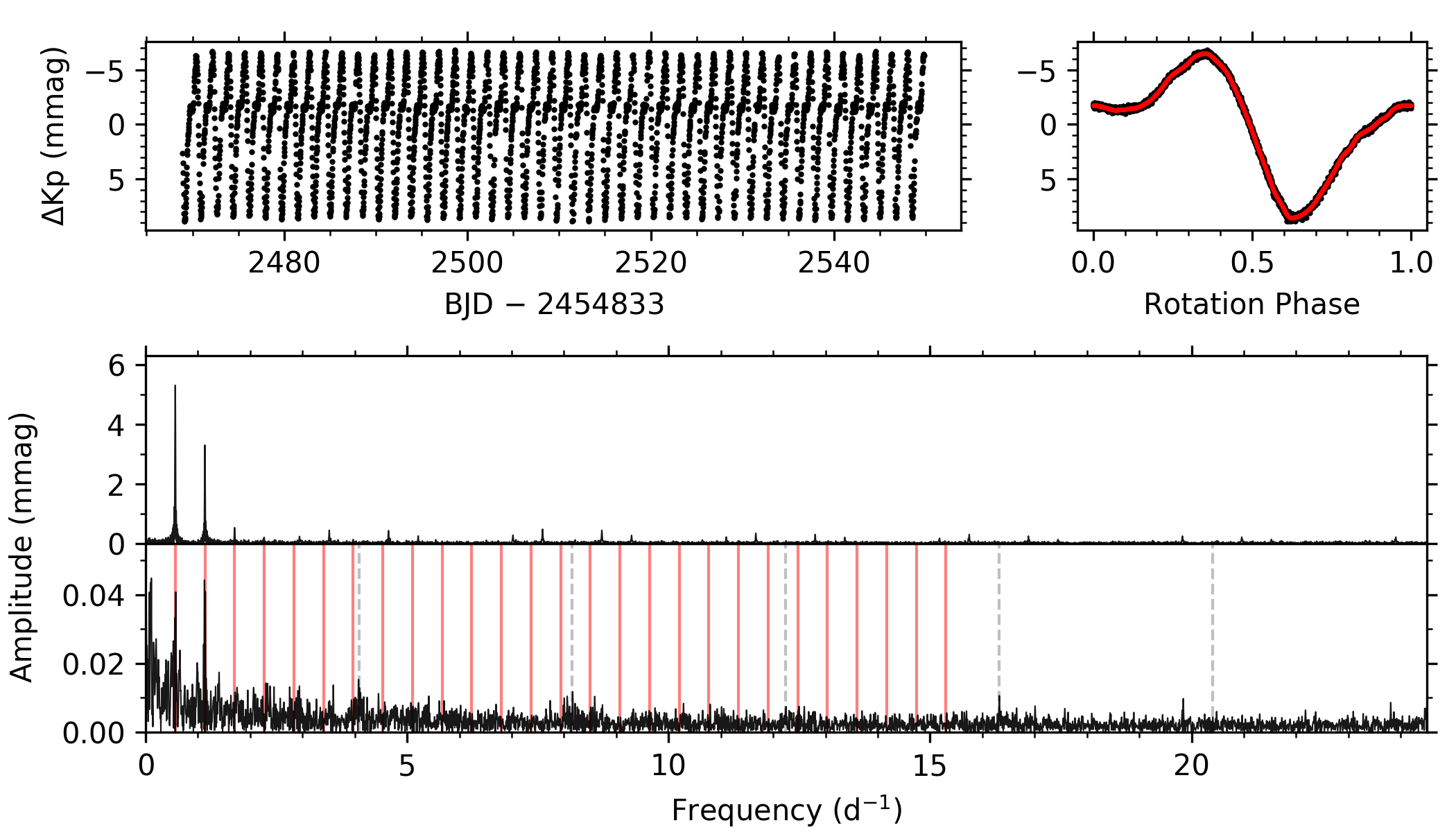}
\caption{Rotational modulation in EPIC~214775703 (HD~178786); same layout shown as in Fig.~\ref{figure: example}.}
\label{figure: EPIC214775703}
\end{figure*}

\clearpage 

\begin{figure*}
\centering
\includegraphics[width=0.95\textwidth]{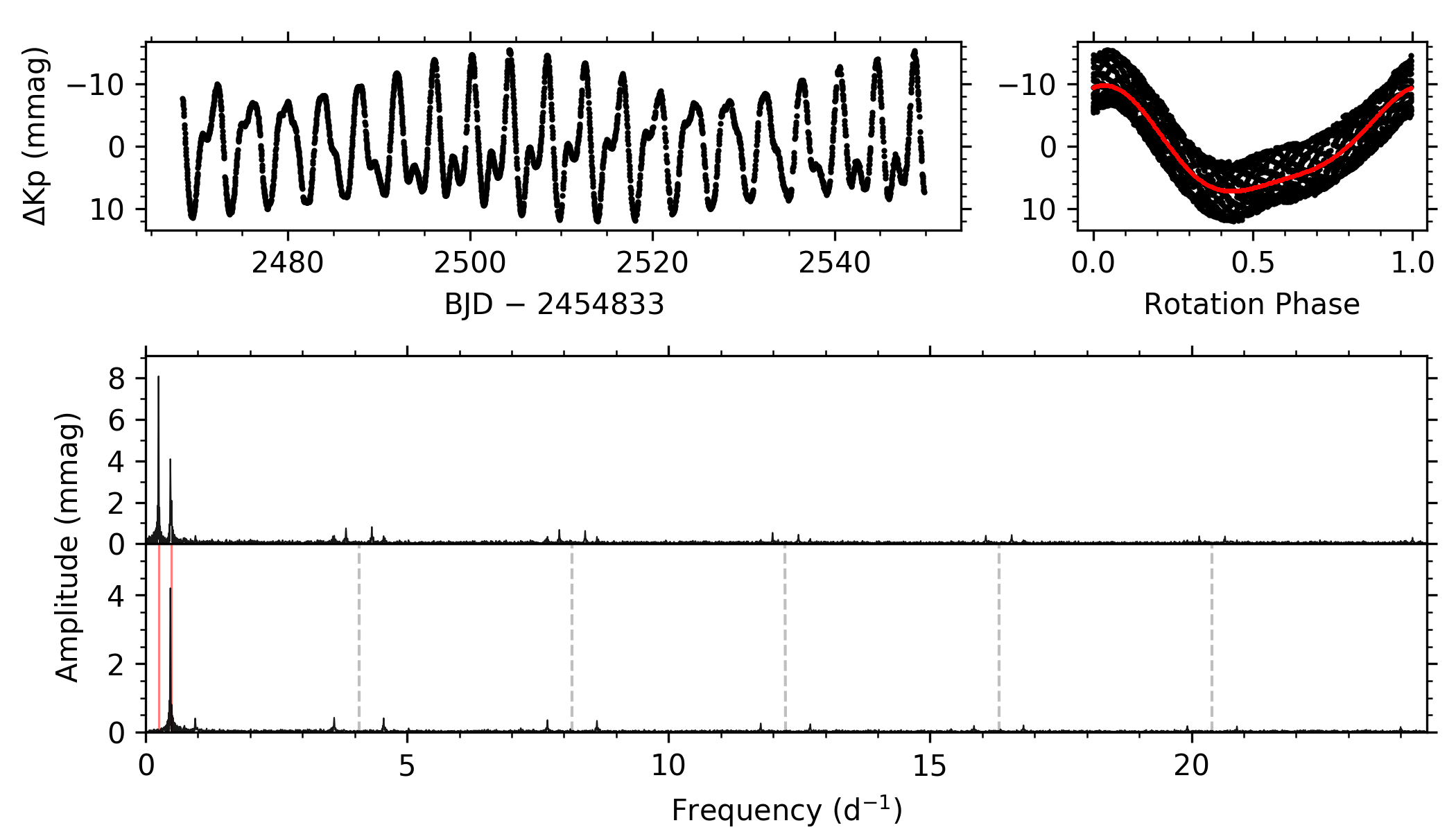}
\caption{Rotational modulation in EPIC~215357858 (HD~174356); same layout shown as in Fig.~\ref{figure: example}.}
\label{figure: EPIC215357858}
\end{figure*}

\begin{figure*}
\centering
\includegraphics[width=0.95\textwidth]{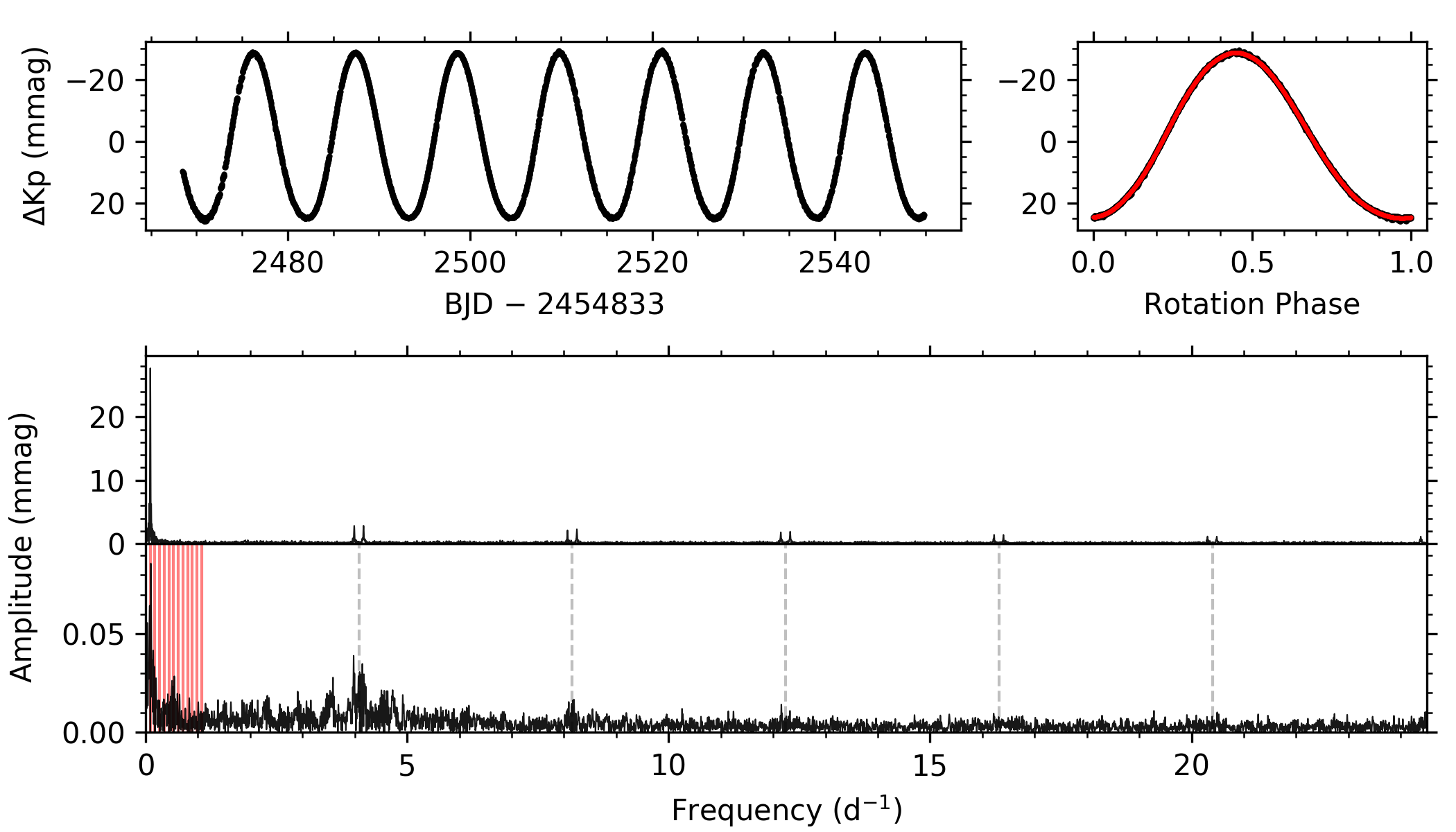}
\caption{Rotational modulation in EPIC~215431167 (HD~174146); same layout shown as in Fig.~\ref{figure: example}.}
\label{figure: EPIC215431167}
\end{figure*}

\clearpage 

\begin{figure*}
\centering
\includegraphics[width=0.95\textwidth]{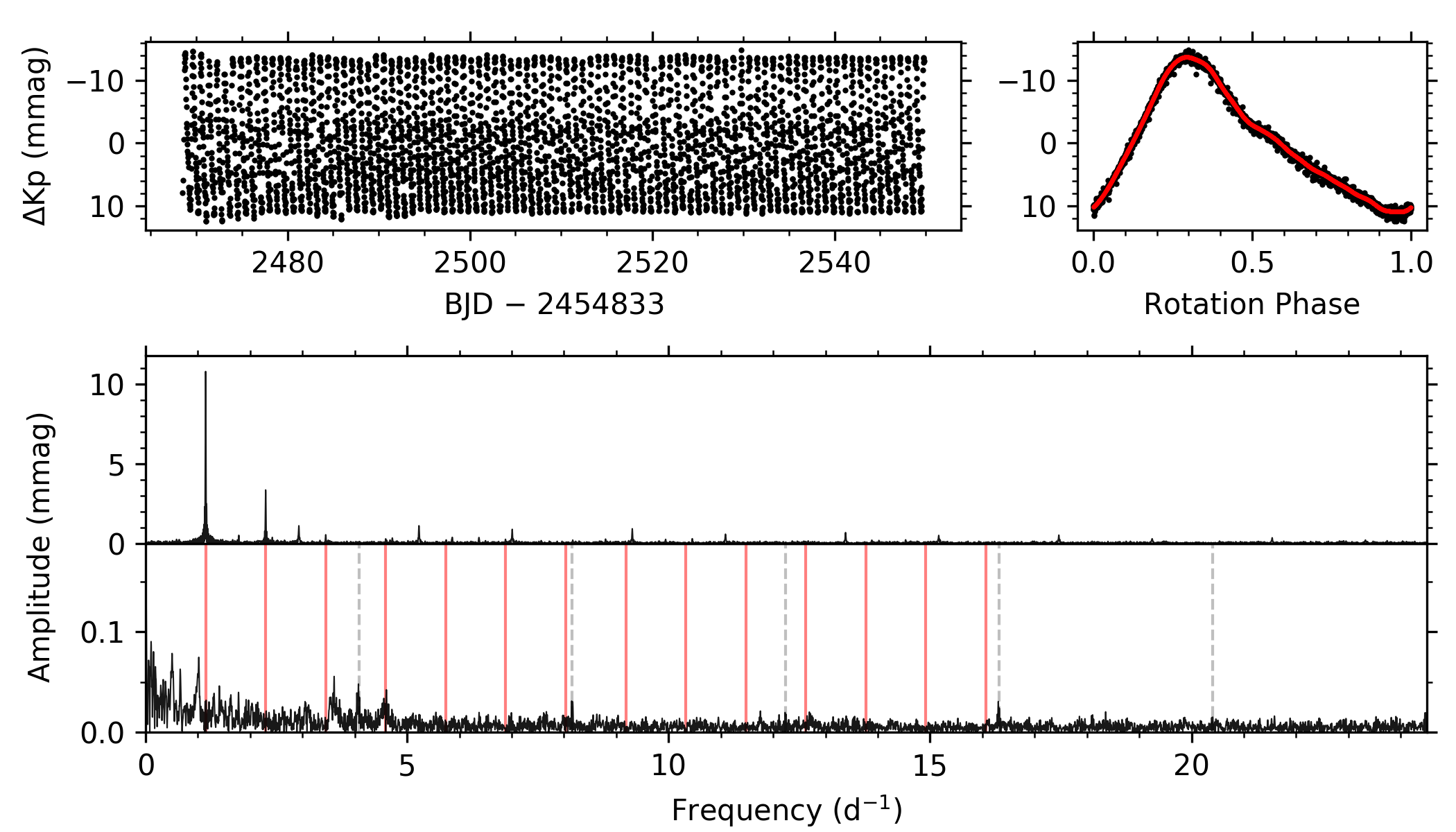}
\caption{Rotational modulation in EPIC~215584931 (HD~172480); same layout shown as in Fig.~\ref{figure: example}.}
\label{figure: EPIC215584931}
\end{figure*}

\begin{figure*}
\centering
\includegraphics[width=0.95\textwidth]{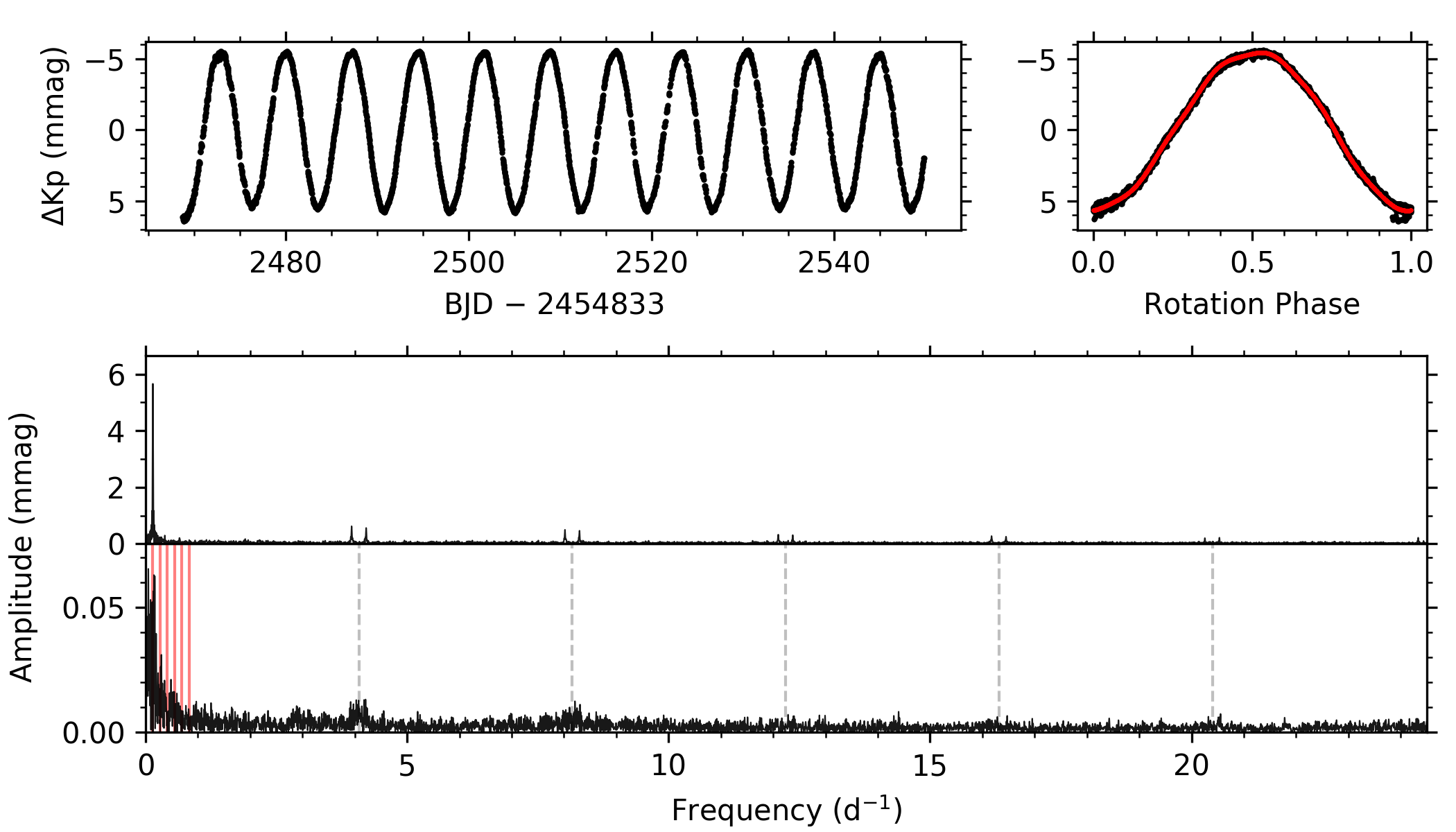}
\caption{Rotational modulation in EPIC~215876375 (HD~184343); same layout shown as in Fig.~\ref{figure: example}.}
\label{figure: EPIC215876375}
\end{figure*}

\clearpage 

\begin{figure*}
\centering
\includegraphics[width=0.95\textwidth]{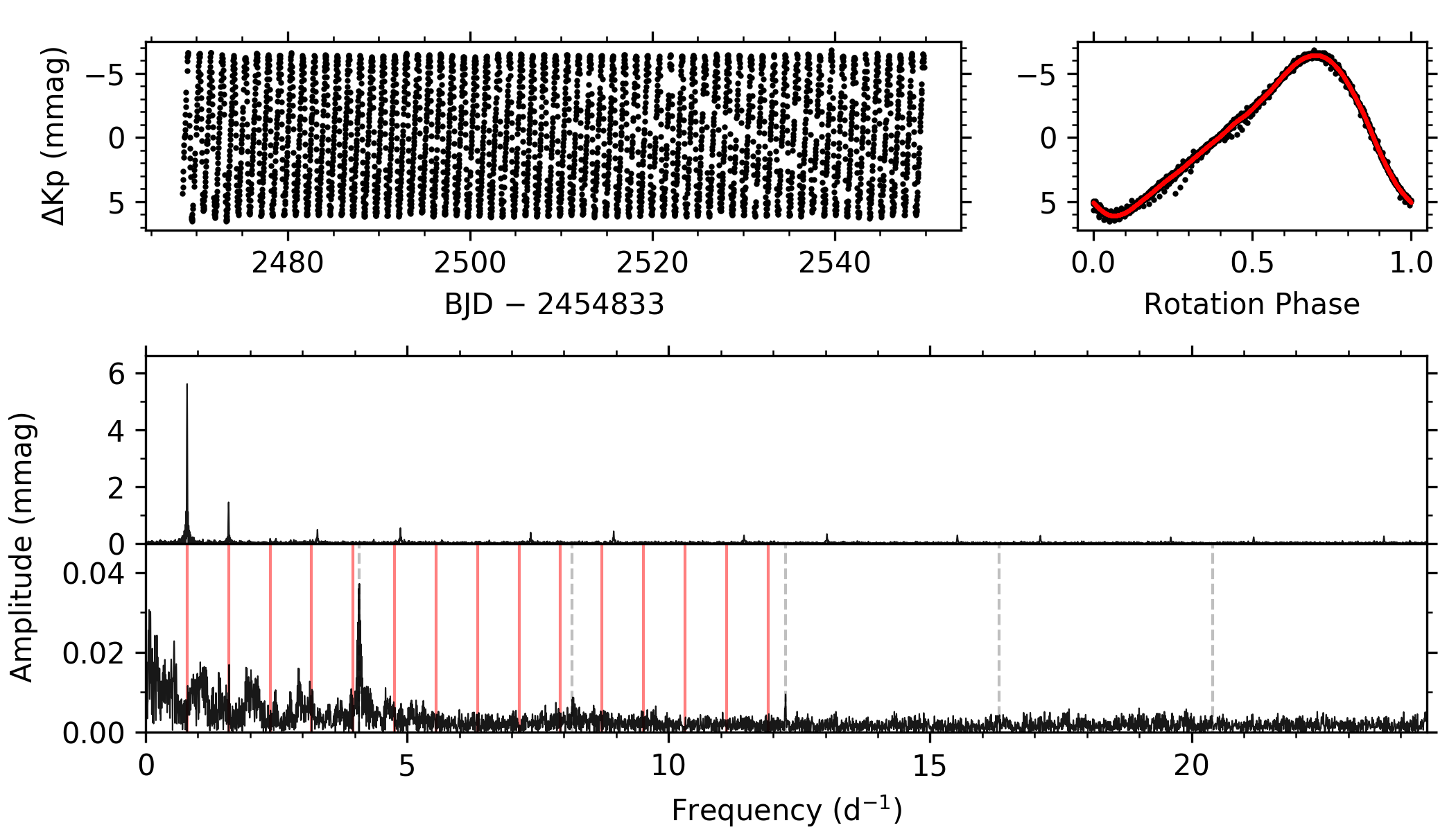}
\caption{Rotational modulation in EPIC~216005035 (HD~182459); same layout shown as in Fig.~\ref{figure: example}.}
\label{figure: EPIC216005035}
\end{figure*}

\begin{figure*}
\centering
\includegraphics[width=0.95\textwidth]{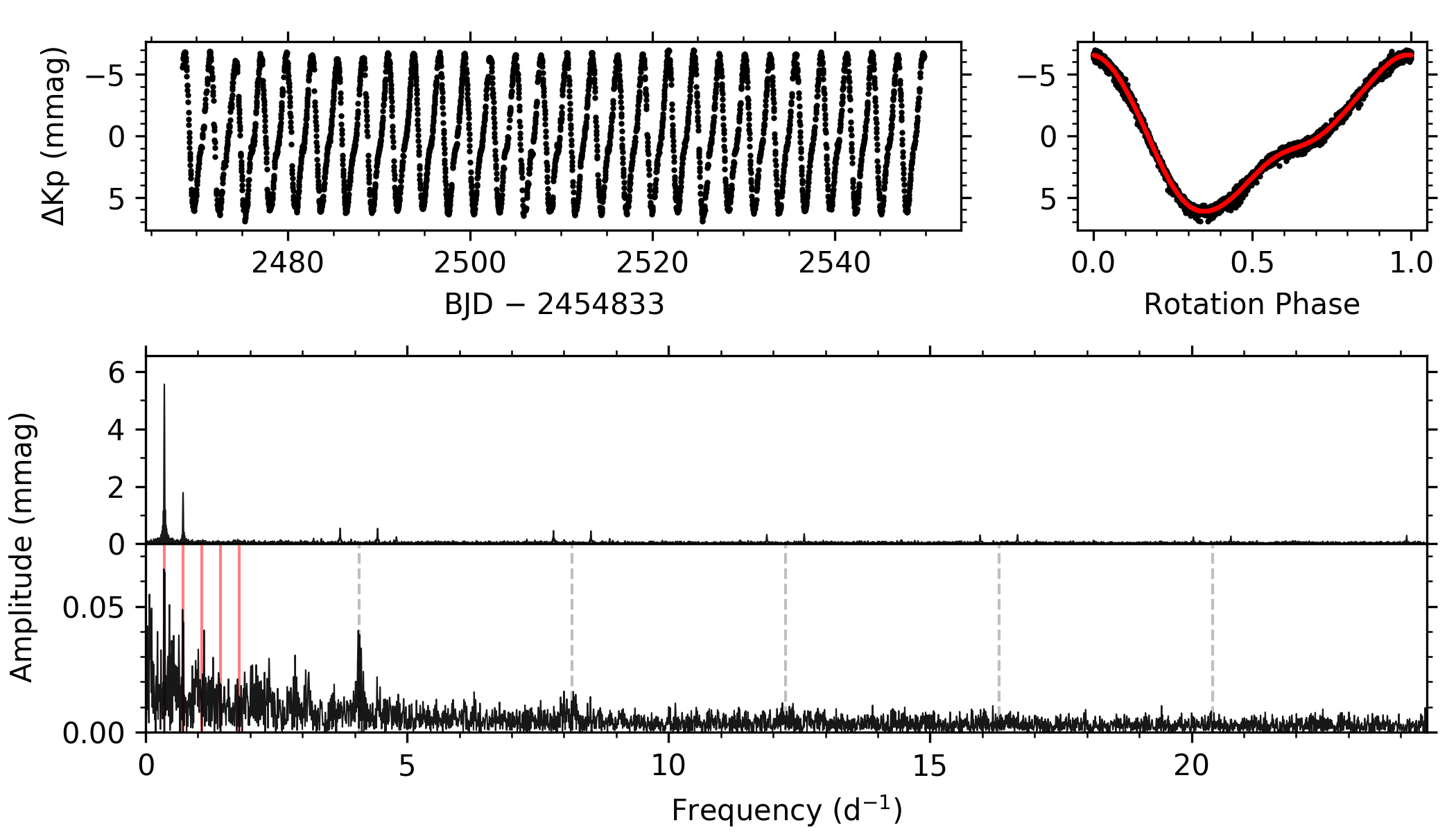}
\caption{Rotational modulation in EPIC~217323447 (HD~180303); same layout shown as in Fig.~\ref{figure: example}.}
\label{figure: EPIC217323447}
\end{figure*}

\clearpage 

\begin{figure*}
\centering
\includegraphics[width=0.95\textwidth]{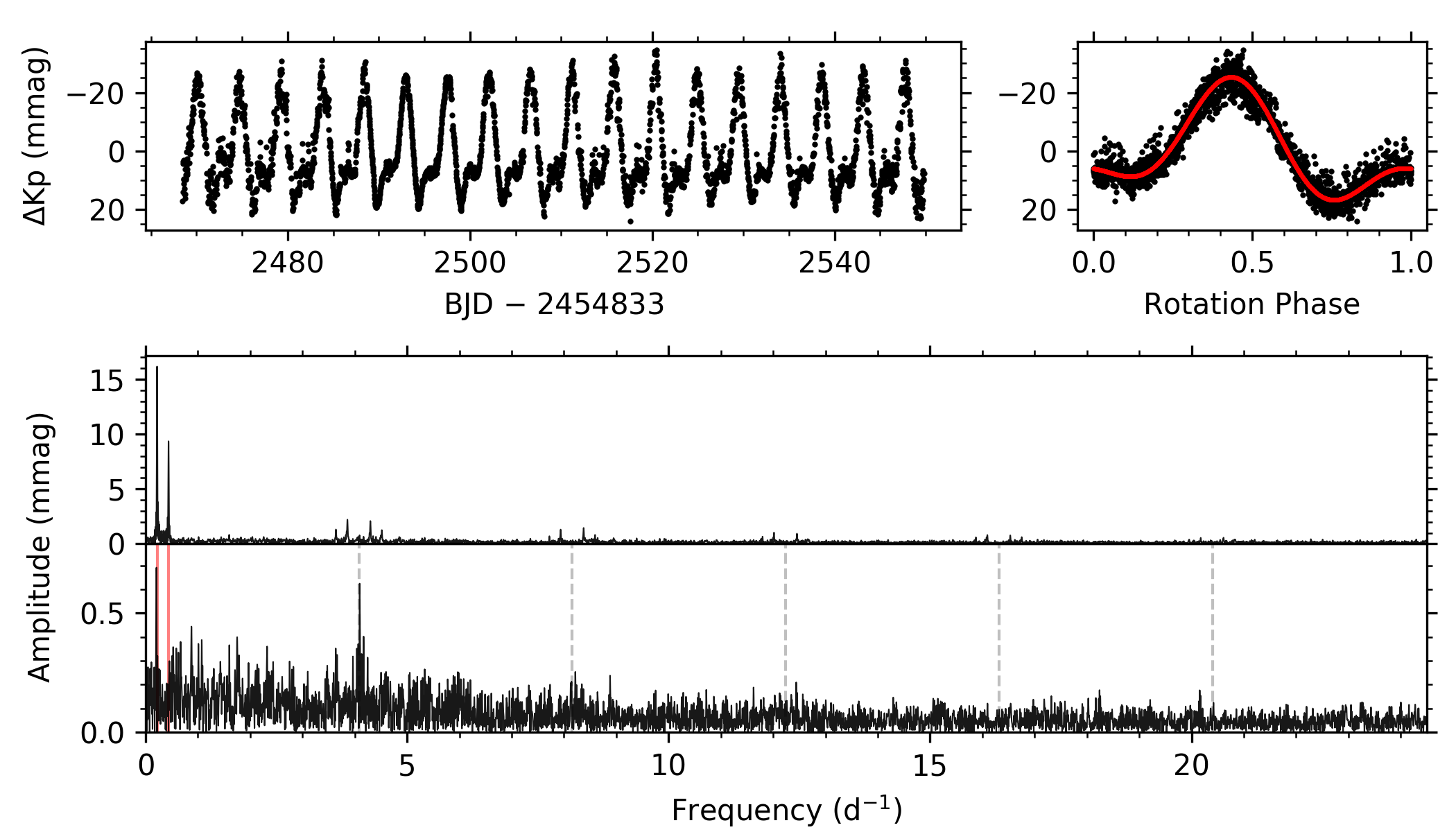}
\caption{Rotational modulation in EPIC~218676652 (HD~173406); same layout shown as in Fig.~\ref{figure: example}.}
\label{figure: EPIC218676652}
\end{figure*}

\begin{figure*}
\centering
\includegraphics[width=0.95\textwidth]{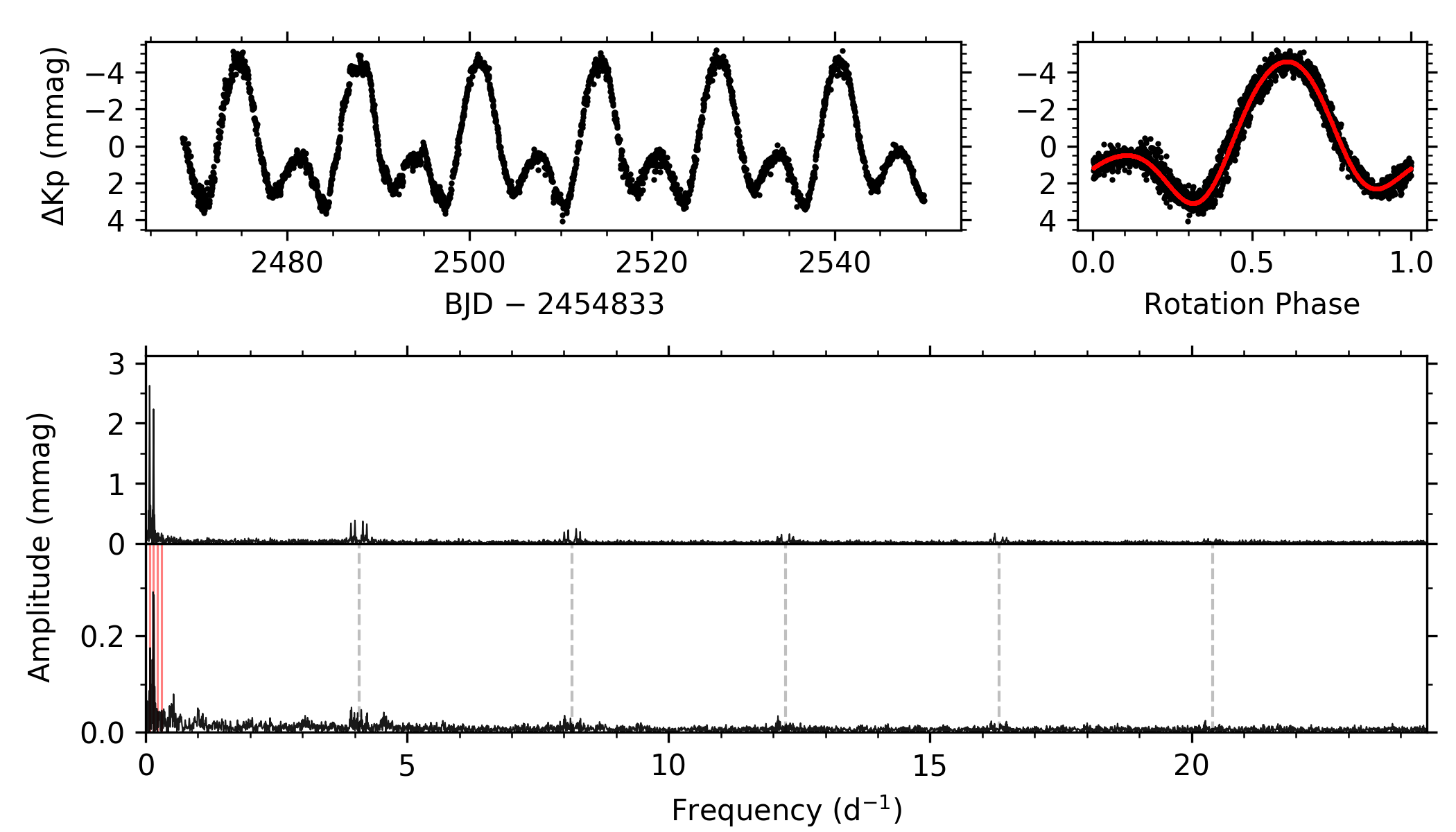}
\caption{Rotational modulation in EPIC~218818457 (HD~176330); same layout shown as in Fig.~\ref{figure: example}.}
\label{figure: EPIC218818457}
\end{figure*}

\clearpage 

\begin{figure*}
\centering
\includegraphics[width=0.95\textwidth]{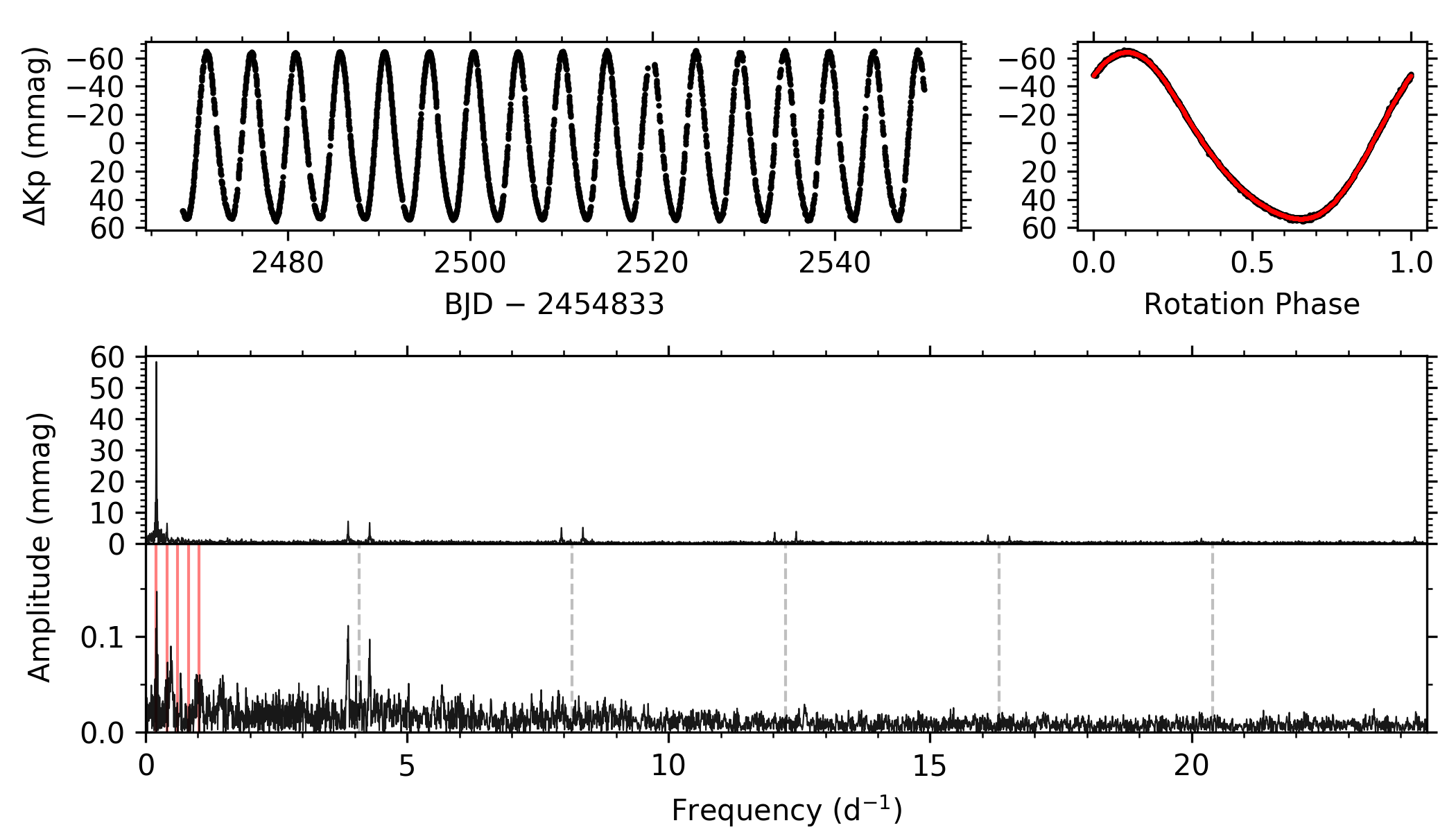}
\caption{Rotational modulation in EPIC~219198038 (HD~177013); same layout shown as in Fig.~\ref{figure: example}.}
\label{figure: EPIC219198038}
\end{figure*}

\begin{figure*}
\centering
\includegraphics[width=0.95\textwidth]{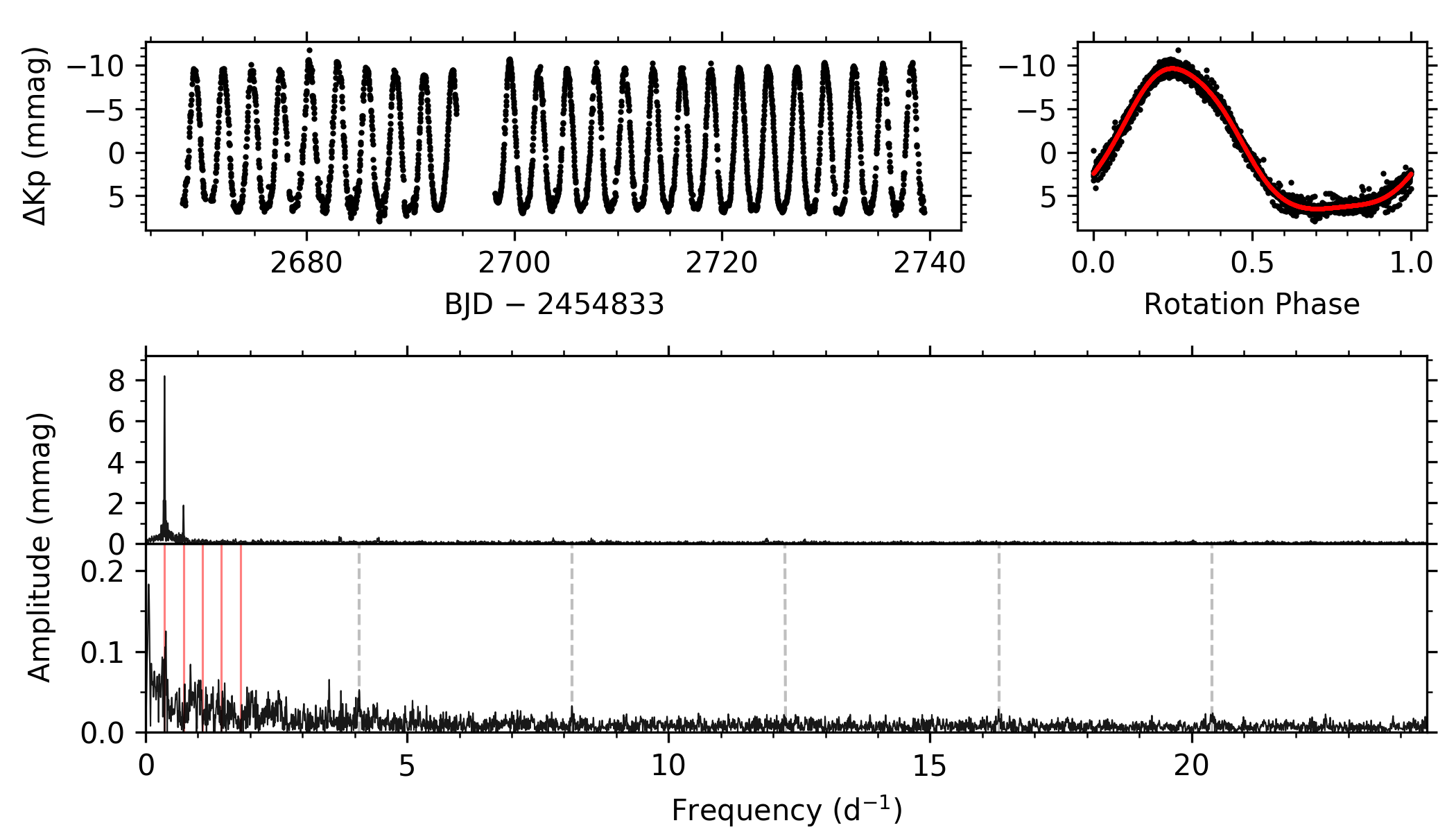}
\caption{Rotational modulation in EPIC~224206658 (HD~165972); same layout shown as in Fig.~\ref{figure: example}.}
\label{figure: EPIC224206658}
\end{figure*}

\clearpage 

\begin{figure*}
\centering
\includegraphics[width=0.95\textwidth]{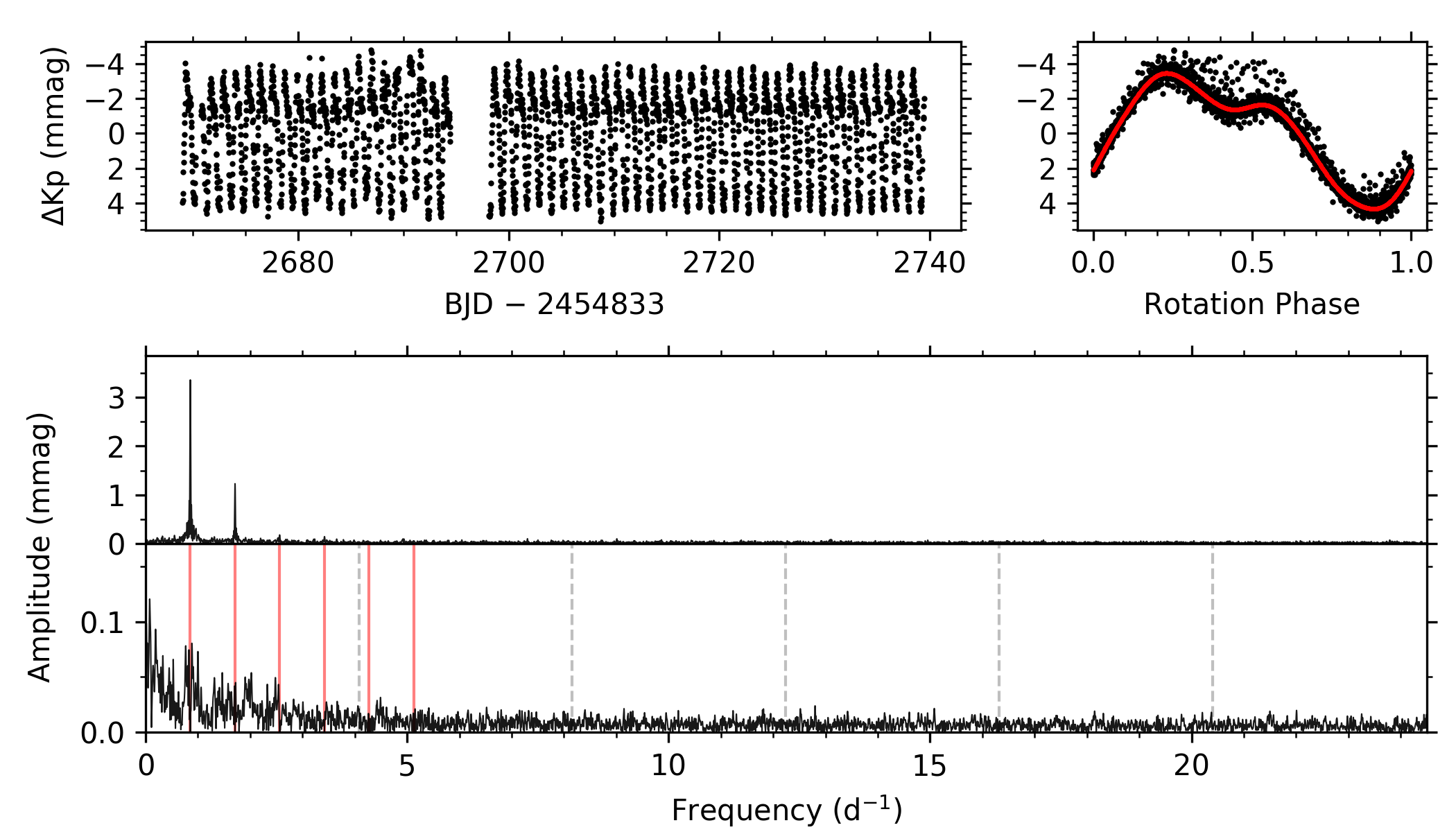}
\caption{Rotational modulation in EPIC~226097699 (HD~166190); same layout shown as in Fig.~\ref{figure: example}.}
\label{figure: EPIC226097699}
\end{figure*}

\begin{figure*}
\centering
\includegraphics[width=0.95\textwidth]{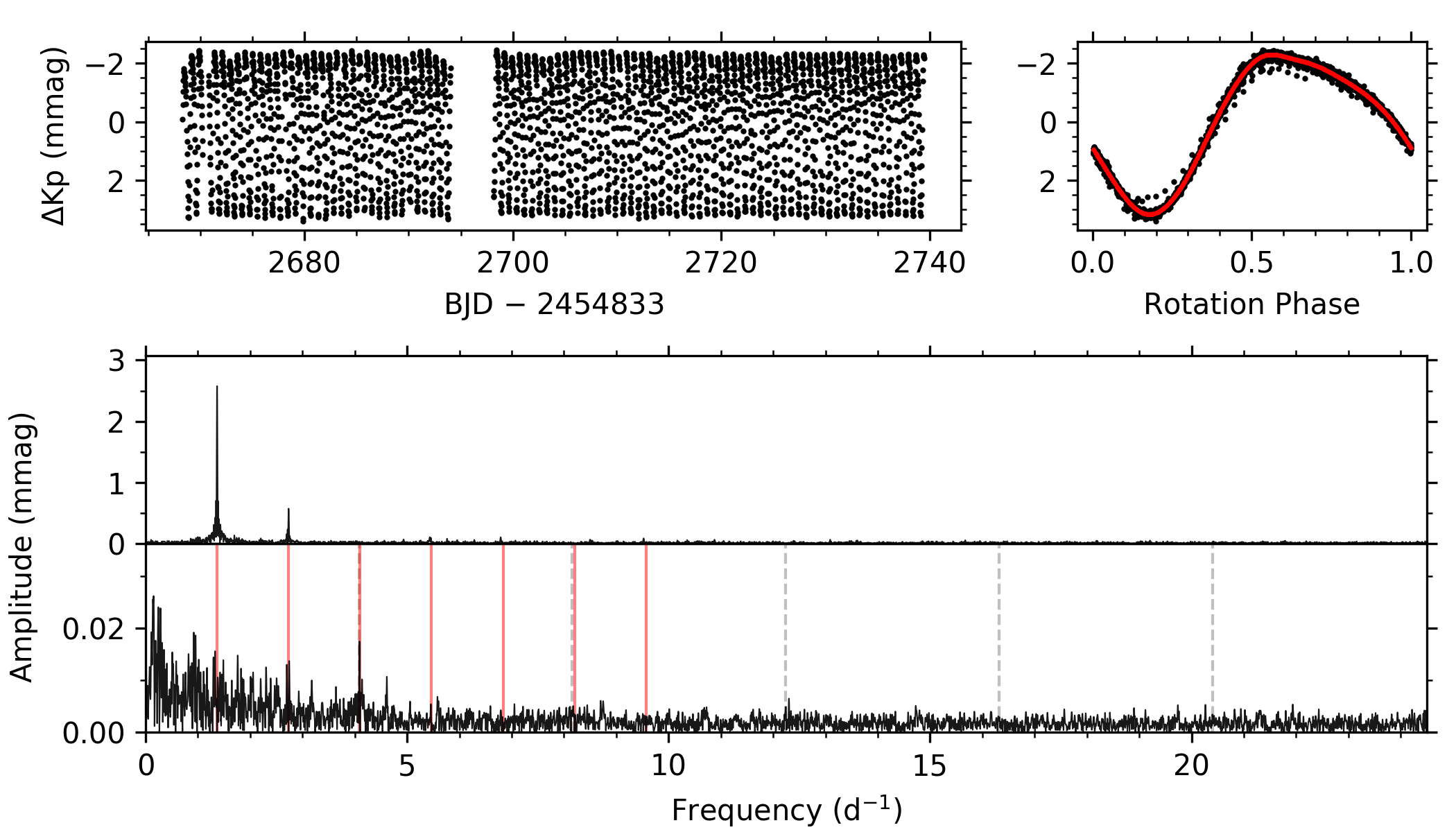}
\caption{Rotational modulation in EPIC~226241087 (HD~164224); same layout shown as in Fig.~\ref{figure: example}.}
\label{figure: EPIC226241087}
\end{figure*}

\clearpage 

\begin{figure*}
\centering
\includegraphics[width=0.95\textwidth]{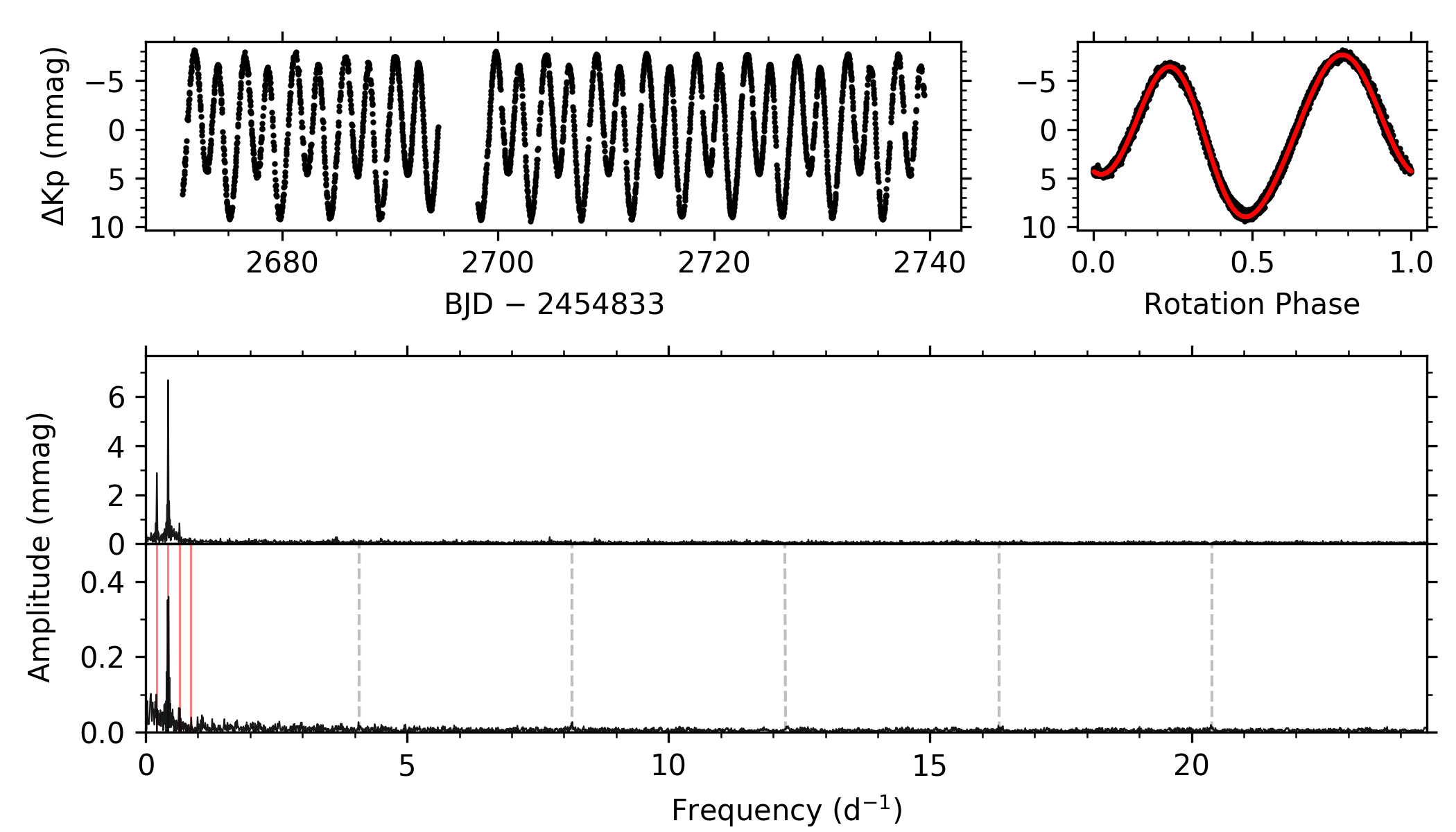}
\caption{Rotational modulation in EPIC~227108971 (HD~164190); same layout shown as in Fig.~\ref{figure: example}.}
\label{figure: EPIC227108971}
\end{figure*}

\begin{figure*}
\centering
\includegraphics[width=0.95\textwidth]{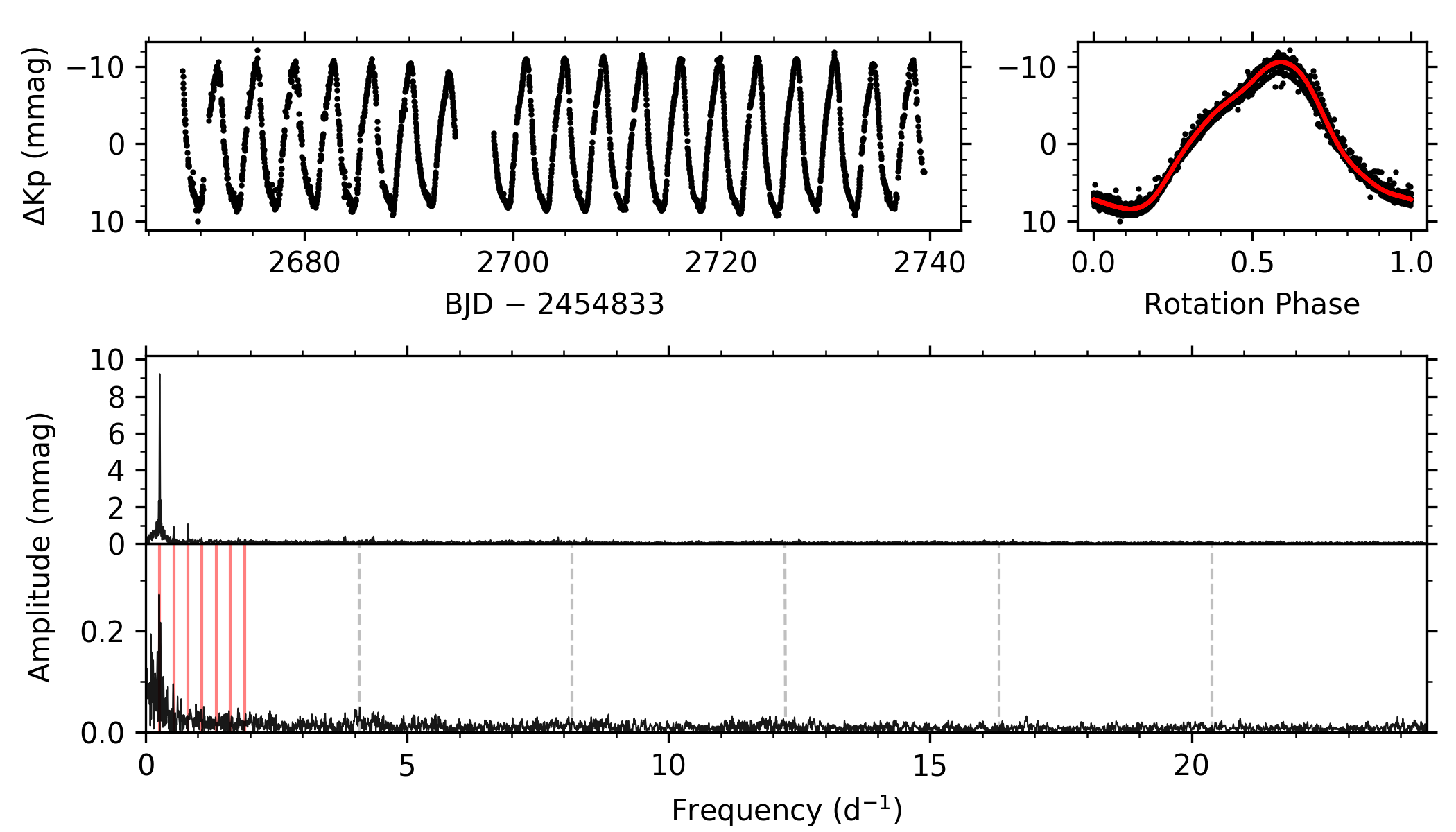}
\caption{Rotational modulation in EPIC~227373493 (HD~166804); same layout shown as in Fig.~\ref{figure: example}.}
\label{figure: EPIC227373493}
\end{figure*}

\clearpage 

\begin{figure*}
\centering
\includegraphics[width=0.95\textwidth]{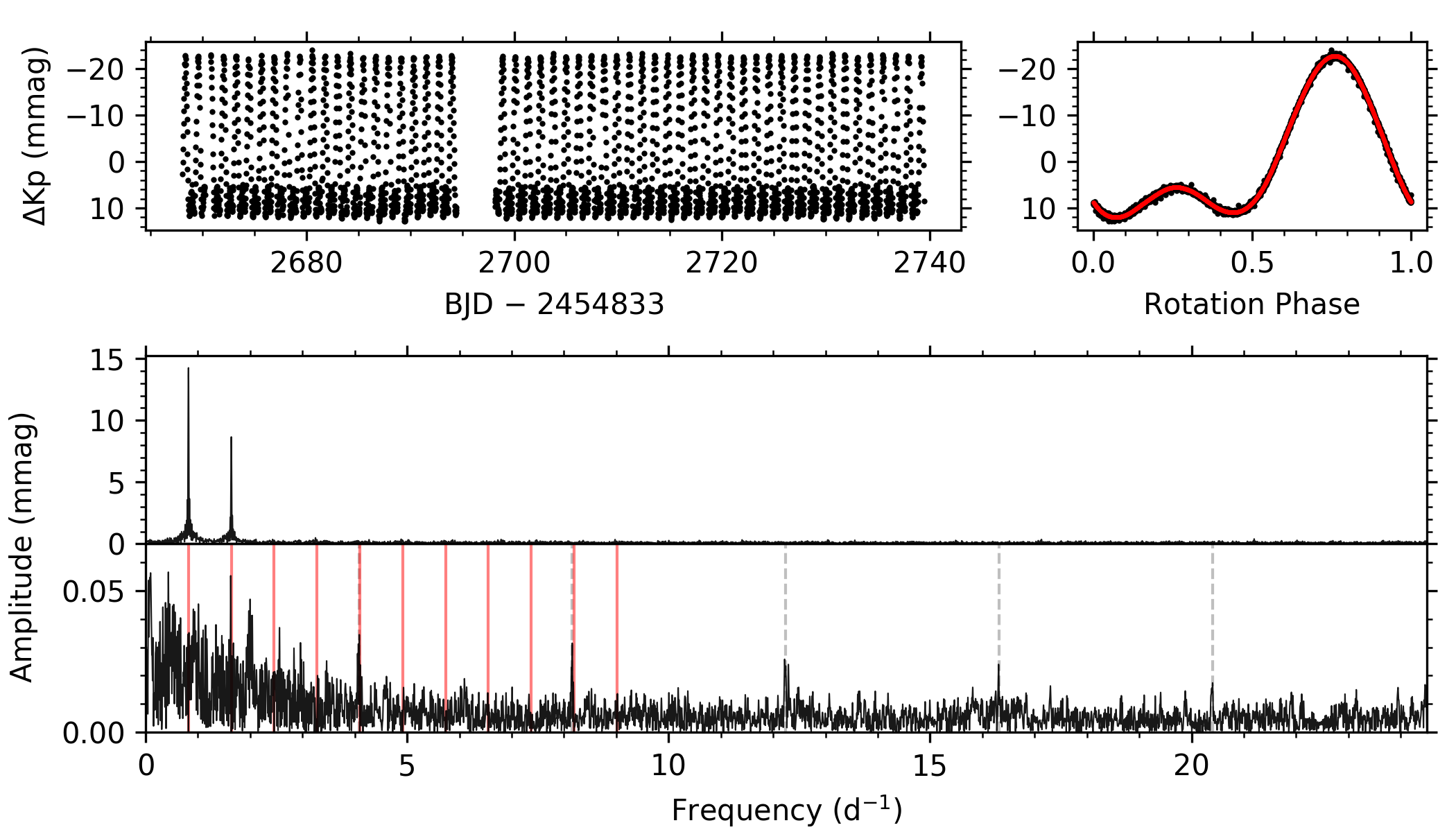}
\caption{Rotational modulation in EPIC~227825246 (HD~164085); same layout shown as in Fig.~\ref{figure: example}.}
\label{figure: EPIC227825246}
\end{figure*}

\begin{figure*}
\centering
\includegraphics[width=0.95\textwidth]{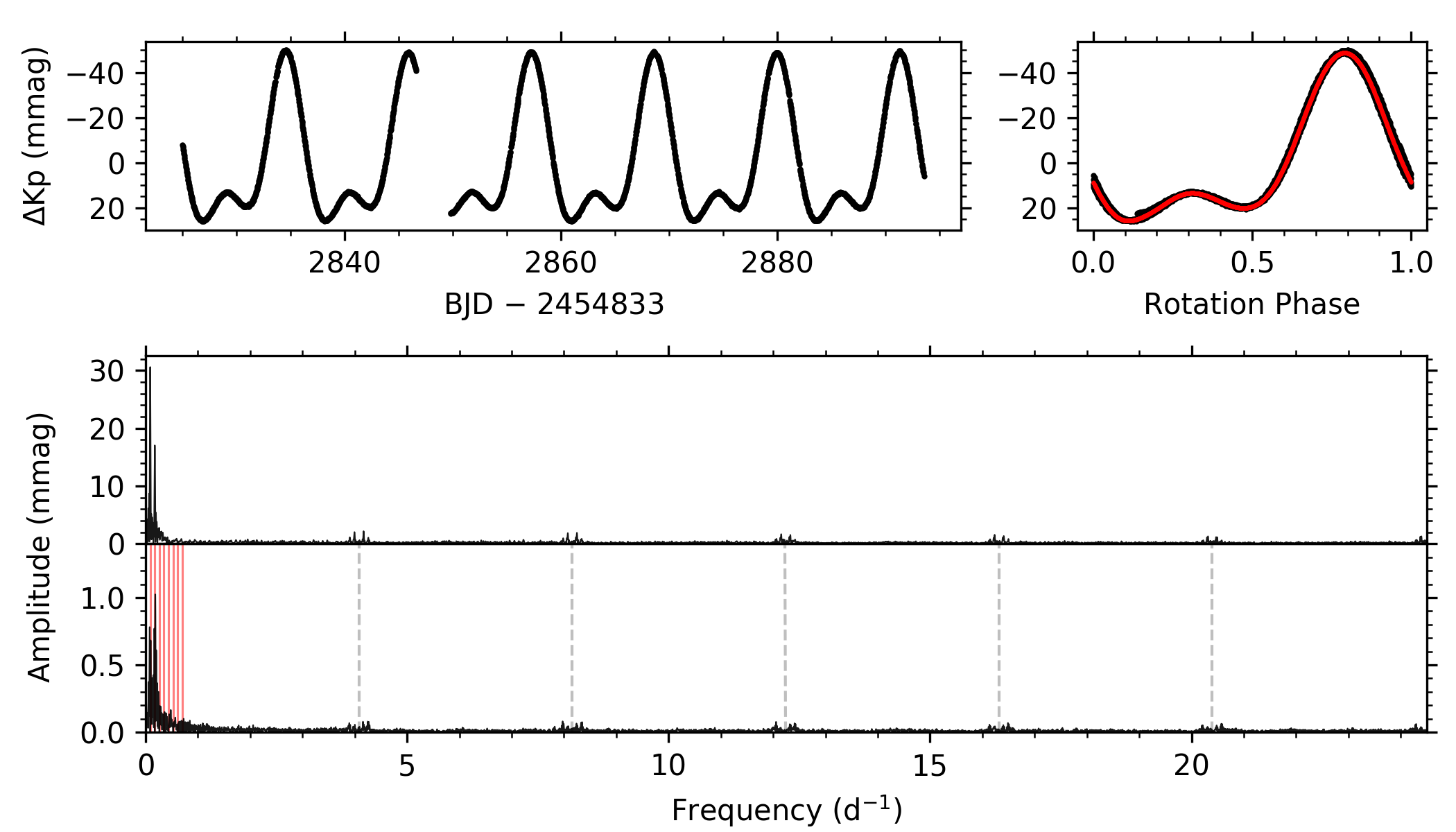}
\caption{Rotational modulation in EPIC~232147357 (HD~153192); same layout shown as in Fig.~\ref{figure: example}.}
\label{figure: EPIC232147357}
\end{figure*}

\clearpage 

\begin{figure*}
\centering
\includegraphics[width=0.95\textwidth]{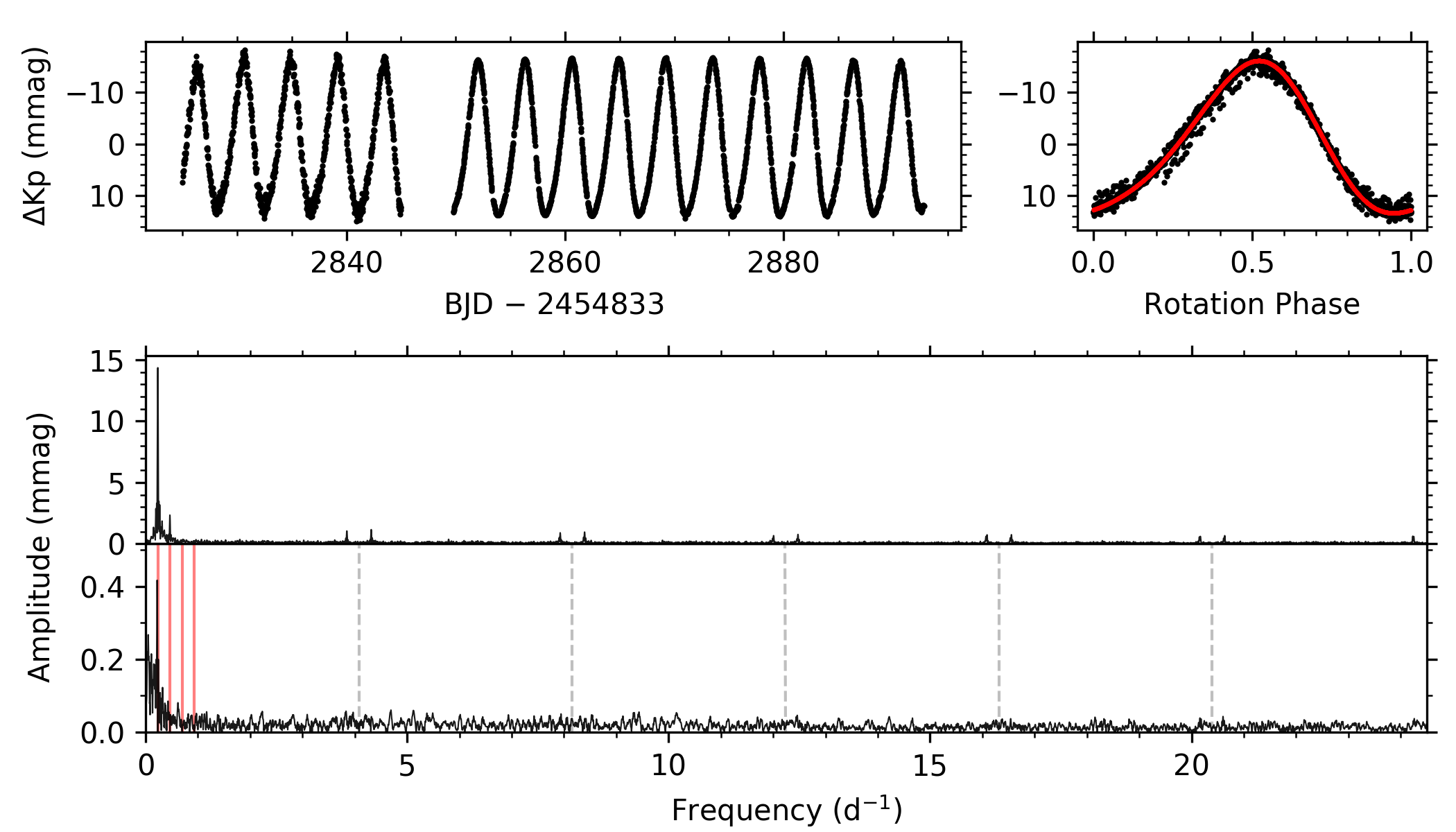}
\caption{Rotational modulation in EPIC~232176043 (HD~152834); same layout shown as in Fig.~\ref{figure: example}.}
\label{figure: EPIC232176043}
\end{figure*}

\begin{figure*}
\centering
\includegraphics[width=0.95\textwidth]{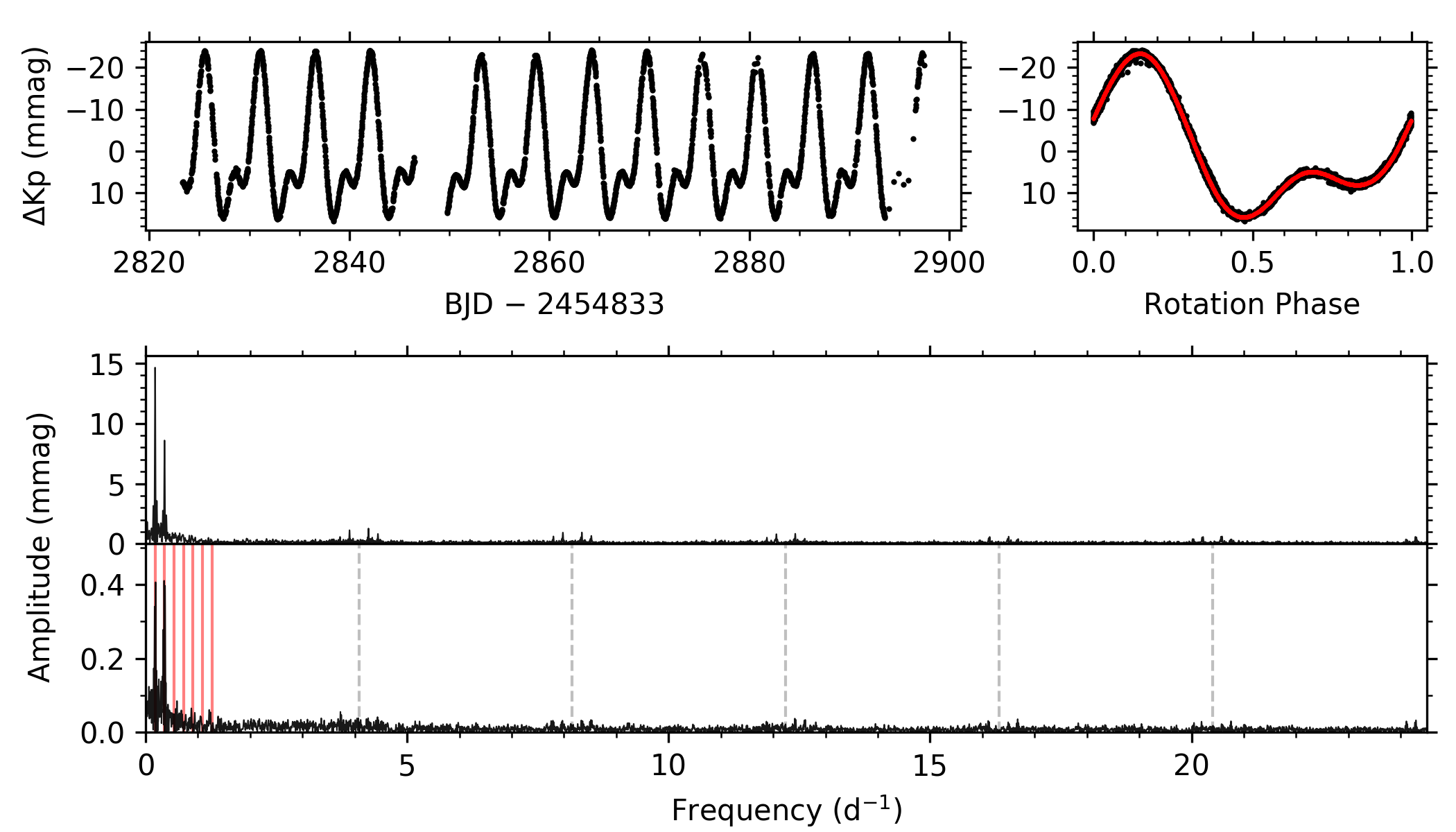}
\caption{Rotational modulation in EPIC~232284277 (HD~155127); same layout shown as in Fig.~\ref{figure: example}.}
\label{figure: EPIC232284277}
\end{figure*}

\clearpage 

\begin{figure*}
\centering
\includegraphics[width=0.95\textwidth]{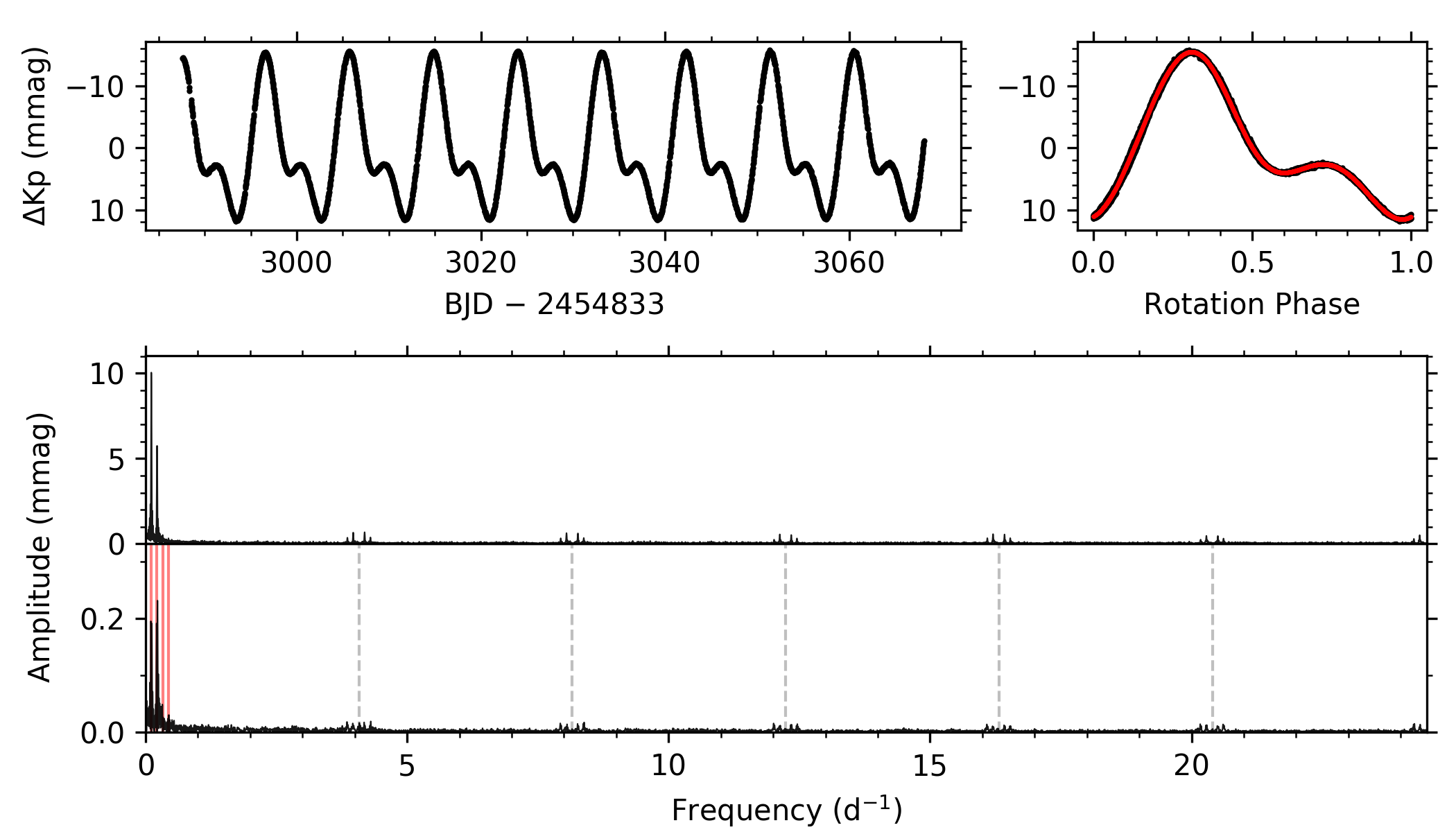}
\caption{Rotational modulation in EPIC~247729177 (HD~284639); same layout shown as in Fig.~\ref{figure: example}.}
\label{figure: EPIC247729177}
\end{figure*}


\clearpage
\section{Candidate pulsating CP stars with rotational modulation}
\label{section: appendix: rotation and pulsation}

In this section, the light curves and amplitude spectra of stars that have measured rotational modulation caused by surface abundance inhomogeneities, but also have additional variability indicative of pulsations, are provided.

\begin{figure*}
\centering
\includegraphics[width=0.95\textwidth]{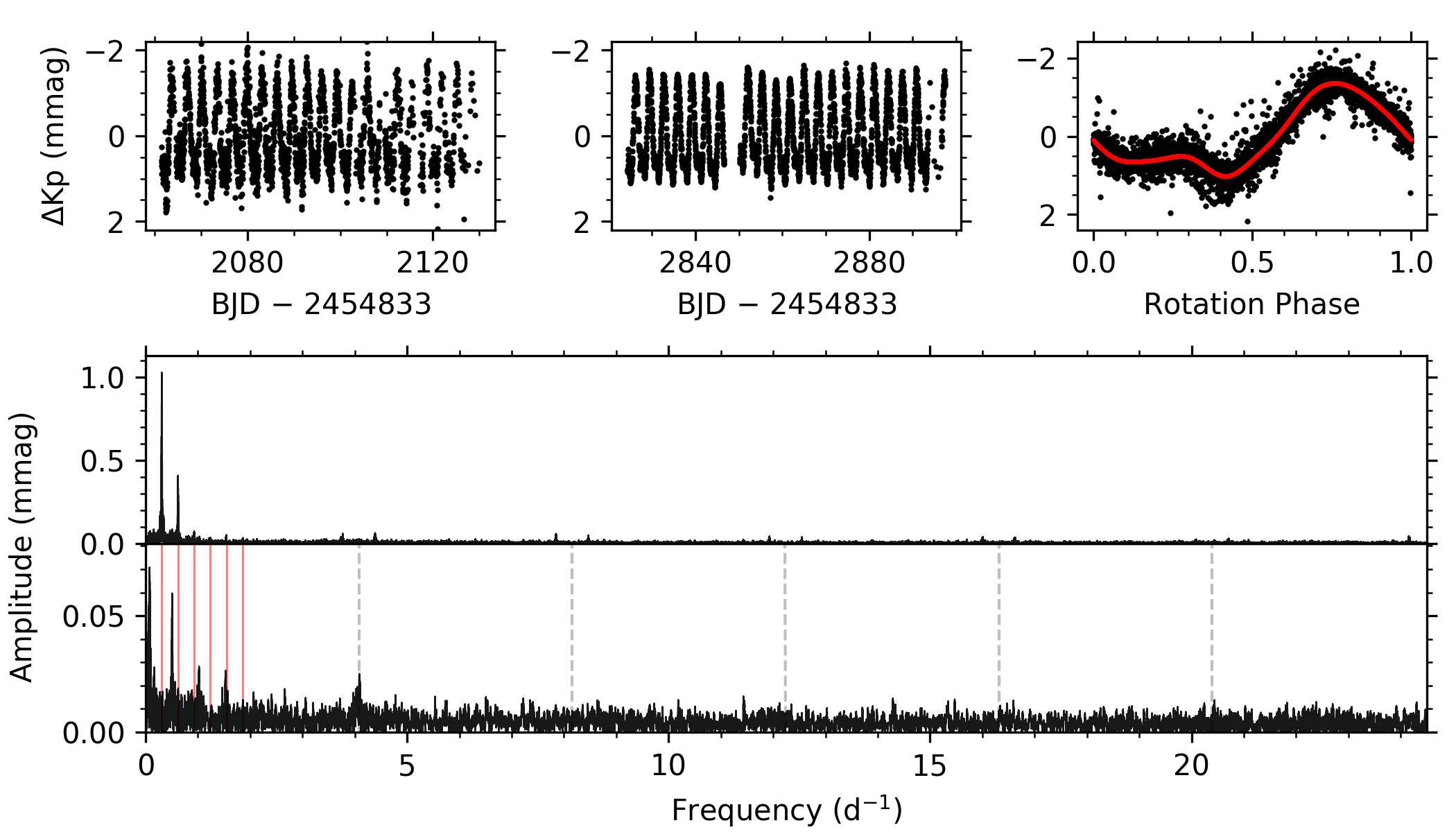}
\caption{Rotational modulation and additional variability indicative of stellar pulsations in EPIC~203749199 (HD~152366); same layout shown as in Fig.~\ref{figure: example}.}
\label{figure: EPIC203749199}
\end{figure*}

\begin{figure*}
\centering
\includegraphics[width=0.95\textwidth]{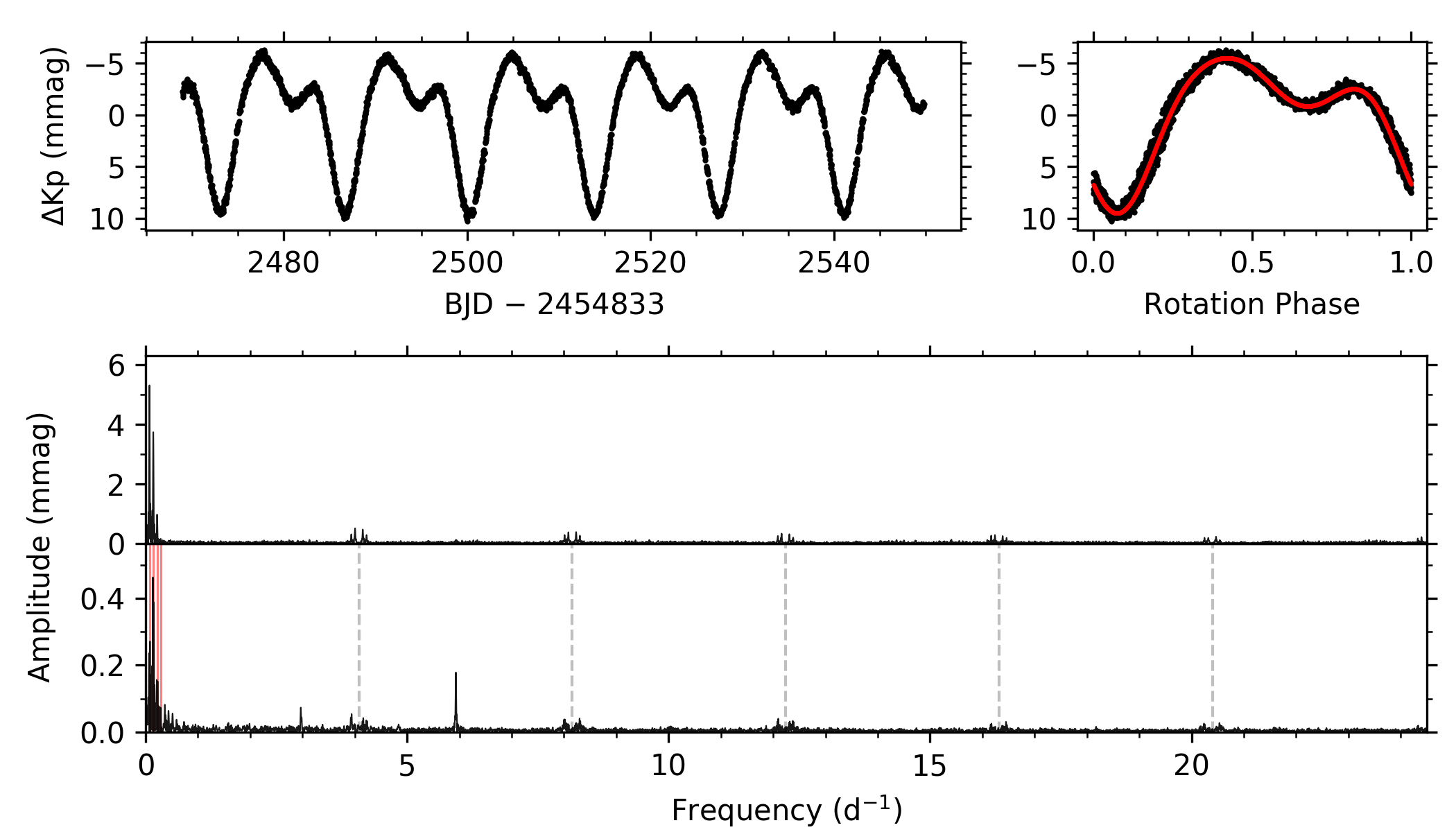}
\caption{Rotational modulation and additional variability indicative of stellar pulsations in EPIC~216956748 (HD~181810); same layout shown as in Fig.~\ref{figure: example}.}
\label{figure: EPIC216956748}
\end{figure*}

\clearpage 

\begin{figure*}
\centering
\includegraphics[width=0.95\textwidth]{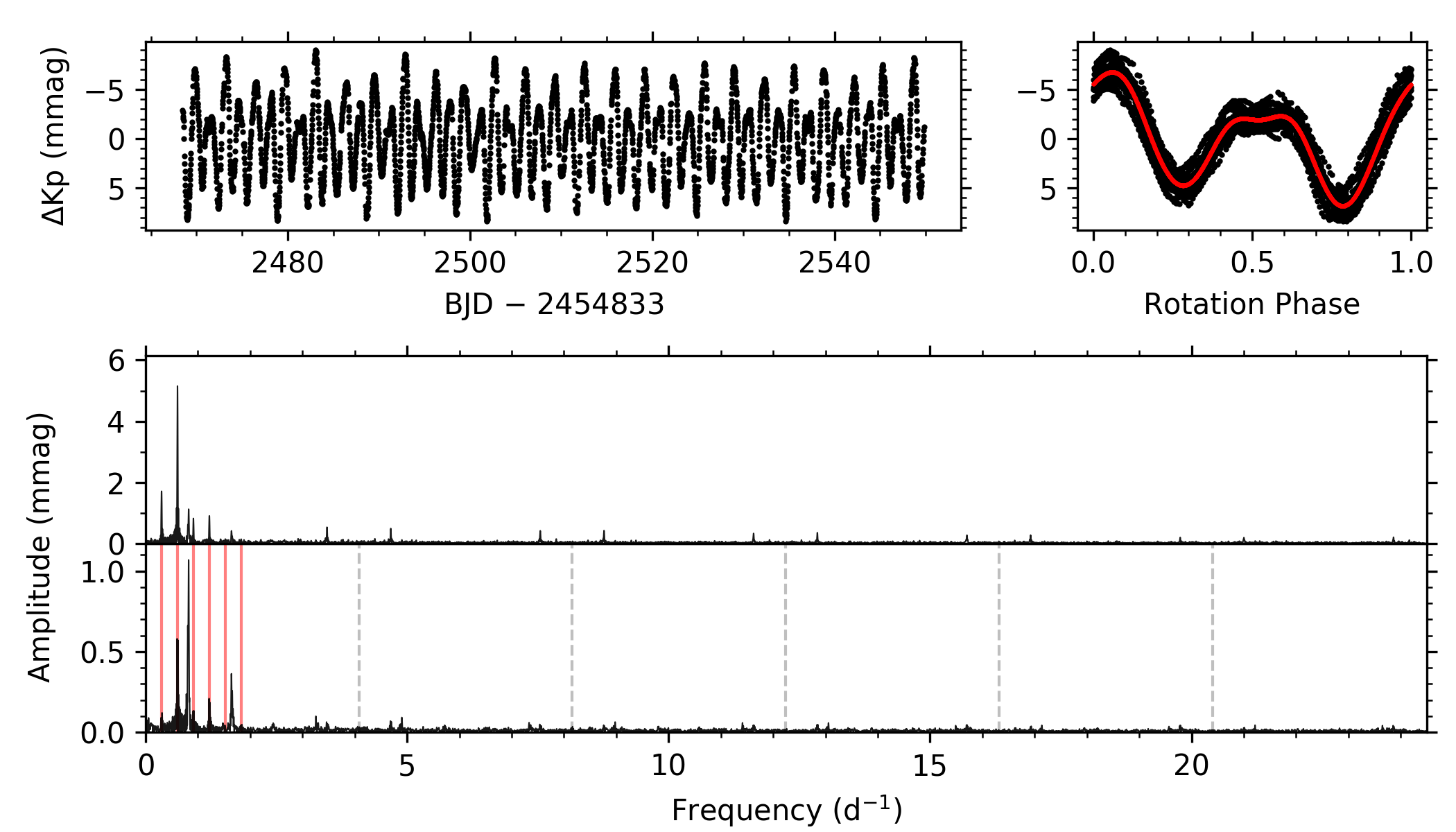}
\caption{Rotational modulation and additional variability indicative of stellar pulsations in EPIC~217437213 (HD~177016); same layout shown as in Fig.~\ref{figure: example}.}
\label{figure: EPIC217437213}
\end{figure*}

\begin{figure*}
\centering
\includegraphics[width=0.95\textwidth]{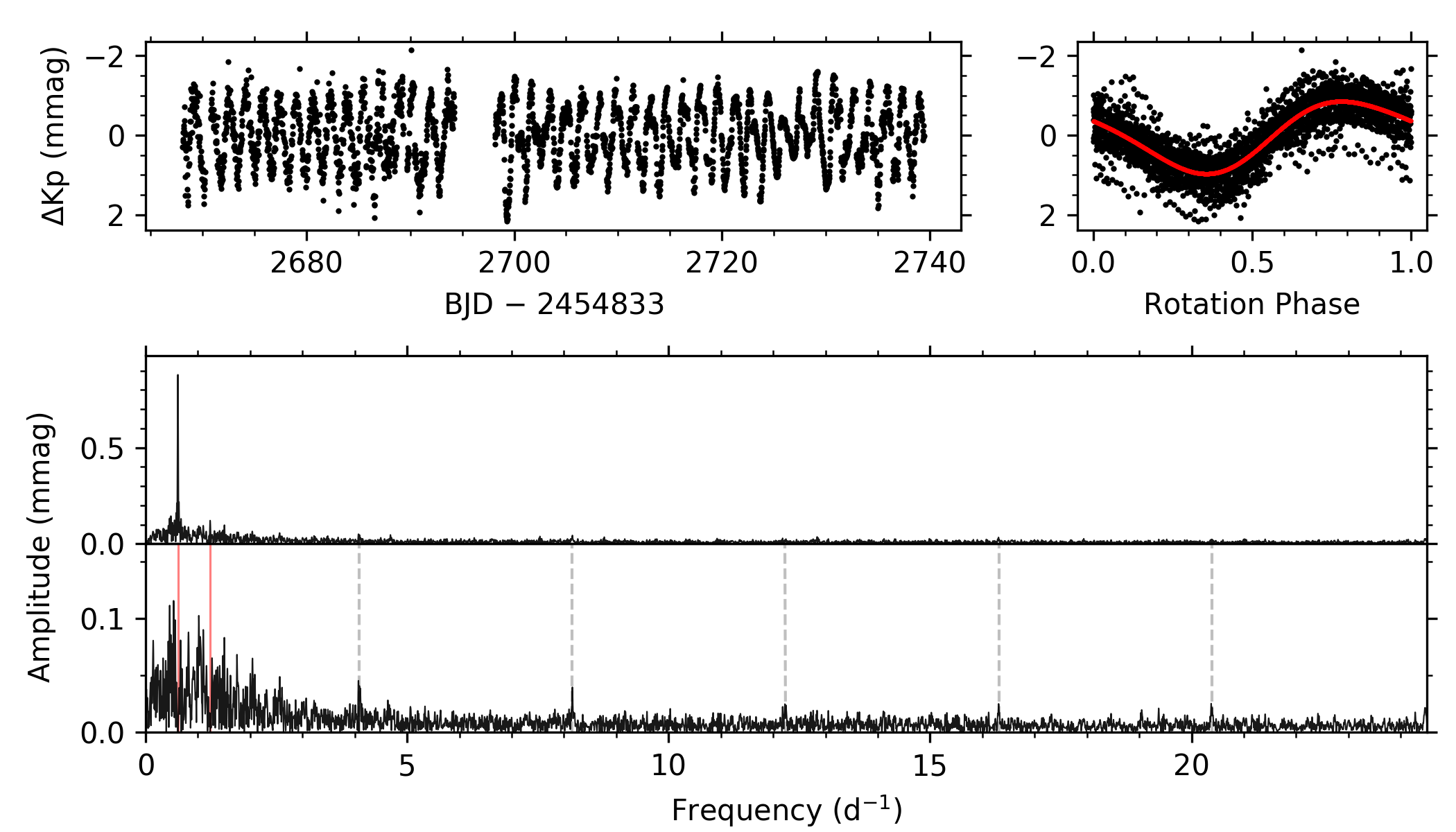}
\caption{Rotational modulation and additional variability indicative of stellar pulsations in EPIC~223573464 (HD~161851); same layout shown as in Fig.~\ref{figure: example}.}
\label{figure: EPIC223573464}
\end{figure*}

\clearpage 

\begin{figure*}
\centering
\includegraphics[width=0.95\textwidth]{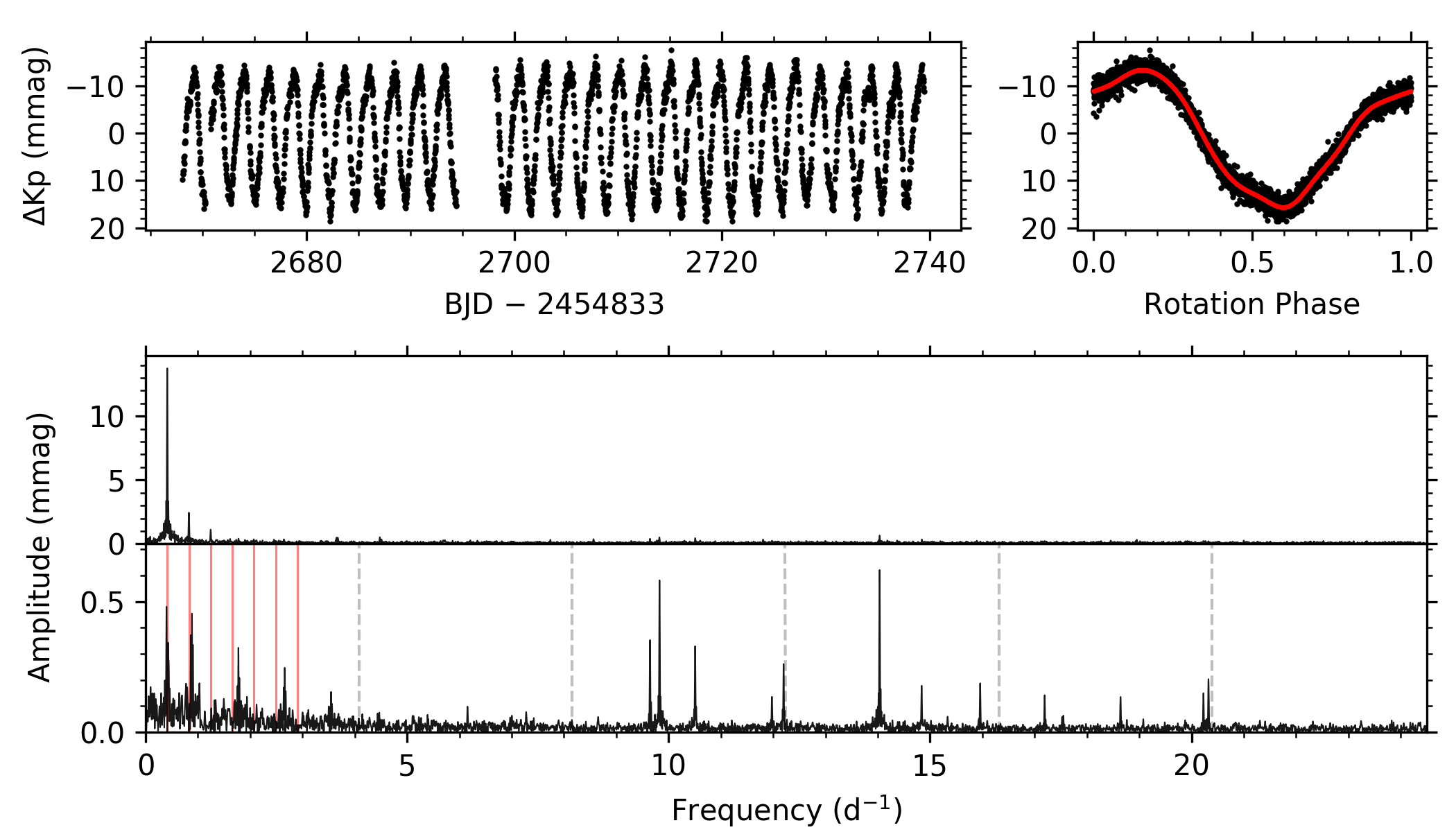}
\caption{Rotational modulation and additional variability indicative of stellar pulsations in EPIC~225191577 (HD~164068); same layout shown as in Fig.~\ref{figure: example}.}
\label{figure: EPIC225191577}
\end{figure*}

\begin{figure*}
\centering
\includegraphics[width=0.95\textwidth]{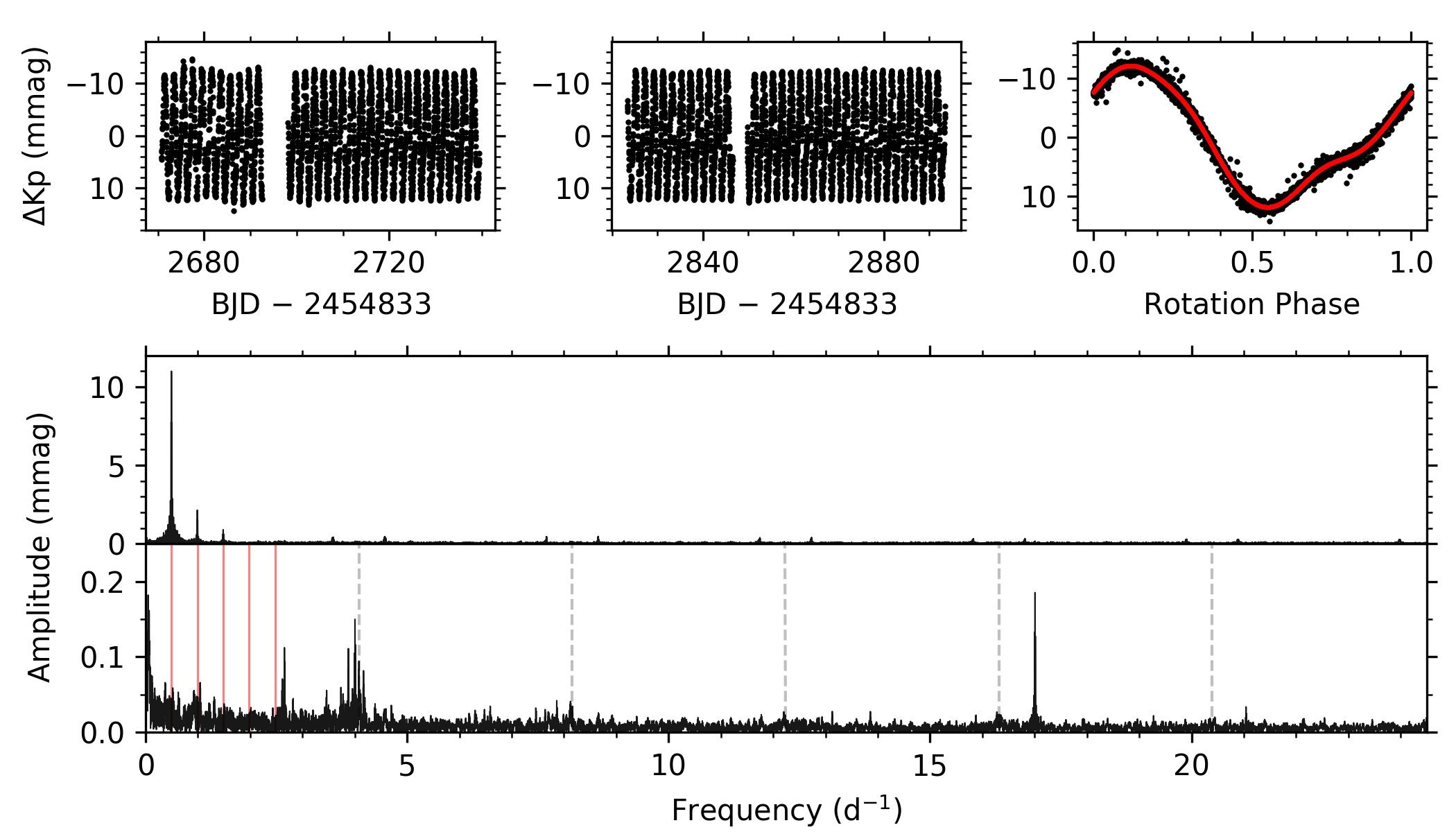}
\caption{Rotational modulation and additional variability indicative of stellar pulsations in EPIC~225990054 (HD~158596); same layout shown as in Fig.~\ref{figure: example}.}
\label{figure: EPIC225990054}
\end{figure*}

\clearpage 

\begin{figure*}
\centering
\includegraphics[width=0.95\textwidth]{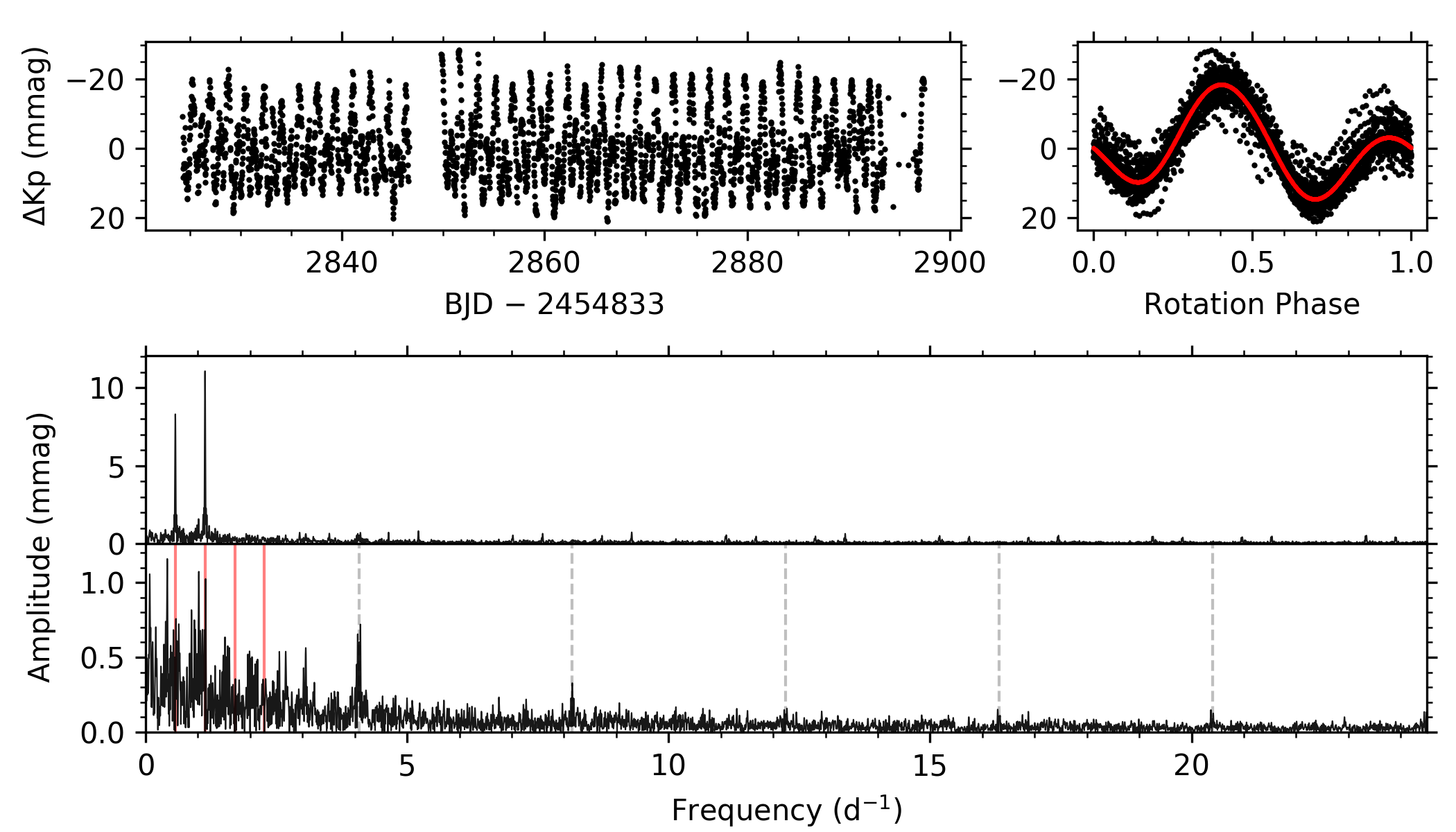}
\caption{Rotational modulation and additional variability indicative of stellar pulsations in EPIC~227231984 (HD~158336); same layout shown as in Fig.~\ref{figure: example}.}
\label{figure: EPIC227231984}
\end{figure*}

\begin{figure*}
\centering
\includegraphics[width=0.95\textwidth]{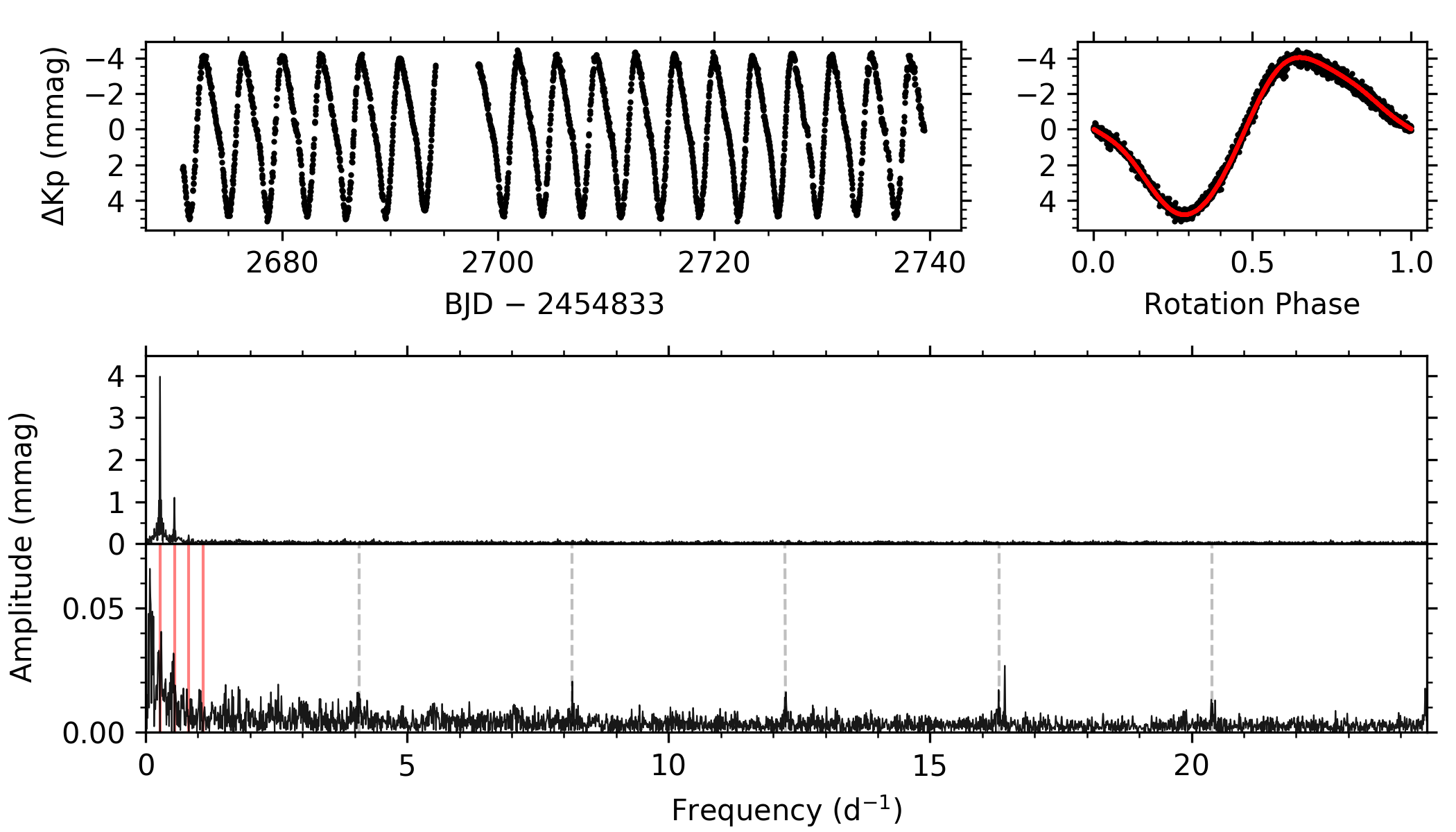}
\caption{Rotational modulation and additional variability indicative of stellar pulsations in EPIC~227305488 (HD~166542); same layout shown as in Fig.~\ref{figure: example}.}
\label{figure: EPIC227305488}
\end{figure*}

\clearpage 

\begin{figure*}
\centering
\includegraphics[width=0.95\textwidth]{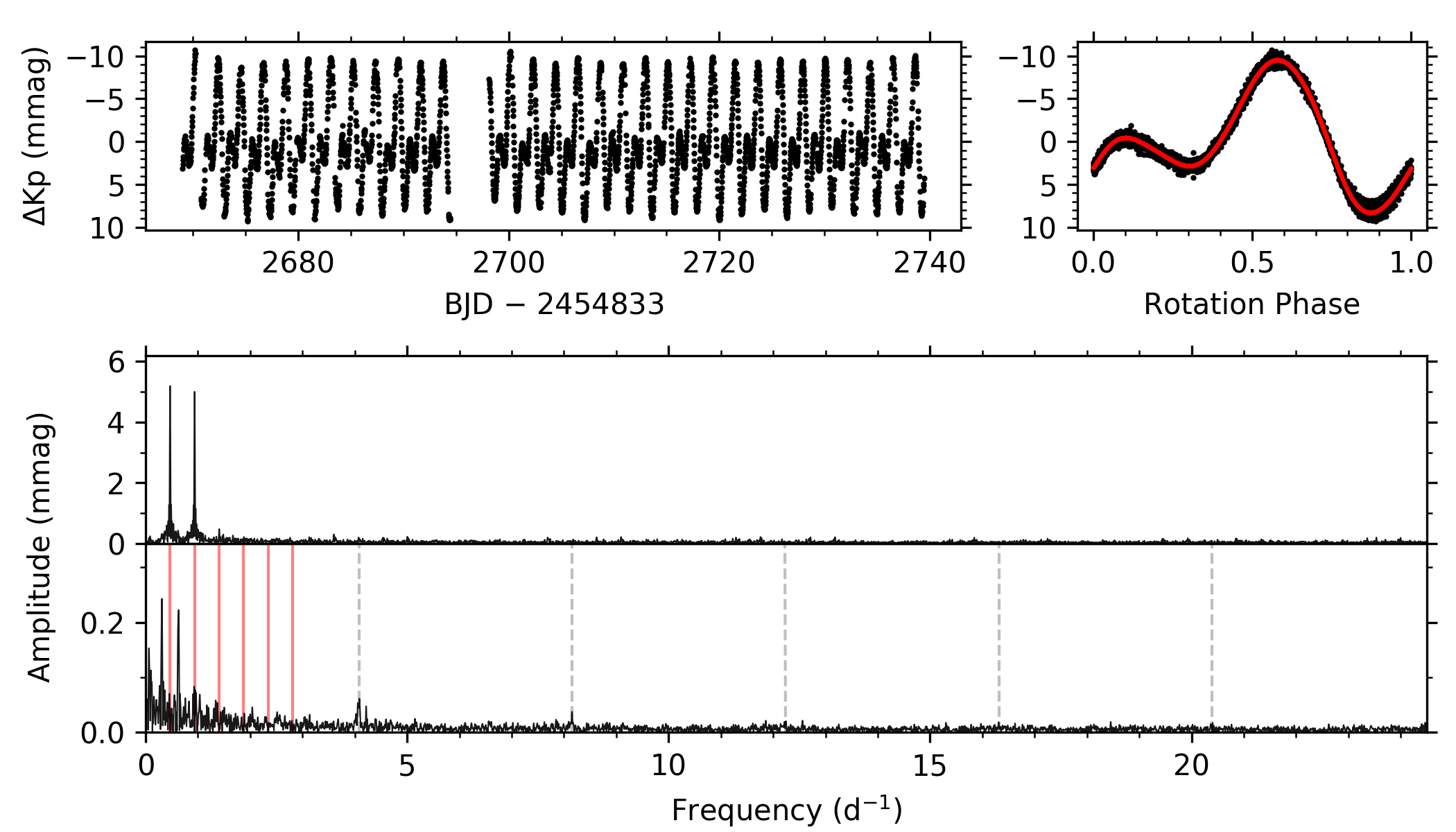}
\caption{Rotational modulation and additional variability indicative of stellar pulsations in EPIC~228293755 (HD~165945); same layout shown as in Fig.~\ref{figure: example}.}
\label{figure: EPIC228293755}
\end{figure*}

\begin{figure*}
\centering
\includegraphics[width=0.95\textwidth]{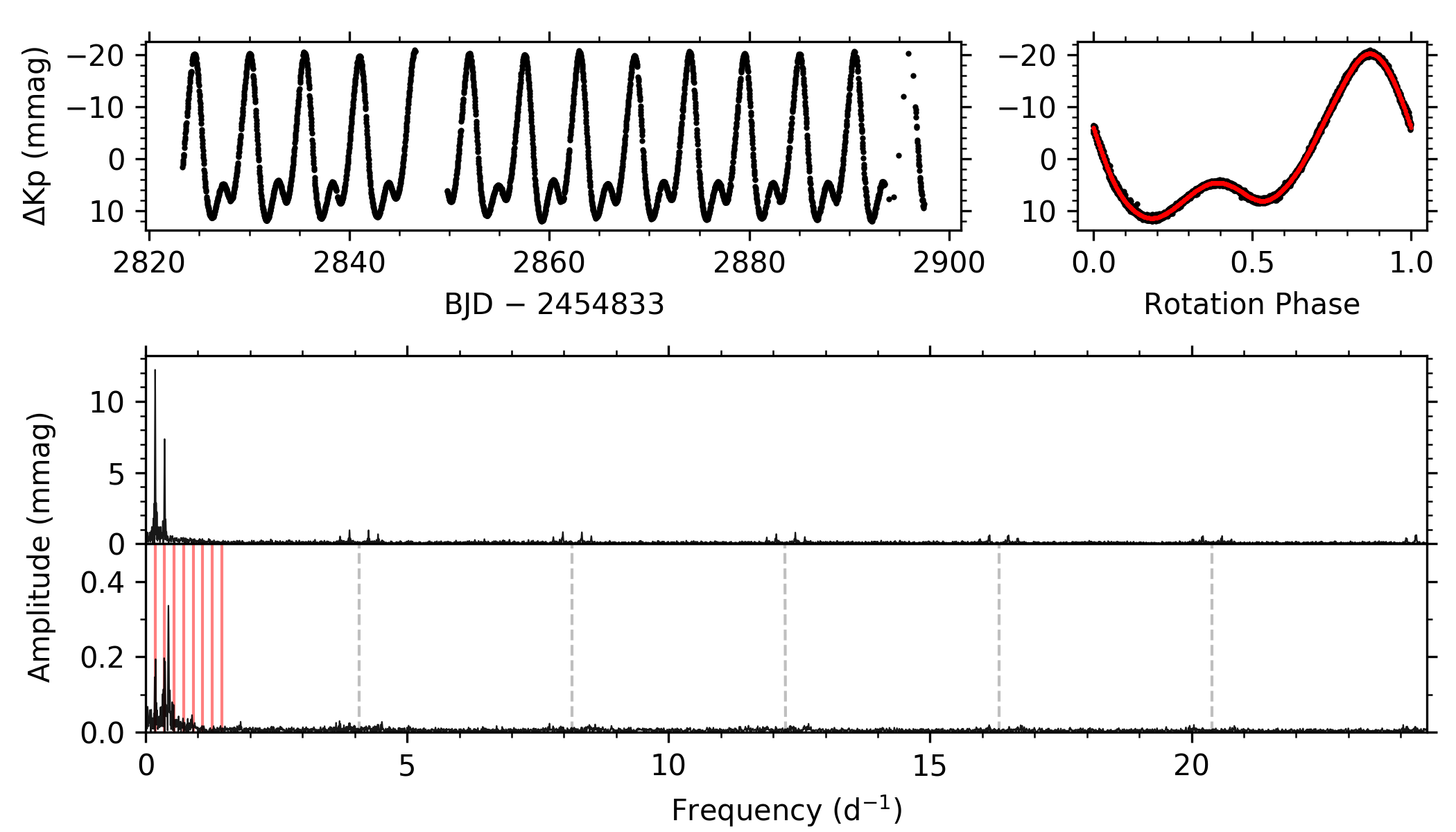}
\caption{Rotational modulation and additional variability indicative of stellar pulsations in EPIC~230753303 (HD~153997); same layout shown as in Fig.~\ref{figure: example}.}
\label{figure: EPIC230753303}
\end{figure*}


\clearpage
\section{Pulsating CP stars that lack rotational modulation}
\label{section: appendix: pulsation}

In this section, the light curves and amplitude spectra of stars that have variability indicative of pulsations, yet lack rotational modulation, are provided.

\begin{figure*}
\centering
\includegraphics[width=0.95\textwidth]{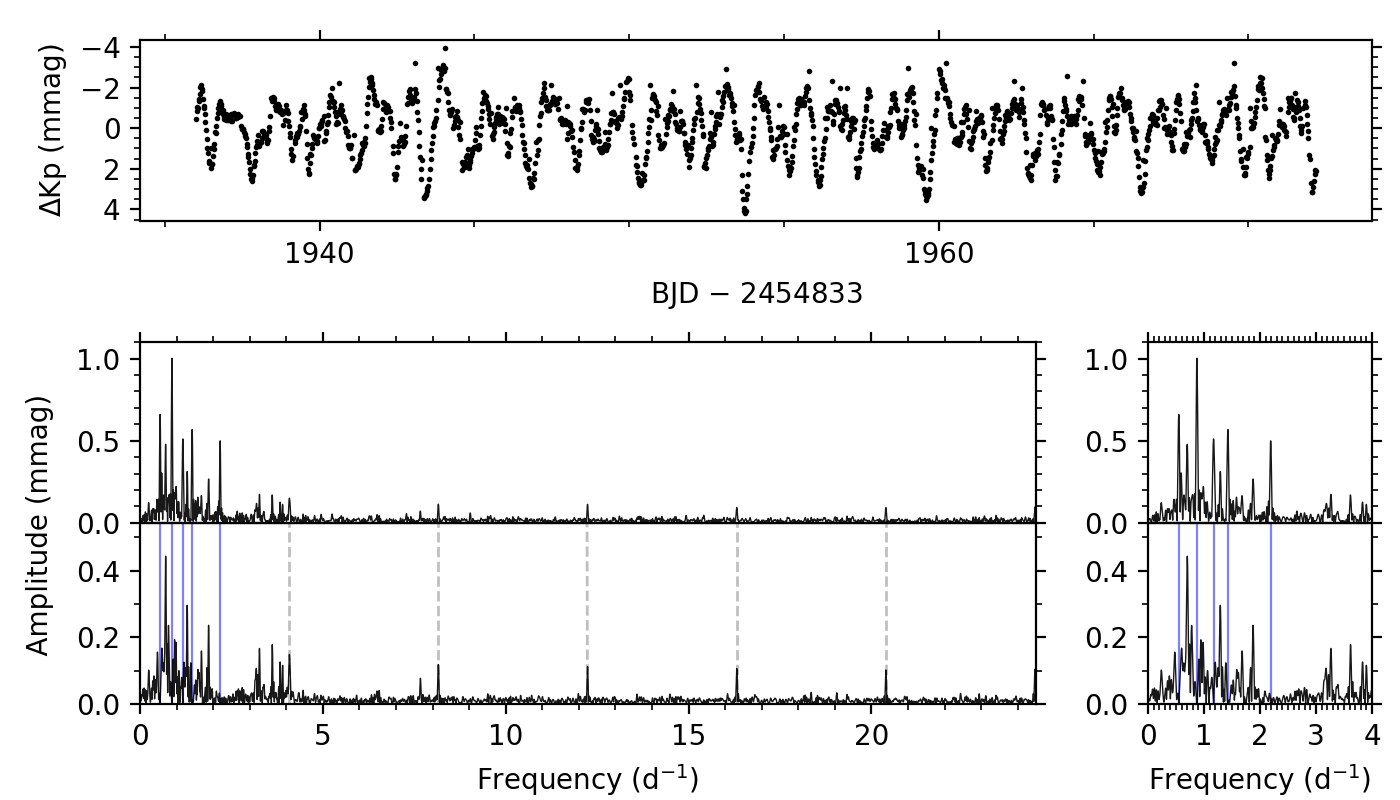}
\caption{Summary figure for EPIC~202060145. The top panel shows the detrended K2 light curve. The bottom left panel shows the amplitude spectrum calculated up to the K2 LC Nyquist frequency of 24.47~d$^{-1}$;  the residual amplitude spectrum is calculated after pulsation mode frequencies (shown as solid blue lines) have been removed. We note the change in ordinate scale. The dashed grey lines indicate multiples of the K2 thruster firing frequency. The bottom right panel shows a zoom-in of the low-frequency range.}
\label{figure: EPIC202060145}
\end{figure*}

\begin{figure*}
\centering
\includegraphics[width=0.95\textwidth]{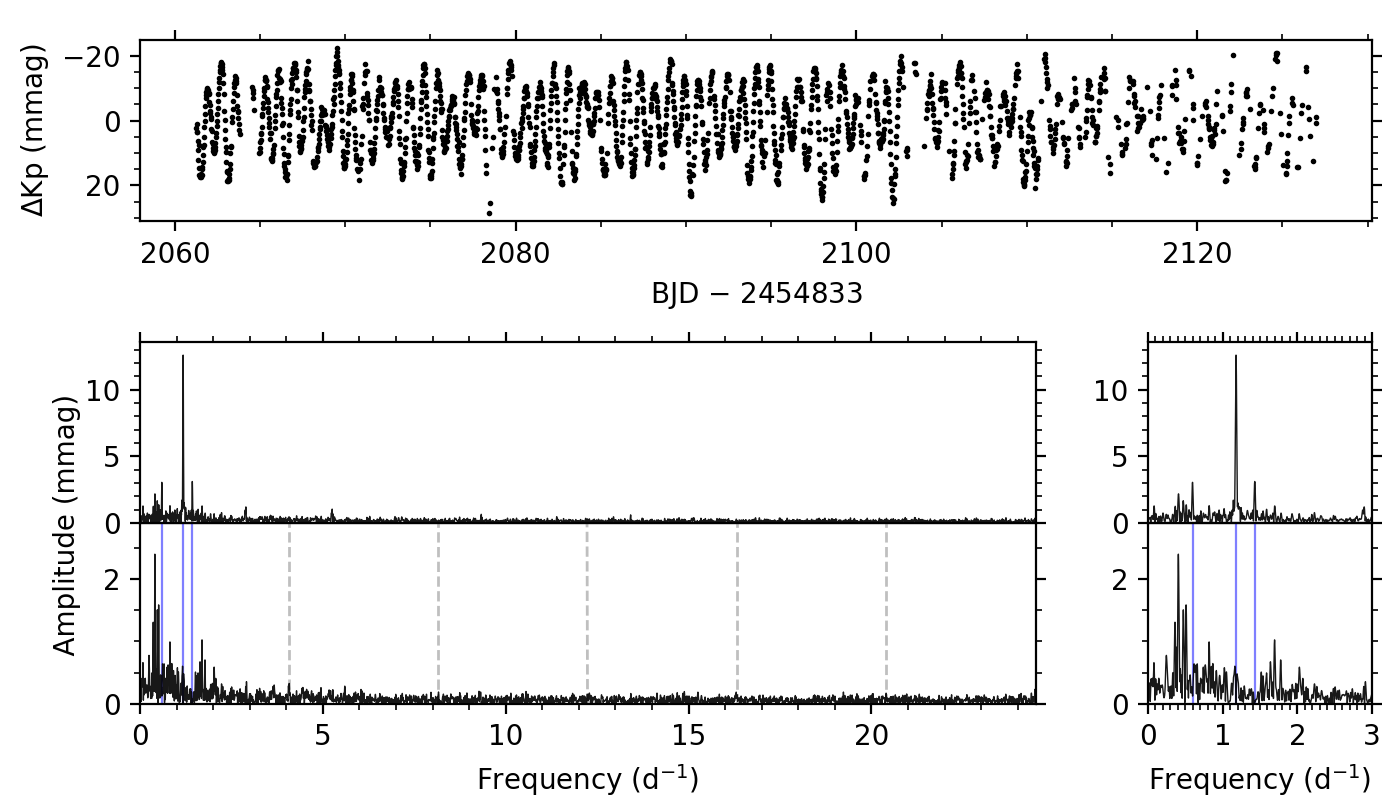}
\caption{Summary figure for EPIC~203917770 (HD~145792); similar layout shown as in Fig.~\ref{figure: EPIC202060145}.}
\label{figure: EPIC203917770}
\end{figure*}

\clearpage 

\begin{figure*}
\centering
\includegraphics[width=0.95\textwidth]{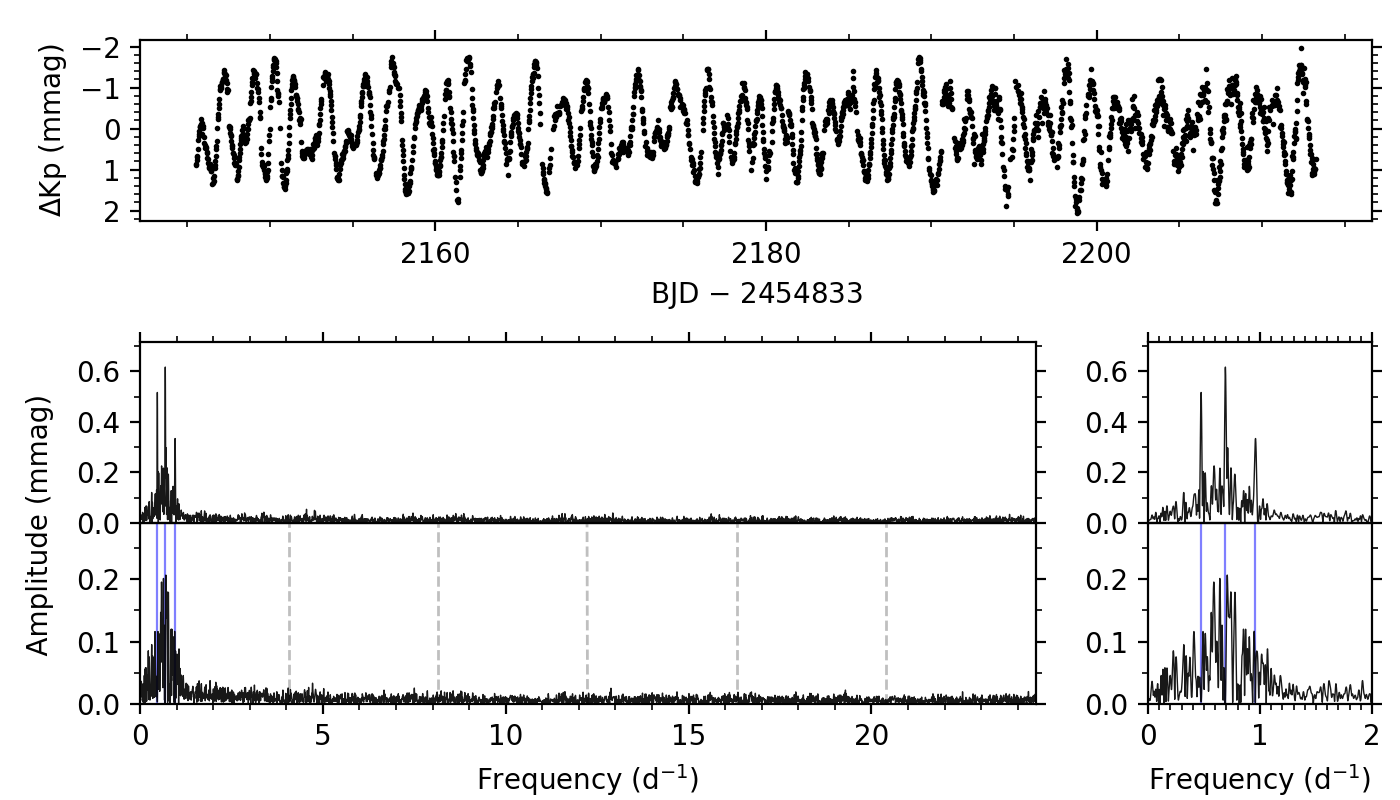}
\caption{Summary figure for EPIC~206120416 (HD~210424); similar layout shown as in Fig.~\ref{figure: EPIC202060145}.}
\label{figure: EPIC206120416}
\end{figure*}

\begin{figure*}
\centering
\includegraphics[width=0.95\textwidth]{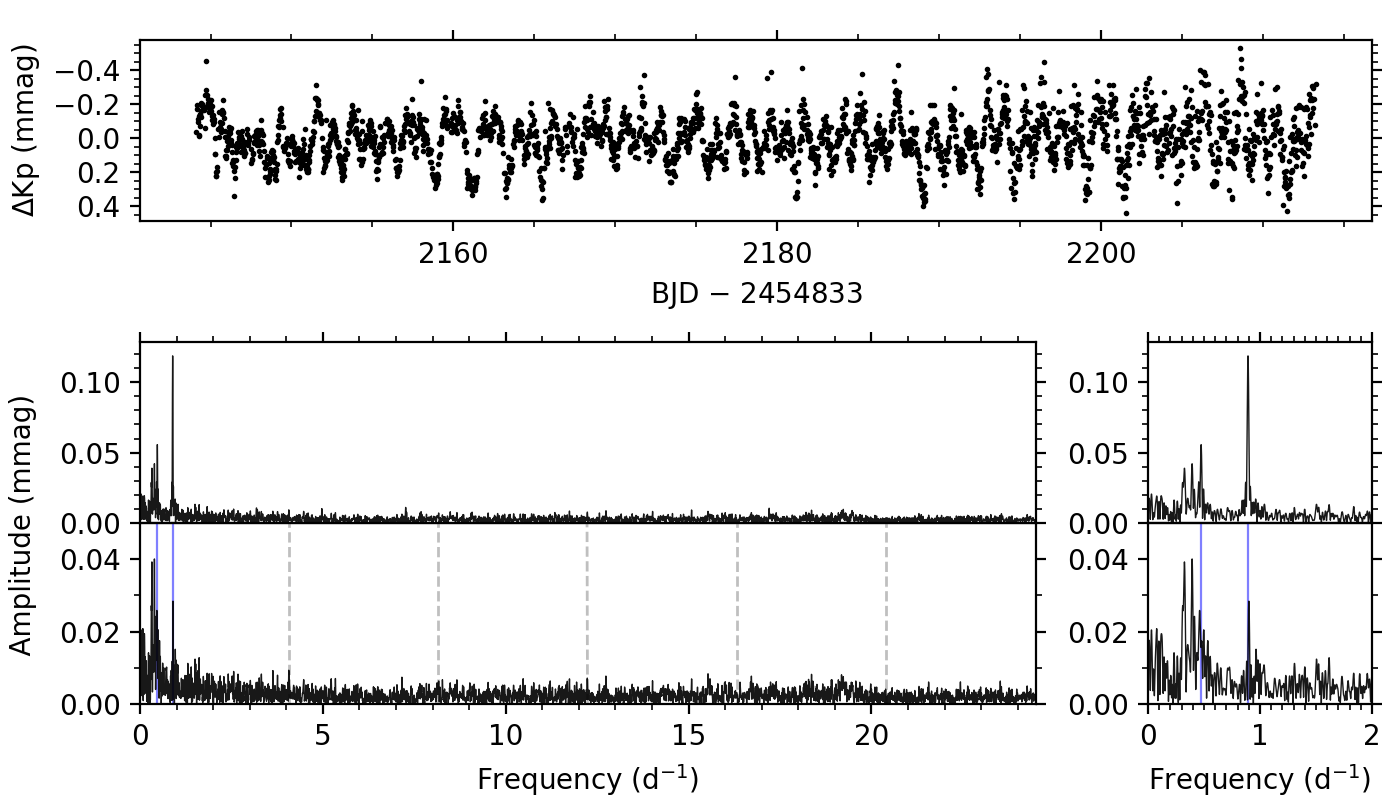}
\caption{Summary figure for EPIC~206326769 (HD~211838); similar layout shown as in Fig.~\ref{figure: EPIC202060145}.}
\label{figure: EPIC206326769}
\end{figure*}

\clearpage 

\begin{figure*}
\centering
\includegraphics[width=0.95\textwidth]{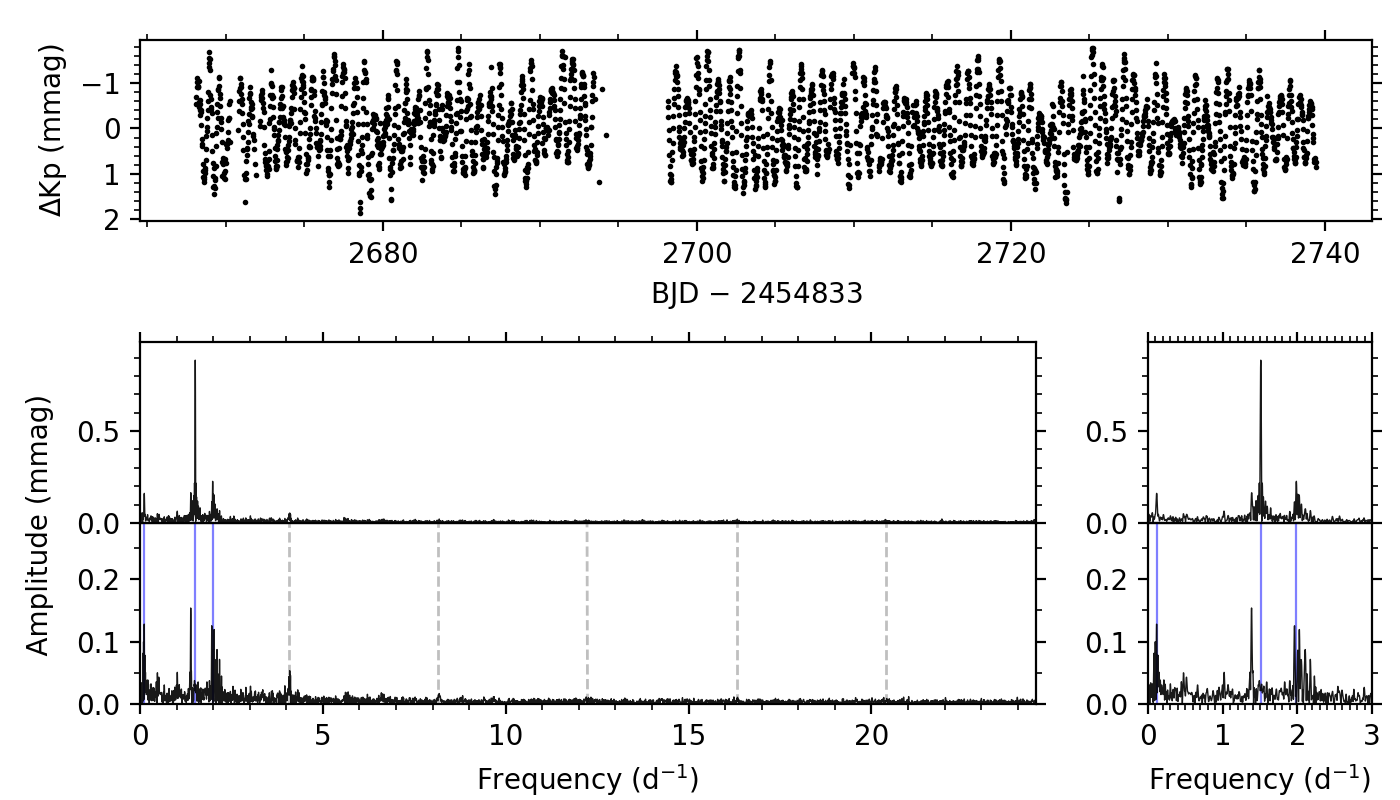}
\caption{Summary figure for EPIC~224947037 (HD~162814); similar layout shown as in Fig.~\ref{figure: EPIC202060145}.}
\label{figure: EPIC224947037}
\end{figure*}

\begin{figure*}
\centering
\includegraphics[width=0.95\textwidth]{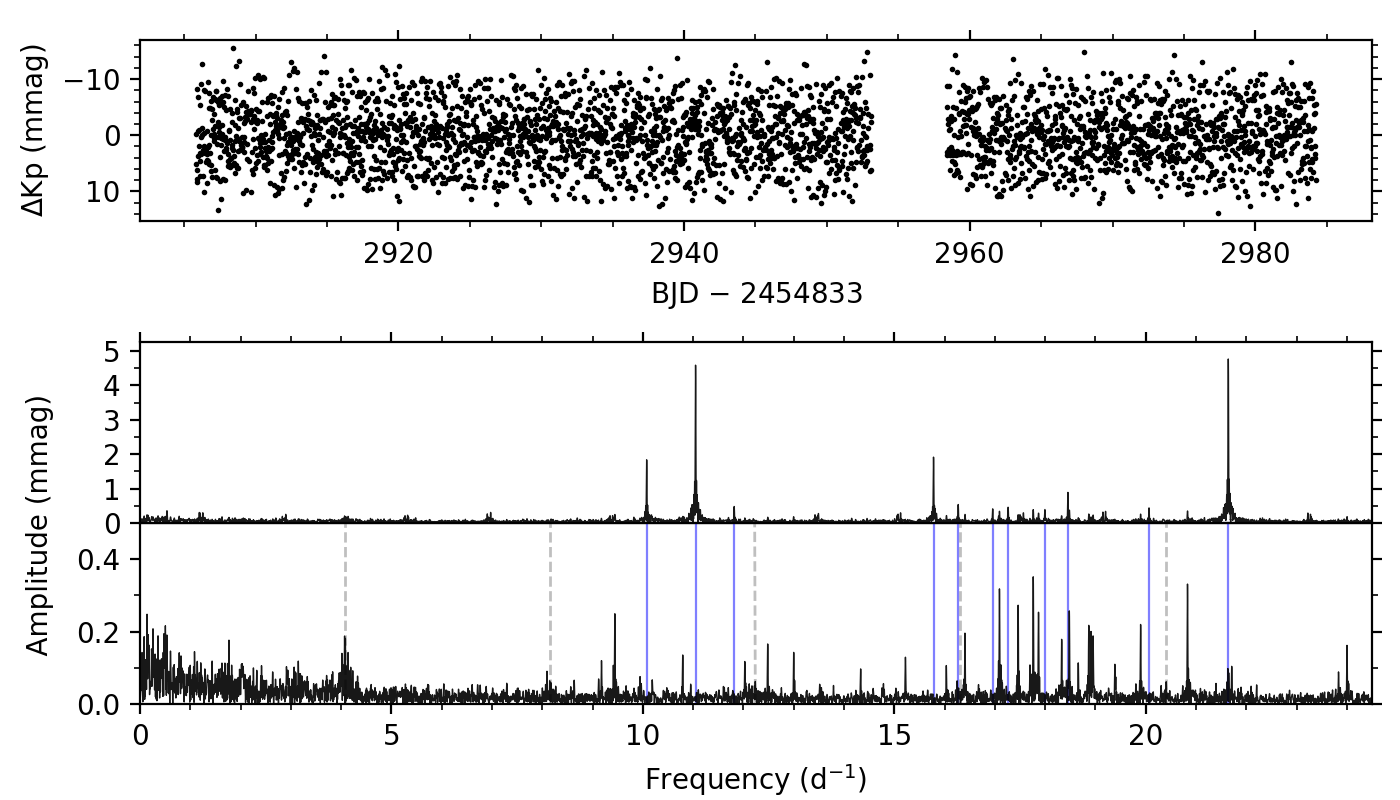}
\caption{Summary figure for EPIC~246152326 (HD~220556); similar layout shown as in Fig.~\ref{figure: EPIC202060145}.}
\label{figure: EPIC246152326}
\end{figure*}


\end{appendix}


\end{document}